# A Near-Optimal Distributed Fully Dynamic Algorithm for Maintaining Sparse Spanners

Michael Elkin [*]

**Abstract**

Currently, there are no known explicit algorithms for the great majority of problems in the *dynamic distributed message-passing model*. Instead, most state-of-the-art dynamic distributed algorithms are constructed by composing a static algorithm for the problem at hand with a simulation technique that converts static algorithms to dynamic ones. We argue that this powerful methodology does not provide satisfactory solutions for many important dynamic distributed problems, and this necessitates developing algorithms for these problems *from scratch*.

In this paper we develop the *first fully dynamic* distributed algorithm for maintaining *sparse spanners*. Our algorithm improves drastically the quiescence time of the state-of-the-art algorithm for the problem. Moreover, we show that the quiescence time of our algorithm is optimal up to a small constant factor. In addition, our algorithm improves significantly upon the state-of-the-art algorithm in *all efficiency parameters*, specifically, it has smaller quiescence message and space complexities, and smaller local processing time.

Finally, we use our technique to improve the state-of-the-art *streaming* algorithm for constructing sparse spanners, and to devise an efficient fully dynamic *centralized* algorithm for maintaining sparse spanners.

[*]Department of Computer Science, Ben-Gurion University of the Negev, Beer-Sheva, Israel, `elkinm@cs.bgu.ac.il`

# 1 Introduction

## 1.1 Discussion and Main Results

In the *message-passing model* of distributed computing a communication network is modeled by an unweighted graph, and each vertex of this graph hosts a processor. The vertices communicate through network links, modeled by graph edges. Devising algorithms for the message-passing model (henceforth, *distributed algorithms*) is an active area of research [1]-[3],[7] - [15], [26, 29, 30, 31, 32]. (See [30] for an excellent survey.)

The main motivation for the study of algorithmic problems in this model is the urge to devise real-life protocols for the growing body of communication networks, particularly the Internet. However, ever since the theoretic study of distributed algorithms started in the early eighties, it was commonly understood [8, 15, 1, 2, 9, 11, 3] that real-life networks are inherently *dynamic*, i.e., the communication links can crash and revive at will. The necessity to devise algorithms that are sufficiently robust to work properly in an unstable environment became even more apparent in view of such recent developments in the world of telecommunications as *ad-hoc, sensor, and wireless networks*, which are all inherently unstable.

Moreover, the resurgence of these novel network architectures dictates certain limitations on algorithms that can be used. Particularly, these algorithms have to use *limited computational resources*, such as local processing time and space, just because the processors in these networks may well be *uncapable* of undertaking a heavy computational task. Even more importantly, these algorithms need to be *simple*, due to both computational and economical limitations of processors that are supposed to execute them. (By economical limitations we mean the necessity to keep the overall manufacturing cost of these processors low.)

Currently, there are no known explicit dynamic distributed algorithms for the great majority of distributed problems. Instead, most dynamic distributed algorithms are constructed by composing a static distributed algorithm for the problem at hand with a simulation technique that converts static algorithms to dynamic ones. (There are important exceptions, particularly, [3, 8].) However, the currently known simulation techniques all suffer from significant drawbacks. The simulation technique of [1, 13] has super-linear in the number of vertices $n$ time and space overheads; the technique of [15] has super-linear in $n$ message and space overheads. Both these techniques entail a large overhead in local processing.

Finally, the simulation technique of Awerbuch et al. [11] has only polylogarithmic time, message, and space overheads, but entails a large overhead in local processing. More importantly, the technique of [11] is extremely complicated, and not self-contained. Particularly, it uses a *reset procedure* from [1] or from [12], and both of them are rather complex by themselves. In addition, it uses an algorithm for constructing *sparse neighborhood covers*, based on the algorithm of Awerbuch and Peleg [13], but more involved than the latter. Also, to adapt the algorithms for constructing these neighborhood covers to an asynchronous network, a *bootstrap* technique is employed [13]. Finally, on top of it, an involved variant of a *local rollback* algorithm of Awerbuch and Sipser [15] is used.

This complex nature of the algorithm of [11] makes it unsuitable for running on simple network devices. Moreover, the paper of [11] was published only as an extended abstract, and no full version of this paper ever appeared. Consequently, whoever would wish to implement the algorithm of [11] in practice will have to fill in a significant body of missing details. From the theoretical viewpoint, though the work of [11] was a conceptual breakthrough in its approach to simulation, using their result as a blackbox is problematic because the only available version of the paper [11] contains only a sketch of the proof of their (very strong and general) result.

These considerations raise the necessity to develop dynamic distributed algorithms for basic algorithmic problems *from scratch*, or to develop alternative simulation techniques. In this paper we follow the first avenue, and develop an *extremely efficient* fully dynamic distributed algorithm for maintaining sparse *spanners*. Informally, graph *spanners* can be thought of as sparse skeletons of communication networks that approximate to a significant extent the metric properties of the respective networks. Spanners serve as an underlying graph-theoretic construct for a great variety of distributed algorithms. Their most prominent applications in this context include *synchronization* [7, 31, 13, 11], *routing* [32, 14, 35], *approximate distance computation* [21, 22], and online load balancing [10].

The state-of-the-art distributed *static* algorithm for constructing sparse spanners is a distributed variant of the algorithm of Baswana and Sen [17]. For an arbitrary positive integer parameter $t$, and arbitrary undirected unweighted $n$-vertex graph, this algorithm constructs $(2t-1)$-spanner with expected $O(t \cdot n^{1+1/t})$ number of edges. (This tradeoff is optimal except for the factor of $t$ in the estimate on the number of edges.) The running time of this algorithm is $O(t)$, its message complexity is $O(|E| \cdot t)$, and its space requirement for a vertex $v$ running the algorithm is $O(deg(v) \cdot \log n)$ bits, where $deg(v)$ is the degree of the vertex $v$.

The common way to measure performance of distributed dynamic algorithms is through *quiescence* complexities. Specifically, one assumes that all topology updates (link crashes and appearances) stop occurring at a certain time $\alpha$, and measures how long is the time period, and how many messages are sent during this time period, during which the algorithm *stabilizes*, that is, reaches a state in which the (distributed) structure that it maintains starts again to satisfy the properties of the problem at hand. This time period is called the *quiescence time complexity* of the the algorithm, and the number of messages as above is called the *quiescence message complexity* of the algorithm.

Running the algorithm of Baswana and Sen [17] on top of the simulation technique of Awerbuch et al. [11] results in a dynamic distributed algorithm for maintaining a $(2t-1)$-spanner of expected size $O(t \cdot n^{1+1/t})$ with quiescence time of $O(t \cdot \log^3 n)$, quiescence message complexity of $O(t \cdot |E| \cdot \log^3 n)$, and space requirement of $O(deg(v) \cdot \log^4 n)$.

We devise a dynamic distributed algorithm for maintaining a $(2t-1)$-spanner of the same expected size that improves the result of [17, 11] described above in *all measures of efficiency*. Particularly, our result reduces *drastically the quiescence time* from $O(t \cdot \log^3 n)$ to $3t$. (In many cases the parameter $t$ is a universal constant, and it is always true that $t = O(\log n)$. Also, although we were not able to explicate the constant hidden by the $O$-notation in the $O(\log^3 n)$ time overhead of the dynamization result of [11], to the best of our understanding it is at least 32, that is, at least ten times larger than the leading constant factor in the quiescence time complexity of our algorithm.) The quiescence message complexity of our algorithm is $O(t \cdot |E|)$, and its space requirement is $O(deg(v) \cdot \log n)$ bits (for a vertex $v$). Note that the latter is optimal up to a constant factor as long as every vertex maintains the identity number of each of its neighbors, and, moreover, since every vertex $v$ must maintain at least one bit for each edge $e$ adjacent to $v$ indicating whether the edge $e$ is up or down, it follows that this space requirement is optimal up to a logarithmic in $n$ factor (under no assumptions).

Moreover, in both algorithms the vertices work in *rounds* or *computational cycles* even in the asynchronous setting (more details about the model appear Section 2), and on each round each vertex $v$ performs a certain local computation for (virtually) every edge adjacent to $v$. In our algorithm this local computation is extremely efficient, and requires $O(1)$ expected time, $O\left(\sqrt{\frac{\log deg(v)}{\log \log deg(v)}}\right)$ time in the worst-case, and $O\left(\sqrt{\frac{\log \log n + (\log n)/t}{\log(\log \log n + (\log n)/t)}}\right)$ time with high probability.

It is not clear to us what is the processing time-per-edge in the algorithm of [17, 11], because no explicit bound is provided in [11] for the local processing time overhead for the simulation algorithm described therein. However, the latter technique employs the so-called *local rollback* method that maintains parts of the communication history, and undoes certain operations from this history upon a topology update. This technique appears to be quite costly in terms of local computation, and consequently, it seems safe to assume that the local processing time-per-edge of the algorithm of [17, 11] is nowhere as low as that of our algorithm. See Table 1 for a concise comparison between the different efficiency measures of our algorithm with that of [17, 11].

|        | Time Complexity | Message Complexity | Space Requirement | Local Processing Time-per-edge |
|--------|-----------------|--------------------|--------------------|-------------------------------|
| [17, 11] | $O(t \cdot \log^3 n)$ | $O(t|E| \cdot \log^3 n)$ | $O(deg(v) \cdot \log^4 n)$ | Not clear |
| **New** | $3t$ | $O(t|E|)$ | $O(deg(v) \cdot \log n)$ | Expected $O(1)$ |
| l.b. | $\lfloor \frac{2}{3}t \rfloor$ | $\Omega(|E|)$ | $\Omega(deg(v))$ | $\Omega(1)$ |

Table 1: A comparison between the algorithm of [17, 11] and our new dynamic distributed algorithm. The space requirement is for a specific vertex $v$. The shortcut ".l.b." stands for "lower bounds".

We also show that the quiescence time of our algorithm is near-optimal. Specifically, we show that for any constant parameter $t$ any distributed algorithm that maintains sparse $(2t-1)$-spanners in a fully dynamic setting has quiescence time greater or equal to $\lfloor \frac{2}{3}t \rfloor$. In fact, our lower bound applies even for *static synchronous* algorithms that are allowed to send messages of an *arbitrarily large* size, while messages sent by our algorithm are all of size $O(\log n)$. Moreover, under Erdős girth conjecture our lower bound can be improved to $t-1$. We also extend the lower bound for super-constant values of $t$. Specifically, the lower bound applies as is for $1 \leq t = o\left(\sqrt{\frac{\log n}{\log \log n}}\right)$. For $t$ in the range $\Omega\left(\sqrt{\frac{\log n}{\log \log n}}\right) = t = o\left(\frac{\log n}{\log \log n}\right)$ we show that any algorithm that has the properties of our algorithm has quiescence time $\left(\frac{2}{3} - o(1)\right) t$ or greater. Finally, for the range $\omega\left(\frac{\log n}{\log \log n}\right) = t = O(\log n)$, the respective lower bound is $\Omega\left(\frac{\log n}{\log \log n}\right)$.

To summarize, our algorithm compares very favorably to the state-of-the-art benchmark algorithm that combines the dynamization technique of [11] with the distributed variant of [17]. No less important is that our algorithm is reasonably simple and self-contained[1], in contrast to the simulation algorithm of [11]. In fact, the incremental variant of our algorithm is *extremely simple*, and can be hard-wired in even the most primitive network devices. The general variant of our algorithm is more complex, but nevertheless, is amenable for implementation on most even rather simple network devices.

## 1.2 Additional Features of our Algorithm

Our algorithm has a number of additional features. First, not only that its quiescence time is at most $3t$, but if at a time $\alpha$ edges stop crashing but are still allowed to appear, still at time $\alpha + 3t$ the spanner maintained by the algorithm will provide a stretch guarantee of $2t - 1$ for all edges of the graph present in the network at time $\alpha$. (Edges that appeared at some time $\beta$, $\beta > \alpha$, will be taken care of by time $\beta + 3t$.)

Second, our algorithm behaves even better in an incremental environment. (This is an environment in which new edges may appear, but no existing edge ever crashes.) The quiescence time of our algorithm becomes $2t$ instead of $3t$ in this setting. In addition, this quiescence time decreases further if the sets of edges that appear in the network possess a certain convenient structure. Particularly, if the set $F$ of edges that appear in the network is a partial matching, then our algorithm takes care of all edges $F$ within *one single round* or *two time units* after these edges appear, depending on whether the network is synchronous or asynchronous, respectively. The size of the spanner keeps to be bounded by $O((t \cdot \log n)^{1-1/t} \cdot n^{1+1/t})$ with high probability all through this process. Note that the set $F$ may contain as many as $n/2$ edges appearing all around the network, and moreover, the topology updates may arrive in *multiple bursts*. Nevertheless, they are processed by the algorithm *on the fly*! This result extends also to more complex topology update edge sets than partial matchings. Specifically, if the update edge set $F$ has maximum degree $\Delta$ then it is processed by the algorithm within $2\Delta$ rounds or time units.

---
[1]Except for a number of data structures of Beame and Fich [18], Andersson and Thorup [5], and Dietzfilbinger et al. [20], that we use to achieve a very low local processing time-per-edge. Without using these data structures our local processing time-per-edge would grow slightly, but it will still be expected $O(1)$, worst-case $O(\log deg(v))$, and with high probability $O((\log n)/t + \log \log n)$. Other complexity measures of the algorithm would stay unchanged.

Finally, we show stronger bounds on the quiescence time than our general bounds even if most, but not all, the edges of the network are allowed to crash. (Edges are still allowed to appear at will though.) Specifically, we identify a certain set of at most $n \cdot (t-1)$ *backbone* edges in the spanner constructed by the algorithm. Suppose that backbone edges do not crash for a reasonably long period of time. Then an arbitrary set of non-backbone edges may crash, and the algorithm takes care of their crashes within *one round* or *two time units* in synchronous and asynchronous networks, respectively. In other words, once again the spanner adapts to crashes of the non-backbone edges *on the fly*! Moreover, we show that for any particular edge of the network, it has only a probability of roughly $n^{-1/t}$ to belong to the backbone. Hence every edge of the original network is much more likely than not to be a non-backbone edge, and this makes the assumption that backbone edges do not crash weaker and more realistic. We remark, however, that the bound of $3t$ on the quiescence time of our algorithm requires no assumptions. The assumptions are needed only if one is interested in yet stronger bounds.

## 1.3 Additional Results

### 1.3.1 The Streaming Model

A variant of our algorithm improves significantly the state-of-the-art algorithm of Feigenbaum et al. [24] for constructing sparse spanners and approximate all-pairs-distance-computation (henceforth, *APDC*) in the streaming model of computation.

The study of the streaming model became an important research area after the seminal papers of Alon, Matias and Szegedy [4], and Feigenbaum et al. [25] were published. More recently, research in the streaming model was extended to traditional graph problems [16, 23, 22, 24]. The input to a graph algorithm in the streaming model is a sequence (or *stream*) of edges representing the edge set $E$ of the graph. This sequence can be an arbitrary permutation of the edge set $E$.

The state-of-the-art streaming algorithm for computing a sparse spanner for an input (unweighted undirected) $n$-vertex graph $G = (V, E)$ was presented in a recent breakthrough paper of Feigenbaum et al. [24]. Their algorithm, with high probability, constructs a $(2t+1)$-spanner with $O(t \cdot \log n \cdot n^{1+1/t})$ edges *in one pass* over the input using $O(t \cdot \log^2 n \cdot n^{1+1/t})$ bits of space. It processes each edge in the stream in time $O(t^2 \cdot \log n \cdot n^{1/t})$. Their result also immediately gives rise to a streaming algorithm for $(2t+1)$-approximate APDC algorithm with the same parameters.

Our algorithm constructs a $(2t-1)$-spanner with $O((t \cdot \log n)^{1-1/t} \cdot n^{1+1/t})$ edges in one pass over the input using $O(t^{1-1/t} \cdot \log^{2-1/t} n \cdot n^{1+1/t})$ bits of space. (The size of the spanner and the number of bits are with high probability.) Most importantly, the *processing time-per-edge* of our algorithm is *drastically smaller* than that of Feigenbaum et al. . Specifically, the expectation of the processing time-per-edge in our algorithm is $O(1)$, it is $O\left(\sqrt{\frac{\log \log n}{\log^{(3)} n}}\right)$ with high probability, and in the worst-case processing an edge $e = (v, u)$ requires $O\left(\sqrt{\frac{\log deg(e)}{\log \log deg(e)}}\right)$, where $deg(e) = \max\{deg(v), deg(u)\}$.

To summarize, our algorithm constructs a spanner with a slightly better stretch guarantee ($2t-1$ instead of $2t+1$), with a slightly smaller number of edges and number of bits of space used (by a factor of $(t \cdot \log n)^{1/t}$, and it does so using a *drastically* reduced (and very close to optimal) *processing time-per-edge*, *at no price whatsoever*. Our result also gives rise to an improved streaming $(2t - 1)$-approximate APDC algorithm with the same parameters. A concise comparison of our result with the state-of-the-art result of Feigenbaum et al. [24] can be found in Table 2.

|  | # passes | Processing time-per-edge | Stretch | # Edges (whp) |
|---|---|---|---|---|
| [24] | 1 | whp $O(t^2 \cdot \log n \cdot n^{1/t})$ | $2t+1$ | $O(t \cdot \log n \cdot n^{1+1/t})$ |
| **New** | 1 | Expected $O(1)$, whp $O\left(\sqrt{\frac{\log \log n}{\log^{(3)} n}}\right)$, worst-case $O\left(\sqrt{\frac{\log deg(e)}{\log \log deg(e)}}\right)$ | $2t-1$ | $O((t \cdot \log n)^{1-1/t} \cdot n^{1+1/t})$ |

Table 2: A comparison between the algorithm of Feigenbaum et al. [24] and our new streaming algorithm. The degree of an edge $e = (u, v)$ is $\max\{deg(v), deg(u)\}$. The word "whp" stands for "with high probability".

### 1.3.2 A Centralized Fully Dynamic Algorithm

Finally, a variant of our algorithm can be seen as a fully dynamic centralized (versus distributed) algorithm for maintaining a $(2t-1)$-spanner with $O((t \cdot \log n)^{1-1/t} n^{1+1/t})$ edges. The incremental update time of this algorithm is exactly the processing time-per-edge of our streaming algorithm, and, in fact, the way that the dynamic algorithm processes incremental updates is identical to the way that our streaming algorithm processes each edge of the stream. The expected decremental update time (the time required to update the data structures of the algorithm when an edge $e$ is deleted) is $O(\frac{m}{n^{1/t}} \cdot (t \log n)^{1/t})$, and moreover, with probability at least $1 - \left(\frac{t \cdot \log n}{n}\right)^{1/t}$, the decremental update time is $O\left(\sqrt{\frac{h}{\log h}}\right)$, where $h = \max\{\log deg(e), \log \log n\}$. The size of the data structures maintained by an incremental variant of our algorithm is $O(t^{1-1/t} \cdot (\log n)^{2-1/t} \cdot n^{1+1/t})$ bits. To cope with decremental updates as well, our algorithm needs to maintain $O(|E| \cdot \log n)$ bits. (Note that the incremental algorithm maintains data structures of overall size *sublinear* in the size of the input $|E|$. This is not really surprising, since this is essentially a streaming algorithm.)

To our knowledge, this is the first fully dynamic centralized algorithm for maintaining sparse spanners that provides non-trivial bounds for a wide range of stretch parameter $t$. The only previously known algorithm of this type is a recent result due to Ausillo et al. [6], who devised a fully dynamic algorithm for maintaining 3- and 5-spanners of optimal size with $O(n)$ amortized time per operation for an intermixed sequence of $\Omega(n)$ edge insertions and deletions.

### 1.4 Our Techniques

Our basic static distributed algorithm combines the techniques of Feigenbaum et al. [24], and of Baswana and Sen [17], with some new ideas. The extension of the algorithm to dynamic setting requires, however, a novel approach to handle edge crashes. The standard approach is a rollback technique [8, 1, 15, 9, 12, 11, 3], which makes each vertex to maintain some part of the history of communication, and to undo certain operations from this history once an edge crash occurs. Our approach is mainly based on looking for a "replacement" for every crashing edge. Our algorithm also undoes certain operations in some cases (when backbone edges crash), but the history of communication is never explicitly maintained, but rather the list of operations to be undone is *deduced from the current state of affairs*.

We believe that this method of undoing operations without maintaining an explicit history of communication is the key to the extreme robustness and efficiency of our algorithm, and is our main technical contribution. In this paper we demonstrate the power of this method by using it to improve the state-of-the-art algorithms for constructing spanners in three different computational models. Specifically, these models are the dynamic distributed, streaming and dynamic centralized ones. It is plausible that this method will be found useful for other dynamic problems.

## 1.5 The Structure of the Paper

Terminology and definitions that are commonly used throughout the paper are presented in Section 2. Section 3 is devoted to the streaming model of computation. In this section we also do some groundwork for the analysis of our algorithm in other computational models. In Section 4 we view the algorithm of Section 3 as a centralized dynamic algorithm, and extend the analysis of Section 3 to this setting. In Section 5 we turn our attention to distributed models of computations, which are the main subject of this paper. Sections 5 - 10 are all devoted to this topic.

In Section 5 we present and analyze a static synchronous distributed algorithm for constructing sparse spanners. This algorithm is very simple, but nevertheless, very efficient, and it illustrates some of the main ideas used in all other algorithms presented in the paper. In Section 6 this algorithm and its analysis are extended to the incremental setting. Section 7 is devoted to the issue of asynchrony, and there we argue that all our algorithms extend to the asynchronous setting. Sections 8 and 9 extend the algorithm of Section 7 to the distributed fully dynamic setting. The main topic of these sections is coping with edge crashes. This is done in two stages. In Section 8 we assume that backbone edges never crash, and in Section 9 this assumption is abandoned. Section 10 is devoted to the lower bounds.

## 2 Preliminaries

In this section we introduce the computational models and basic terminology.

In the message-passing model a communication network at hand is modeled by an undirected unweighted $n$-vertex graph $G = (V, E)$. Each vertex $v \in V$ hosts a processor. The processors share no common memory, and they communicate via edges of the graph $G$. Specifically, a processor $v$ can send and receive messages over each edge $e = (v, u)$ adjacent to the vertex $v$ in $G$. Communication links of the network are assumed to have a limited capacity, and this is modeled by assuming that the size of each single message is at most $O(\log n)$. Vertices are assumed to have distinct identifiers from the range $\{1, 2 \ldots, n\}$.

In the *synchronous model* the communication occurs in *discrete rounds*. On each round each vertex $v$ is allowed to send messages to all its neighbors in $G$. A message sent by a vertex $v$ over an edge $e = (v, u)$ on round $R$ arrives to its destination $u$ before round $R + 1$ starts. In the *asynchronous model* each vertex maintains its own clock, and clocks of different vertices may disagree. Nevertheless, the vertices are assumed to work locally in a manner similar to the one in which they work in the synchronous model. Specifically, on each cycle of its clock ("round") each vertex $v$ processes the messages that it has received since the last time it was in the receiving mode (the beginning of the previous clock cycle), and sends new messages that result from this processing. Messages sent from a vertex to its neighbor arrive within finite but unpredictable time period. The maximum duration of this time period is called the *time unit*.

The *time complexity* of a synchronous (respectively, asynchronous) algorithm $\mathcal{A}$ is the maximum number of rounds (resp., time units) that an execution of $\mathcal{A}$ on lasts. The *message complexity* is the maximum overall number of messages sent during an execution of $\mathcal{A}$. The *space complexity* is the maximum number of bits used by a certain vertex at a single moment during an execution of $\mathcal{A}$. The *local processing time-per-round* is the maximum duration of a time period that some vertex spends on processing data *locally* on a single clock cycle (or round). In algorithms that we will consider each vertex $v$ processes (almost) each edge adjacent to $v$ on each round, and thus it makes sense to consider also the *local processing time-per-edge*. The latter is defined as the maximum duration of a time period that some vertex spends on processing some single edge adjacent to it locally on a certain round.

In a dynamic setting edges of the graph are allowed to appear and disappear (henceforth, *crash*) at any time. The challenge in this context is to design algorithms that are robust to this ever-changing environment. An *incremental* (respectively, *decremental*) dynamic algorithm is an algorithm that can handle edge appearances (resp., crashes) but not crashes (resp., appearances). A *fully dynamic* algorithm is an algorithm that can handle both appearances and crashes. The usual way to measure performance of distributed dynamic algorithms is through *quiescence time and message* complexities, that is, the time and message requirements after the last topological change. See also the beginning of Section 1.1.

Throughout the paper we consider topology updates that involve only appearances and crashes of *edges*. Our algorithm can cope with *vertex* appearances and crashes as well. The algorithm handles a vertex crash exactly as it handles the crash of all edges adjacent to the crashing vertex. Vertex appearances are handled analogously.

We assume that the number $L$ of crashes of backbone edges is at most polynomial in $n$. If this assumption does not hold, our guarantee on the size of the spanner maintained by the algorithm would grow by a factor $\frac{\log L}{\log n}$. We believe that this limitation is not really restrictive because one can use fresh coin tosses after, say, $n^{10}$ rounds elapse. Our algorithm is sufficiently robust to guarantee that this switch to new coins does not even require to coordinate the processors (in the asynchronous scenario).

We also assume that a message is lost only if the edge through which it was sent crashes. Moreover, in our model messages sent through a fixed edge $e = (v, u)$ in a fixed direction arrive to their destination in the First-In-First-Out (FIFO) manner. This assumption can be eliminated at the expense of using a more complicated analysis.

In addition, we assume that the number of vertices $n$ is known in advance, in the beginning of computation. This assumption can be weakened so that only an upper bound $\hat{n}$ on the number of vertices wil be known at the beginning of computation, but then $\hat{n}$ would have to be plugged in all our bounds instead of $n$. Similarly to the algorithm of Feigenbaum et al. [24], there is an important special case of $t = \log n$ for which the algorithm does not need to know $n$ in advance. In this case the probability $p$ can be set as $\frac{1}{2}$ independently of $n$.

Finally, we assume that the input graph, the topology updates and their order, are all chosen by a *non-adaptive adversary* obliviously of the coin tosses of the algorithm. Although this assumption is not completely realistic, we believe that it captures many of the practical situations rather truthfully. In addition, this assumption is significantly weaker than an assumption of some stochastic model for topology updates, and the latter is very common in the application-oriented literature (see [33], and the references therein). Also, our assumption appears to be implicit in the previous work of [11, 24].

We end this section with the definition of *spanner*. For a parameter $\alpha$, $\alpha \geq 1$, a subgraph $G'$ of the graph $G = (V, E)$ is called an $\alpha$-*spanner* of $G$ if for every pair of vertices $x, y \in V$, $dist_{G'}(x, y) \leq \alpha \cdot dist_G(x, y)$, where $dist_G(u, w)$ denotes the distance between $u$ and $w$ in $G$. The parameter $\alpha$ is called the *stretch* or *distortion* parameter of the spanner. Also, for a fixed value of $t = 1, 2, \ldots$, we say that a subgraph $G'$ *spans* an edge $e = (v, u) \in E$, if $dist_{G'}(v, u) \leq 2t - 1$.

# 3 The Streaming Model

In this section we present and analyze the version of our algorithm that constructs spanners in the streaming model of computation.

## 3.1 The Algorithm

The algorithm accepts as input a *stream* of edges of the input graph $G = (V, E)$, and an integer positive parameter $t$, and constructs a $(2t-1)$-spanner $G' = (V, H)$, $H \subseteq E$, of $G$ with $O((t \cdot \log n)^{1-1/t} \cdot n^{1+1/t})$ edges using only $O(|H| \cdot \log n) = O((t \cdot \log n)^{1-1/t} \cdot n^{1+1/t})$ bits of storage space, and processing each edge in $O(1)$ expected time, in one pass over the stream. Note that the space used by the algorithm is linear in the size of the representation of the spanner. Regarding the processing time-per-edge, processing the edge $e$ requires $O\left(\sqrt{\frac{\log deg(e)}{\log \log deg(e)}}\right)$ time in the worst-case, and moreover, with high probability, the processing time-per-edge is $O\left(\sqrt{\frac{\log \log n}{\log^{(3)} n}}\right)$.

At the beginning of the execution (before the first edge of the stream arrives), the vertices of $V$ are assigned unique identifiers from the set $\{1, 2, \ldots, n\} = [n]$, $n = |V|$. (Henceforth, for any positive integer $k$, the set $\{1, 2, \ldots, k\}$ is denoted $[k]$, and the set $\{0, 1, \ldots, k\}$ is denoted $[(k)]$.) Let $I(v)$ denote the identifier of the vertex $v$. Also, as a part of preprocessing, the algorithm picks a non-negative integer *radius* $r(v)$ for every vertex $v$ of the graph from the *truncated geometric probability distribution* given by $\mathbb{P}(r = k) = p^k \cdot (1-p)$, for every $k \in [(t-2)]$, and $\mathbb{P}(r = t-1) = p^{t-1}$, with $p = \left(\frac{t \log n}{n}\right)^{1/t}$. Note that this distribution satisfies $\mathbb{P}(r \geq k+1 \mid r \geq k) = p$ for every $k \in [(t-2)]$.

We next introduce a few definitions that will be useful for the description of our algorithm. During the execution, the algorithm maintains for every vertex $v$ the variable $P(v)$, called the *label* of $v$, initialized as $I(v)$. The labels of vertices may grow as the execution proceeds, and they accept values from the set $\{1, 2, \ldots, n \cdot (t-1)\}$. A label $P$ in the range $i \cdot n + 1 \leq P < (i+1)n$, for $i \in [(t-1)]$ is said to be a label of *level* $i$; in this case we write $L(P) = i$. The value $B(P)$ is given by $B(P) = n$ if $n$ divides $P(v)$, and by $B(P) = P(v) \pmod{n}$, otherwise. This value is called the *base value* of the label $P$. The vertex $w = w_P$ such that $I(w) = B(P)$ is called the *base vertex* of the label $P$. A label $P$ is said to exist if the level $L(P)$ of $P$ is no greater than the radius of the base vertex $w_P$, i.e., $L(P) \leq r(w_P)$. The label $P$ is called *selected* if $L(P) < r(w_P)$. Note that for a label $P$ to be selected, it must satisfy $L(P) \leq t - 2$.

One of the basic primitives of the algorithm is comparing the labels. We say that the labels $P(v)$ and $P(v')$ of the vertices $v$ and $v'$, respectively, satisfy the relation $P(v) \succ P(v')$ if and only if either $P(v) > P(v')$ or $(P(v) = P(v')$ and $I(v) > I(v'))$. Note that for every two vertices $v$ and $v'$, either $P(v) \succ P(v')$ or $P(v') \succ P(v)$.

For a label $P$ of level $t-2$ or smaller,

$$\mathbb{P}(P \text{ is selected}) = \mathbb{P}(r(w_P)) \geq L(P) + 1 \mid r(w_P) \geq L(P)) = p \ .$$

**Lemma 3.1** *With high probability, the number of distinct labels of level $t-1$ that occur in the algorithm is $O(n^{1/t} \cdot (t \cdot \log n)^{1-1/t})$.*

**Proof:** Let $i \in [n]$, and let $u \in V$ be the vertex with $I(u) = i$. The probability that the label $n \cdot (t-1) + i$ occurs is equal to the probability that $r(u) = t-1$, that is, $p^{t-1}$. Hence the expected number of labels of level $t-1$ is $n \cdot p^{t-1} = O(n^{1/t} \cdot (t \cdot \log n)^{1-1/t})$. The lemma follows by Chernoff bound. ∎

We remark that the way that we define and manipulate labels is closely related to the way it is done in Feigenbaum et al. [24].

For every vertex the algorithm maintains an edge set $Sp(v)$, initialized as an empty set. During the execution the algorithm inserts some edges into $Sp(v)$, and never removes them. In other words, the sets $Sp(v)$ grow monotonely during the execution of the algorithm. It is useful to think of the sets $Sp(v)$ as divided into two disjoint subsets $T(v)$ and $X(v)$, $Sp(v) = T(v) \cup X(v)$. The set $T(v)$ is called the set of the *tree edges* of the vertex $v$, and the set $X(v)$ is called the set of the *cross edges* of the vertex $v$. During the execution the algorithm constructs implicitly a *tree cover* of the graph. The edges of this tree cover are (implicitly) maintained in the sets $T(v)$. In addition, the spanner will also contain some edges that connect different trees of the tree cover; these edges are (implicitly) maintained in the sets $X(v)$. Each edge $e$ that is (implicitly) inserted into the set $T(v)$ will also be labeled by a label of $v$ at the time of the insertion. An insertion of an edge $e = (v, u)$ into the set $T(v)$ will cause $v$ to change its label to the label of $u$ plus $n$, that is $P(u) + n$. The edge $e$ will also be labeled by this label.

In addition, for every vertex $v$ a table $M(v)$ is maintained. These tables are initially empty. Each table $M(v)$ is used to store all the base values of levels $P$ such that there exists at least one neighbor $z$ of $v$ that was labeled by $P$ at some point of the execution of the algorithm, and such that the edge $(v, z)$ was inserted into the set $X(v)$ at that point of the execution.

The algorithm itself is very simple. It iteratively invokes the Procedure *Read_Edge* on every edge of the stream, until the stream is exhausted. At this point it outputs the set $\bigcup_v Sp(v) = \bigcup_v T(v) \cup \bigcup_v X(v)$ as the resulting spanner. The Procedure *Read_Edge* accepts as input an edge $e = (u, v)$ that it is supposed to "read". The procedure finds the endpoint $x$ of the edge $e$ that has a greater label $P(x)$ (with respect to the order relation $\succ$). Suppose without loss of generality that $x = u$, i.e., $P(u) \succ P(v)$. Then the procedure tests whether $P(u)$ is a selected label. If it is, the edge $e$ is inserted into the set of tree edges $T(v)$ of $v$, and $v$ adapts the label $P(u) + n$. If $P(u)$ is not a selected label, then the procedure tests whether the base value $B(P(u))$ of the label $P(u)$ is stored in the table $M(v)$. If it is not, then the edge $e$ is inserted into the set $X(v)$ of the cross edges of $v$, and the label of $v$ does not change. If $P(u)$ is already stored in $M(v)$ then nothing needs to be done.

The pseudo-code of the Procedure *Read_Edge* is provided below. Its main difference from the description above is that the sets $X(v)$ and $T(v)$ are not maintained explicitly, but rather instead there is just one set $Sp(v)$ maintained. The reason for this difference is that we aim to present the simplest version of the algorithm for which we can prove the desired bounds. However, it is more convenient to reason about the sets $X(v)$ and $T(v)$ explicitly, rather than about the set $Sp(v)$ as a whole, and thus in the analysis we will analyze the version of the algorithm that maintains the sets $T(v)$ and $X(v)$ explicitly. (It is obvious that the two versions are equivalent.)

1: **for** all the edges $e$ of the input stream **do**
2:    invoke $Read\_Edge(e)$
3: **end for**

---

**Algorithm 1** Procedure $Read\_Edge(e = (u, v))$: the streaming algorithm for constructing a sparse $(2t - 1)$-spanner.

---

1: Let $u$ be the vertex s.t. $P(u) \succ P(v)$
2: **if** $P(u)$ is a selected label **then**
3:    $P(v) \leftarrow P(u) + n$
4:    $Sp(v) \leftarrow Sp(v) \cup \{e\}$
5: **else if** $B(P(u)) \notin M(v)$ **then**
6:    $M(v) \leftarrow M(v) \cup \{B(P(u))\}$
7:    $Sp(v) \leftarrow Sp(v) \cup \{e\}$
8: **end if**

The set $T(v)$ (resp., $X(v)$) is the set of edges inserted into the set $Sp(v)$ on line 4 (resp., 7) of the Procedure *Read_Edge*. We will say that an edge $e$ *is inserted into* $T(v)$ *(resp., $X(v)$)* if it is inserted into $Sp(v)$ on line 4 (resp., 7) of the algorithm.

Note that the Procedure *Read_Edge* is extremely simple, and the only operations that might require a super-constant time are lines 5 and 6, which require testing a membership of an element in a data structure, and an insertion of an element into the data structure if it is not there already. These operations can be implemented very efficiently in a general scenario via a balanced search tree, or a hash table. Moreover, we will also show later that with high probability, the size of each table is quite small, specifically $\tilde{O}(n^{1/t})$, and thus, in our setting these operations can be implemented even more efficiently.

## 3.2 The Size of the Spanner

We start with showing that the resulting spanner is sparse. For this end we show that both sets $\bigcup_{v \in V} T(v)$ and $\bigcup_{v \in V} X(v)$ are sparse. We start with arguing that the set $\bigcup_{v \in V} T(v)$ is sparse.

**Lemma 3.2** *For every vertex $v \in V$, $|T(v)| \leq t - 1$.*

**Proof:** Each time an edge $e = (v, u)$ is inserted into $T(v)$, the label of $v$ grows from $P(v)$ to $P(u) + n$. Moreover, note that $P(u) \geq P(v)$ for such an edge. Consequently, the level of $P(v)$ grows at least by 1. Hence at any given time of an execution of the algorithm, $L(P(v))$ is an upper bound on the number of edges currently stored in $T(v)$. Since $L(P(v))$ never grows beyond $t - 1$, it follows that $|T(v)| \leq t - 1$. ∎

Consequently, the set $\bigcup_v T(v)$ contains at most $n \cdot (t - 1)$ edges, i.e.,

$$\left| \bigcup_{v \in V} T(v) \right| \leq n \cdot (t - 1) \ . \tag{1}$$

We next argue that the set $\bigcup_{v \in V} X(v)$ is sparse as well. First, by Lemma 3.1, the number of distinct labels of level $t - 1$ that occur during the algorithm is, with high probability, $O(n^{1/t} \cdot (t \cdot \log n)^{1 - 1/t})$. Fix a vertex $v \in V$. Since, by line 5 of Algorithm 1, for each such a label $P$ at most one edge $(u, v)$ with $P(u) = P \succ P(v)$ is inserted into $X(v)$, it follows that the number of edges $(u, v)$ with $P(u) \succ P(v)$, $L(P(u)) = t - 1$, inserted into $X(v)$, is, with high probability, at most $O(n^{1/t} \cdot (t \cdot \log n)^{1 - 1/t})$.

For an index $i \in [(t - 1)]$, let $X^{(i)}(v)$ denote the set of edges $(u, v)$, with $L(P(u)) < t - 1$, inserted into $X(v)$ during the period of time that $L(P(v))$ was equal to $i$.

**Lemma 3.3** $X^{(t-1)}(v) = \emptyset$.

**Proof:** Suppose for contradiction that there exists an edge $e = (v, u) \in X^{(t-1)}(v)$. By definition of $X^{(t-1)}(v)$, the label of $v$ at the time when the algorithm read the edge $e$ satisfied $L(P(v)) = t - 1$. Since $e$ was inserted into $X(v)$, it follows that $P(u) \succ P(v)$, where $P(u)$ is the label of $u$ at that time. Consequently, $L(P(u)) = t - 1$ as well. This is a contradiction to the definition if $X^{(t-1)}(v)$. ∎

We next show that the cardinalities of the sets $X^{(i)}(v)$, $0 \leq i \leq t - 2$, are small as well. (Though these sets may be not empty.)

**Lemma 3.4** *For every input sequence of edges $(e_1, e_2, \ldots, e_m)$ determined obliviously of the coin tosses of the algorithm, for every vertex $v$, and index $i \in [(t-2)]$, with high probability, $|X^{(i)}(v)| = O\left(n^{1/t} \cdot \frac{\log^{1-1/t} n}{t^{1/t}}\right)$.*

**Proof:** The value $L(P(v))$ grows each time that the algorithm encounters an edge $(u, v)$ with $P(u) \succ P(v)$ and such that $P(u)$ is a selected label. On the other hand, for a fixed index $i \in [(t - 2)]$, during the time period when the condition $L(P(v)) = i$ holds, the size of the set $X^{(i)}(v)$ is incremented each time the algorithm encounters an edge $(u, v)$ with $P(u) \succ P(v)$, such that $P(u)$ is not a selected label, and such that $B(P(u))$ does not belong to $M(v)$.

Fix an execution of the algorithm, and an index $i$, $i \in [(t-2)]$. Consider the sequence $\eta$ of all edges $\eta = (e_1 = (u_1, v), e_2 = (u_2, v), \ldots, e_k = (u_k, v))$, for some integer $k \geq 0$, that arrived during the time period when the condition $L(P(v)) = i$ holds, and such that $P(v)$ did not grow as a result of processing these edges. Let $\sigma = (P_1, P_2, \ldots, P_k)$, $P_j = P(u_j)$, $j \in [k]$, be the sequence of labels of vertices $u_j$ such that the edge $(u_j, v)$ appears in $\eta$. Note that the edge $e_j = (u_j, v)$ is inserted into $X(v)$ only if the base value $B_j$ of the label $P_j$ appears in $\sigma$ for the first time. Moreover, the edge contributes to $X^{(i)}(v)$ only if $L(P_j) < t - 1$.

Let $\sigma' = (P_{j_1}, P_{j_2}, \ldots, P_{j_\ell})$, for some integer $0 \leq \ell \leq k$, be the subsequence of $\sigma$ that contains only labels $P_{j_q} = P(u_{j_q})$, $q \in [\ell]$, of level at most $t - 2$ such that no other label with the same base value appears in $\sigma$ with an index smaller than $j_q$. It follows that $|X^{(i)}(v)| = \ell$. Moreover, all labels that appear in $\sigma'$ are not selected, as the algorithm increases $L(P(v))$ whenever it encounters an edge $(u_j, v)$ as above with a selected label $P(u_j)$.

Since all labels of $\sigma'$ have distinct base values, and they are of level at most $t - 2$, each of these labels has a probability exactly $p$ to be selected independently of other labels in the sequence. Hence the probability that $\ell$ or more unselected labels of level $t - 2$ or less with distinct base values will appear *in a row* (with no selected label in-between them) is at most

$$(1 - p)^\ell = \left(1 - \left(\frac{t \cdot \log n}{n}\right)^{1/t}\right)^\ell.$$

In other words,

$$\mathbb{P}(|X^{(i)}(v)| \geq \ell) \leq \left(1 - \left(\frac{t \cdot \log n}{n}\right)^{1/t}\right)^\ell.$$

For $\ell = c \log n \cdot \left(\frac{n}{t \cdot \log n}\right)^{1/t}$ for a sufficiently large value of $c$, this probability is at most $\frac{1}{n^c}$.

Hence, with high probability,

$$|X^{(i)}(v)| = O\left(\frac{n^{1/t}}{t^{1/t}} \cdot \log^{1 - 1/t} n\right). \quad \blacksquare$$

We are now ready to prove the desired upper bound on $|\bigcup_{v \in V} X(v)|$.

**Corollary 3.5** *Under the assumption of Lemma 3.4, for every vertex $v \in V$, with high probability, the overall number of edges inserted into $X(v)$ is $O(n^{1/t} \cdot (t \cdot \log n)^{1 - 1/t})$.*

**Proof:** By Lemma 3.4, with high probability, the number of edges $(u, v)$ with $L(P(u)) < t - 1$ inserted into $X(v)$ is at most

$$\sum_{i=0}^{t-2} |X^{(i)}(v)| = O(n^{1/t} \cdot (t \cdot \log n)^{1 - 1/t}).$$

As was argued in the discussion preceding Lemma 3.4, the number of edges $(v, u)$ with $L(P(u)) = t - 1$ inserted into $X(v)$ is, with high probability, $O(n^{1/t} \cdot (t \cdot \log n)^{1 - 1/t})$ as well. $\blacksquare$

We summarize the size analysis of the spanner constructed by Algorithm 1 with the following corollary.

**Corollary 3.6** *Under the assumptions of Lemma 3.4, with high probability, the spanner $H$ constructed by the algorithm contains $O(n^{1+1/t} \cdot (t \cdot \log n)^{1-1/t})$ edges. Moreover, each table $M(v)$, $v \in V$, stores, with high probability, at most $O(n^{1/t} \cdot (t \cdot \log n)^{1-1/t})$ values, and consequently, overall the algorithm uses $O(|H| \cdot \log n) = O(n^{1+1/t} \cdot t^{1-1/t} \cdot (\log n)^{2-1/t})$ bits of space.*

**Proof:** The resulting spanner is $(\bigcup_{v \in V} T(v) \cup \bigcup_{v \in V} X(v))$. By the inequality (1), $|\bigcup_{v \in V} T(v)| \leq n \cdot (t-1)$. By Corollary 3.5, with high probability, $|\bigcup_{v \in V} X(v)| = O(n^{1+1/t} \cdot (t \cdot \log n)^{1-1/t})$, and so the first assertion of the corollary follows.

For the second assertion recall that a new value is added to $M(v)$ only when a new edge $(u, v)$ is introduced into the set $X(v)$. By Corollary 3.5, with high probability, $|X(v)| = O(n^{1/t} \cdot (t \cdot \log n)^{1-1/t})$, and therefore the same bound applies for $|M(v)|$ as well.

To calculate the overall size of the data structures used by the algorithms we note that

$$\left| \bigcup_{v \in V} M(v) \right| \leq \left| \bigcup_{v \in V} X(v) \right| \leq \left| \bigcup_{v \in V} X(v) \right| + \left| \bigcup_{v \in V} T(v) \right| = O(|H|) = O(n^{1+1/t} \cdot (t \cdot \log n)^{1-1/t}) \,.$$

Since each label and edge requires $O(\log n)$ bits to represent, the desired upper bound on the size of the data structures follows. ∎

### 3.3 The Stretch Guarantee of the Spanner

We next show that the subgraph constructed by the algorithm is a $(2t-1)$-spanner of the original graph $G$.

For an integer $k \geq 1$, and a vertex $v \in V$, let $P_k(v)$ denote the label of $v$, $P(v)$, before reading the $k$th edge of the input stream.

**Lemma 3.7** *Let $v, v' \in V$ be a pair of vertices such that there exist positive integers $k, k' \geq 1$ such that $B(P_k(v)) = B(P_{k'}(v'))$. Then there exists a path of length at most $L(P_k(v)) + L(P_{k'}(v'))$ between $v$ and $v'$ in the (final) set $\bigcup_{v \in V} T(v)$.*

**Proof:** The proof is by induction of $L(P_k(v)) + L(P_{k'}(v'))$. The induction base is when $L(P_k(v)) = L(P_{k'}(v')) = 0$. Hence $B(P_k(v)) = B(P_{k'}(v'))$. It follows that $P_k(v) = B(P_k(v)) = B(P_{k'}(v')) = P_{k'}(v')$, and so $I(v) = I(v')$. Hence $v = v'$, and the statement follows. (The path that starts and ends in $v$, and contains no other vertex is said to have length 0.)

For the induction step we first observe that the only way for a vertex $v$ to be labeled by a label of level greater than 0 is by "borrowing" the label from a neighboring vertex (and incrementing its level).

Since $L(P_k(v)) + L(P_{k'}(v')) > 0$, and $L(P_k(v)), L(P_{k'}(v')) \geq 0$, it follows that either $L(P_k(v)) > 0$ or $L(P_{k'}(v')) > 0$. Suppose without loss of generality that $L(P_k(v)) > 0$. Let $u$ be the vertex such that when the edge $(u, v)$ was read, the vertex $v$ adapted the label $P(u)+n = P_k(v)$. Let $k''$ be the time step on which this happened. It follows that on time step $k''$ the edge $(u, v)$ was inserted into $T(v)$ (on line 4 of Algorithm 1), and on this time step the label of $u$, $P_{k''}(u)$, was equal to $P_k(v) - n$, and so $L(P_{k''}(u)) = L(P_k(v)) - 1$, and $B(P_{k''}(u)) = B(P_k(v)) = B(P_{k'}(v'))$. Hence $L = L(P_{k''}(u)) + L(P_{k'}(v')) = L(P_k(v)) + L(P_{k'}(v')) - 1$, and so the induction hypothesis is applicable to the pair of vertices $\{u, v'\}$. Hence there exists a path of length $L$ between $u$ and $v'$ in $\bigcup_{v \in V} T(v)$, and so there exists a path of length $L+1 = L(P_k(v))+L(P_{k'}(v'))$ connecting $v$ and $v'$ in the edge set $\bigcup_{v \in V} T(v)$. ∎

The next lemma shows that the edge set $H = \bigcup_{v \in V} T(v) \cup \bigcup_{v \in V} X(v)$ is a $(2t-1)$-spanner.

**Lemma 3.8** *Let $e = (v, v') \in E$ be an edge. Then there exists a path of length at most $2t - 1$ between $v$ and $v'$ in the edge set $H$.*

**Proof:** If there exist indices $k, k' \geq 1$ such that $B(P_k(v)) = B(P_{k'}(v'))$ then by Lemma 3.7 there exists a path of length at most $L(P_k(v)) + L(P_{k'}(v')) \leq 2t - 2$ between $v$ and $v'$ in the spanner.

Otherwise, consider the time step $k$, $k \geq 1$, of the execution of the algorithm on which the edge $e = (v, v')$ was read. Let $P = P_k(v)$, $P' = P_k(v')$ be the labels of $v$ and $v'$, respectively, at time $k$. Suppose without loss of generality that $P' \succ P$. If $P'$ is a selected label, then the edge $e$ was added to $T(v)$, and so there exists a path of length 1 connecting $v$ and $v'$ in the spanner.

Otherwise, consider the case that $P'$ is not a selected label. Let $M_k(v)$ (respectively, $X_k(v)$) denote the set $M(v)$ (resp., $X(v)$) at time $k$. If $B(P') \notin M_k(v)$ then the edge $e = (v, v')$ was added to $X(v)$ (line 7 of Algorithm 1), and again the distance in the spanner between $v$ and $v'$ is equal to 1.

We are left with the case that the label of $v'$, $P'$, is not a selected label, and $B(P') \in M_k(v)$. In this case, by construction, there exists an edge $(u', v) \in X(v)$ such that $u'$ was labeled by $P''$, $B(P'') = B(P')$, at some earlier time $k'$, $k' \leq k$. (This can be seen by a straightforward induction on the time moment $k$.) Hence, by Lemma 3.7, since $B(P_k(v')) = B(P_{k'}(u')) = B(P')$, there exists a path of length at most $2t - 2$ connecting the vertices $v'$ and $u'$ in the set $\bigcup_{v \in V} T(v)$. Since $(u', v) \in X(v)$, it follows that there exists a path of length at most $2t - 1$ between $v$ and $v'$ in the spanner. ∎

## 3.4 The Processing Time-per-edge

To conclude the analysis of our streaming algorithm for constructing sparse spanners, we show that it has a very small processing time-per-edge. For this purpose we now fill in a few implementation details that have so far been unspecified. Specifically, on lines 5 and 6 of the Procedure *Read_Edge* the algorithm tests whether an element $B$ belongs to a set $M(v)$, and if it does not, the algorithm inserts it there. The set $M(v)$ is a subset of the universe $[n]$, and by Corollary 3.6, its size is, with high probability, $O((t \cdot \log n)^{1-1/t} \cdot n^{1/t})$. Moreover, since $|M(v)| \leq |X(v)|$, it follows that $|M(v)| \leq deg(v)$.

Let $N = c \cdot (t \cdot \log n)^{1-1/t} \cdot n^{1/t}$, for a sufficiently large constant $c$. (The probability that $|M(v)| \leq c \cdot (t \cdot \log n)^{1-1/t} \cdot n^{1/t}$ for every vertex $v \in V$ is at least $1 - \frac{1}{n^{c-2}}$. Hence choosing $c = 4$ is sufficient.) As a part of preprocessing the algorithm computes a random hash function $h : [n] \to [N]$. Specifically, for each number $i \in [n]$, the algorithm picks a value $j \in [N]$ uniformly at random, and sets $h(i) = j$. The table representation of this hash function is written down, and is used throughout the execution of the algorithm for the tables $M(v)$ for all the vertices $v \in V$. This representation requires $O(n \cdot \log n)$ space, and can be computed in $O(n)$ time during the preprocessing.

For every vertex $v$ the algorithm maintains a hash table $M(v)$ of size $N$. Every base value $B$ for which the algorithm needs to test its membership in $M(v)$ on line 5 of the Procedure *Read_Edge* is hashed to $h(B)$ using the hash function $h$. To resolve collisions, for each entry of the hash table $M(v)$ we use a dynamic dictionary data structure of Beame and Fich [18] (with the dynamization result of Andersson and Thorup [5]). This data structure maintains a dynamic set of $q$ keys from an arbitrary universe using $O\left(\sqrt{\frac{\log q}{\log \log q}}\right)$ time per update (insertion or deletion) and membership queries. This completes the description of the implementation details of the algorithm.

Note that the preprocessing of the algorithm requires $O(n)$ time. We next argue that implemented this way, the algorithm enjoys an extremely low processing time-per-edge.

First, note that as with high probability $|M(v)| \leq N$, and each element of $M(v)$ is hashed uniformly at random into one of the $N$ entries of the hash table, it follows that the expected number of elements hashed into each specific entry of the hash table is $O(1)$, and thus the expected processing time-per-edge is $O(1)$.

We next provide an upper bound on the processing time-per-edge that holds with high probability (rather than on expectation). We will use the following fact that can be easily verified using standard calculus.

**Fact 3.9** *For positive integers $N, r$, $N \geq r > 0$,*

$$\left(\frac{N}{r}\right)^r \leq \binom{N}{r} \leq \left(\frac{N \cdot e}{r}\right)^r .$$

We also need some additional definitions. For an execution $\varphi$ of the algorithm, we say that a vertex $v$ *encounters* the base value $B$ during the execution $\varphi$ of the algorithm if there exists an edge $e = (u, v) \in E$ that satisfies (when it is read by the algorithm) that $P(u) \succ P(v)$, $B(P(u)) = B$, and $P(u)$ is not selected.

Fix a vertex $v$, and a specific entry $j$, $j \in [N]$, in the hash table $M(v)$. For every integer $r$, $1 \leq r \leq N$, let $X(r)$ denote the event that at least $r$ of the base values encountered by $v$ during an execution of the algorithm are hashed into the entry $j$ of $M(v)$. We will analyze the probability of this event for every input graph and every ordering of the edges of the input graph. The probability space is determined by the coin tosses of the algorithm. Note that

$$\mathbb{P}(X(r)) \leq \left(\frac{1}{N}\right)^r \cdot \binom{N}{r} \leq \left(\frac{e}{r}\right)^r .$$

For a fixed vertex $v$, let $Y(r)$ denote the event that there exists an entry in $M(v)$ with at least $r$ values hashed to this entry during an execution of the algorithm. By the union-bound,

$$P(Y(r)) \leq N \cdot (e/r)^r .$$

Finally, let $Z(r)$ denote the event that there exists a vertex $v$, and an entry $j$ in its hash table $M(v)$ with at least $r$ values hashed to this entry during an execution of the algorithm. Using the union-bound again, we conclude that

$$P(Z(r)) \leq n \cdot N \cdot (e/r)^r .$$

Observe that this inequality holds for every (possibly chosen adversarially) input graph $G$ and ordering $\rho$ in which the algorithm reads the edges of $G$.

Set $r = c \cdot \frac{\log n}{\log \log n}$, for a sufficiently large constant $c$. Since $N = o(n)$, it follows that with probability at least $1 - \frac{1}{n^{c-2}}$, for every vertex $v$ and every entry $j$ of the hash table $M(v)$, at most $r$ values were hashed to this entry.

Recall that the insert and query operations of the dynamic dictionary data structure of [18, 5] require $O\left(\sqrt{\frac{\log r}{\log \log r}}\right)$ time per operation. Hence, with high probability, this expression is bounded by $O\left(\sqrt{\frac{\log \log n}{\log^{(3)} n}}\right)$. (We use the notation $\log^{(i)} n$, for an integer $i \geq 0$, to denote the $i$-iterated logarithm of $n$.) Let $\lambda(n)$ (respectively, $\sigma(n)$) denote the function $\sqrt{\frac{\log \log n}{\log^{(3)} n}}$ (resp., $\sqrt{\frac{\log n}{\log \log n}}$).

We summarize the analysis of the processing time-per-edge of our streaming by the following corollary.

**Corollary 3.10** *The processing time-per-edge of our algorithm is, with high probability, $O(\lambda(n))$. The expected processing time-per-edge is $O(1)$.*

We note also that for every edge $e = (u, v)$ adjacent to a vertex $v$, at most one value $B = B(P(u))$ may be inserted into the set $M(v)$. Thus, it is always true that $|M(v)| \leq deg(v)$, and so it follows that the processing time of an edge $e$ in our algorithm is *always* (with probability 1) at most $O(\sigma(deg(e)))$.

**Remark:** An improved data structure that supports insertion and membership queries would enable to improve the processing time-per-edge of our algorithm even further. (The data structure that we use supports also deletion, which is completely unnecessary for the streaming version of our algorithm.)

**Remark:** It is easy to verify that setting $p = n^{-1/t}$ in our algorithm guarantees the expected size of $O(t \cdot n^{1+1/t})$ for the spanner, but slightly increases the upper bound on the number of edges of the spanner that holds with high probability. The latter becomes $O(t \cdot n^{1+1/t} \cdot \log n)$.

The properties of our streaming algorithm are summarized in the following theorem. This theorem follows directly from Corollary 3.6, Lemma 3.8, and Corollary 3.10.

**Theorem 3.11** *Let $n, t$, $n \geq t \geq 1$, be positive integers. Consider an execution of our algorithm in the streaming model on an input (unweighted undirected) $n$-vertex graph $G = (V, E)$ such that both the graph and the ordering $\rho$ of its edges are chosen by a non-adaptive adversary obliviously of the coin tosses of the algorithm. The algorithm constructs a $(2t-1)$-spanner $H$ of the input graph. The expected size of the spanner is $O(t \cdot n^{1+1/t})$ or the size of the spanner is $O((t \cdot \log n)^{1-1/t} \cdot n^{1+1/t})$ with high probability (depending on the choice of $p$; in the first case the guarantee on the size that holds with high probability is $O(t \cdot \log n \cdot n^{1+1/t})$). The algorithm does so in one pass over the input stream, and requires $O(1)$ expected processing time-per-edge, $O(\lambda(n))$ processing time-per-edge with high probability, and $O(\sigma(deg(e)))$ processing time-per-edge in the worst-case, for an edge $e = (v, u)$. The space used by the algorithm is $O(|H| \cdot \log n) = O(t^{1-1/t} \cdot \log^{2-1/t} n \cdot n^{1/t})$ bits with high probability. The preprocessing of the algorithm requires $O(n)$ time.*

**Remark:** Our algorithm can be easily adapted to construct a $(2t-1)(1+\epsilon)$-spanners for weighted graphs, for an arbitrary $\epsilon > 0$. The size of the obtained spanner becomes $O(\log_{(1+\epsilon)} \hat{\omega} \cdot (t \cdot \log n)^{1-1/t} \cdot n^{1+1/t})$, where $\hat{\omega}$ is the aspect ratio of the network. The algorithm still works in one pass, and has the same processing time-per-edge. This adaptation is achieved in a standard way (see, e.g., [24]), by constructing $\log_{(1+\epsilon)} \hat{\omega}$ different spanners in parallel. For completeness, we next overview this adaptation.

The edge weights can be scaled so that they are all greater or equal to 1 and smaller or equal to $\hat{\omega}$. All edges are partitioned logically into $\lceil \log_{(1+\epsilon)} \hat{\omega} \rceil$ categories, indexed $i = 1, 2, \ldots, \lceil \log_{(1+\epsilon)} \hat{\omega} \rceil$, according to their weights, with the category $i$ containing the edges with weights greater of equal to $(1+\epsilon)^{i-1}$ and smaller than $(1+\epsilon)^i$. When an edge $e = (u, v)$ is read, it is processed according to its category, and it is either inserted into the spanner for the edges of category $i$, or discarded.

Obviously, after reading all the edges, we end up with $\lceil \log_{(1+\epsilon)} \hat{\omega} \rceil$ subgraphs, with the $i$th subgraph being a $(2t-1)(1+\epsilon)$-spanner for the edges of the category $i$. Consequently, the union of all these edges is a $(2t-1)(1+\epsilon)$-spanner for the entire graph. The cardinality of this union is at most $\lceil \log_{(1+\epsilon)} \hat{\omega} \rceil$ times the maximum cardinality of one of these subgraphs, which is, in turn, at most $O((t \cdot \log n)^{1-1/t} \cdot n^{1+1/t})$ with high probability.

## 4 A Centralized Dynamic Algorithm

Our streaming algorithm can be seen as an incremental dynamic algorithm for maintaining a $(2t-1)$-spanner of size $O((t \cdot \log n)^{1-1/t} \cdot n^{1+1/t})$ for unweighted graphs, where $n$ is an upper bound on the number of vertices that are allowed to appear in the graph.

The initialization of the algorithm is as follows. Given a graph $G = (V, E)$, we run our streaming algorithm with the edge set $E$, where the order in which the edge set is read is arbitrary. As a result, the spanner, and the satellite data structures $\{M(v), Sp(v) \mid v \in V\}$ are constructed. This requires $O(|E|)$ expected time, $O(|E| \cdot \lambda(n))$ time with high probability, and $O(|E| \cdot \sigma(\Delta))$ in the worst-case, where $\Delta$ is the maximum degree of a vertex in $G$.

Each edge that is added to the graph is processed using our streaming algorithm. The spanner and the satellite data structures are updated in expected $O(1)$ time-per-edge, $O(\lambda(n))$ time-per-edge with high probability, and $O(\sigma(\Delta))$ time in the worst-case.

We next make the algorithm robust to decremental updates as well. Note that for an edge $e = (v, u)$ to become a $T$-edge (an edge of $\bigcup_{x \in V} T(x)$), it must hold that at the time that the algorithm reads the edge, the greater of the two labels $P(u)$ and $P(v)$ (with respect to the order relation $\succ$) is selected. The probability of a label to be selected is at most $p = \left(\frac{t \cdot \log n}{n}\right)^{1/t}$ if its level is smaller than $t-1$, and is 0 otherwise. Hence the probability of $e$ to become a $T$-edge is at most $p$.

In Section 8.3.5 (right after Corollary 8.17) we will show that a crash of an edge $e = (v, u)$ that does not belong to $\bigcup_{x \in V} T(x)$ can be processed in expected time $O(1)$. Moreover, with high probability the processing of such a crash requires $\sigma(h)$ time, where $h = \max\{deg(e), \log n\}$. Since the entire spanner can be recomputed in expected time $O(|E|)$ by our algorithm, it follows that the expected decremental update time of our algorithm is $O(\frac{|E|}{n^{1/t}} \cdot (t \cdot \log n)^{1/t})$. The size of the data structure maintained by the incremental variant of the algorithm is $O(t^{1-1/t} \cdot (\log n)^{2-1/t} \cdot n^{1+1/t})$ bits, and the fully dynamic algorithm maintains a data structure of size $O(|E| \cdot \log n)$.

We summarize this discussion with a following corollary.

**Corollary 4.1** *For positive integer $n, t$, $n \geq t \geq 1$, the algorithm is a fully dynamic algorithm for maintaining $(2t-1)$-spanners with expected $O(t \cdot n^{1+1/t})$ number of edges (or $O((t \log n)^{1-1/t} \cdot n^{1+1/t})$ edges with high probability, depending on the choice of $p$) for graphs with at most $n$ vertices. If $G = (V, E)$ is the initial graph, then the initialization of the algorithm requires $O(|E|)$ expected time, $O(|E| \cdot \lambda(n))$ time with high probability, and $O(|E| \cdot \sigma(\Delta))$ in the worst-case. The expected incremental update time of the algorithm is $O(1)$, with high probability it is $O(\lambda(n))$, and in the worst-case it is $O(\sigma(deg(e)))$ (for an edge $e$ that joins the graph). The expected decremental update time is $O(\frac{|E|}{n^{1/t}} \cdot (t \cdot \log n)^{1/t})$, and with probability at least $1 - \left(\frac{t \cdot \log n}{n}\right)^{1/t}$ the decremental update time is $O(\sigma(h))$, where $h = \max\{deg(e), \log n\}$.*

To our knowledge, this is the first fully dynamic algorithm for maintaining sparse spanners for a wide range of values of the stretch parameter $t$ with non-trivial guarantees on both the incremental and decremental update times. (Ausillo et al. [6] devised such an algorithm for 3- and 5-spanners.)

Using Corollary 4.1 in conjunction with the dynamic All-Pairs-Almost-Shortest-Paths (henceforth, APASP) algorithm of Roditty and Zwick [34] we obtain tradeoffs for the incremental APASP problem that are advantageous for a certain range of parameters. Specifically, we use the following result.

**Theorem 4.2** *[34] For every $\epsilon, \delta > 0$, and $z \leq m^{1/2-\delta}$, there exists a fully dynamic $(1 + \epsilon)$-approximate APASP algorithm with an amortized update time $\tilde{O}(|E|n/z)$, and query time $O(z)$.*

Consequently, by maintaining an incremental $(2t-1)$-spanner with $\tilde{O}(n^{1+1/t})$ edges, and maintaining the dynamic data structure of [34] computed for the maintained spanner, we obtain the following result.

**Theorem 4.3** *For every $\epsilon, \delta > 0$, positive integer $t \geq 1$, and $z \leq n^{\frac{1}{2}(1+1/k)-\delta}$, there exists a fully dynamic $(2t-1)(1+\epsilon)$-approximate APASP algorithm for unweighted undirected $n$-vertex graphs with amortized expected update time of $\tilde{O}\left(\frac{n^{2+1/t}}{z} + \frac{|E|}{n^{1/t}} \cdot (t \cdot \log n)^{1/t}\right)$, and query time at most $O(z)$.*

Note that when choosing $z = \omega(n^{1/t})$, this algorithm has a *sublinear* amortized update time in terms of the number of edges even when the number of edges $m = \Theta(n^2)$. To our knowledge, this is the first dynamic APASP algorithm that achieves an amortized update of $o(n^2)$ for *all graphs* with a non-trivial query time is non-trivial ($z = n^{1/t+\eta}$, $\eta > 0$ is a constant).

# 5 The Static Distributed Algorithm

In this section we present and analyze a static synchronous distributed algorithm for constructing $(2t-1)$-spanners with $O((t \cdot \log n)^{1-1/t} \cdot n^{1+1/t})$ edges. The running time of our algorithm is $2t$ rounds or time units depending on whether the setting is synchronous or asynchronous, respectively. The algorithm is extremely simple, and is, consequently, very well suited for ad-hoc or sensor networks in which the processors possess only very limited computational resources.

For simplicity we assume that every vertex $v$ has a unique identifier $I(v) \in [n]$. If this is not the case, each vertex $v$ can pick its $I(v)$ uniformly at random from the set $[n^4]$. With probability at least $(1 - 1/n^2)$ the identifiers will be all distinct. The fact that the range of the identifiers becomes $n^4$ instead of $[n]$ would necessitate only minor straightforward changes to the algorithm and its analysis.

## 5.1 The Algorithm

Like in the streaming version of the algorithm, each vertex $v$ maintains the variables $I(v)$, $P(v)$, $r(v)$, $Sp(v)$ and $M(v)$, that have the same functionality as in the centralized algorithm. Also, as in the centralized setting, each set $Sp(v)$ is logically partitioned into the set $T(v)$ and $X(v)$, where $T(v)$ (respectively, $X(v)$) is the set of the tree (resp., cross) edges of the algorithm.

In addition, each vertex $v$ maintains a variable $ttl(v)$, standing for the "time-to-live", which reflects the maximum distance to which the current label of $v$, $P(v)$, is allowed to propagate. Initially, $ttl(v)$ is set to $r(v)$. However, later when the vertex $v$ adapts a label of one of its neighbors $u$, the value of $ttl(v)$ becomes equal to $ttl(u) - 1$, reflecting the requirement that the label of $u$ is not allowed to propagate beyond radius $ttl(u)$.

The initial value of $r(v)$ is set according to the same truncated geometric distribution as in the centralized version of the algorithm. As before, we assume that the number of vertices $n$, or at least an upper bound on $n$, is known to all the vertices prior to the beginning of the execution. The order relation $\succ$ between the labels of labels $P(v)$ of different vertices is defined exactly as in Section 3.1.

We next describe the algorithm itself. The vertex $v$ that runs the algorithm invokes the Procedure *Round* for $2t$ rounds in a row. During these rounds it computes its set $Sp(v)$ which is the output of the algorithm. In other words, $\bigcup_{v \in V} Sp(v)$ is the sparse spanner returned by the algorithm. Note that the Procedure *Round* accepts no input parameters, and consequently, it runs in the same way on every round of the algorithm. In addition, there is just one possible type of messages that can be sent by vertices that run the algorithm. (These are messages $(P(v), ttl(v))$, where $v$ is the sender of the message.) This uniformity of the algorithm makes it extremely robust to dynamic changes of the networks, as we will see in Sections 6, 8 and 9.

The Procedure *Round* is very similar to the streaming algorithm from the Section 3 (the Procedure *Read_Edge*). The most significant difference is that while the Procedure *Read_Edge* processes one single edge, the Procedure *Round* processes the set of edges adjacent to a particular vertex. However, the Procedure *Round* can be seen as, essentially, invoking the Procedure *Read_Edge* on all the edges adjacent to $v$, sequentially, one after another, in an arbitrary order.

We remark that the Procedure *Round* goes over all the messages received on the previous round, and for each message $(P(u), ttl(u))$, reads the edge $(v, u)$. Observe that on round 1 there are still no messages received, and thus the only processing done on this round is sending out the messages $(P(v), ttl(v))$ on line 9.

The formal description of the algorithm follows.

**for** rounds $1, 2 \ldots, 2t$ **do**
  Invoke Procedure *Round*
**end for**

**Algorithm 2** A distributed static synchronous algorithm for constructing a sparse $(2t-1)$-spanner (Procedure *Round*). The pseudo-code is for a fixed vertex $v \in V$.

1: Go over all received messages in an abitrary order and do
2: **while** $\exists$ message $(P(u), ttl(u))$ with $P(u) \succ P(v)$ **do**
3:   **if** $(ttl(u) > 0)$ **then**
4:     $P(v) \leftarrow P(u) + n$; $ttl(v) \leftarrow ttl(u) - 1$; $Sp(v) \leftarrow Sp(v) \cup \{e\}$
5:   **else if** $B(P(u)) \notin M(v)$ **then**
6:     $M(v) \leftarrow M(v) \cup \{B(P(u))\}$; $Sp(v) \leftarrow Sp(v) \cup \{e\}$
7:   **end if**
8: **end while**
9: Send to all your neighbors the message $(P(v), ttl(v))$

## 5.2 The Size of the Spanner

We first introduce a few definitions. We say that an edge $e = (w, w') \in E$ is *scanned* by the algorithm if at some point of the execution, either the algorithm run by $w$ (meaning that $v = w$) substitutes $u = w'$ and passes the condition of the while loop (and gets into the loop), or the algorithm run by $w'$ (meaning that $v = w'$) does the same with $u = w'$. In the former (respectively, latter) case we say that the edge is scanned by $w$ (resp., $w'$). We say that a vertex $v$ *reads* an edge $e = (v, v')$ on round $j$, if either the vertex $v$ scans it on round $j$ or if it discovers that $P(v') \prec P(v)$ (and thus does not scan it). In Algorithm 2 every vertex $v$ reads all edges that are adjacent to it on every round of the algorithm.

Obviously, no edge is scanned (or read) on round 1, because the first messages arrive on round 2. Interestingly, due to the distributed nature of our algorithm, it may happen that on a certain round $i$, $i \in \{2, 3, \ldots, 2t\}$, the edge $e = (w, w')$ is scanned neither by the vertex $w$ nor by by $w'$. We will soon describe a possible scenario of this kind, but first we need the following simple observation.

Let $P_i(w)$ be the label of $w$ in the beginning of the round $i$. Note that $w$ had the same label on line 9 of its execution of the Procedure *Round* on round $i - 1$, and consequently, the label of $w$ known to $w'$ at the beginning of the round $i$ is precisely $P_i(w)$. Symmetrically, the label of $w'$ known to $w$ at the beginning of the round $i$ is $P_i(w')$.

One scenario in which an edge $e = (w, w')$ is scanned neither by $w$ nor by $w'$ on a round $i$, $i \geq 2$, is the following one. Suppose that $P_i(w) \succ P_i(w')$, and so based on the message that $w$ received from $w'$ on round $i$, the vertex $w$ decides that $w'$ is in charge of this edge. However, $w'$ starts with scanning another edge $e' = (w', u)$, and scanning $e'$ increases the label of $w'$ to $P(u) + n \succ P(w)$, and when $w'$ gets to consider the edge $e = (w, w')$ it discovers that $P_i(w)$ is smaller than its current label, and decides to skip $e$.

**Lemma 5.1** *It is impossible for an edge $e = (w, w')$ to be scanned by* both *its endpoints on the same round $i$.*

**Proof:** Consider the time moment when the vertex $w$ scans the edge $e$ on round $i$. Let $P(w)$ be the label of the vertex $w$ at this moment. Then $P(w) \prec P_i(w')$. Note, however, that $P_i(w) \preceq P(w)$. Moreover, the label of $w'$, $P(w')$, is greater or equal than $P_i(w')$ at any moment of the execution on round $i$. In other words, $P(w') \succeq P_i(w') \succ P(w) \succeq P_i(w)$, and thus the vertex $w'$ cannot possibly scan the edge $e$ on round $i$. ∎

We next argue that the level of a vertex is at most $t - 1$.

For the sake of the following lemma we assume that the line 4 of the algorithm is executed as an atomic operation. (This assumption is not necessary for the lemma to hold, and we will later indicate how to get rid of it.)

**Lemma 5.2** *For every vertex $v \in V$, and for every moment of the execution of the algorithm after the variables $P(v)$ and $ttl(v)$ were initialized,*

$$L(P(v)) + ttl(v) = r(z) , \qquad (2)$$

*where $z = z(P(v))$ is the base vertex of the label $P(v)$, and $r(z)$ is the radius chosen by the vertex $z$.*

**Proof:** For a vertex $v$, the variables $P(v)$ and $ttl(v)$ are local to $v$, and hence only the vertex $v$ has access to them. Hence these variables change sequentially, and so, there exists a sequence of discrete time moments $j_1, j_2, \ldots$ such that these variables change at moments $j_1, j_2, \ldots$. (The time moment when $P(v)$ and $ttl(v)$ were initialized is not included in this sequence.)

Order all the time moments $\{j_i(v) \mid v \in V, i = 1, 2, \ldots\}$ as a monotonely increasing sequence $\ell_1 < \ell_2 < \ldots$, and let $\ell_0$ be a time moment smalller than $l_1$ at which all the variables $P(v)$ and $ttl(v)$, for all $v \in V$, were already initialized. Note that for a fixed moment $\ell_i$, it may happen that more than one vertex $v$ changes its variables $P(v)$ and $ttl(v)$ at time $\ell_i$.

We now prove by induction on $i = 0, 1, \ldots$ that at time $\ell$, $\ell \in [\ell_i, \ell_{i+1})$, every vertex $v \in V$ satisfies the equation (2), where $z = z(P(v))$ is the base vertex of the label $P(v)$.

The induction base holds because each each $P(v)$ is initialized as $I(v)$, and $ttl(v)$ is initialized as $r(v)$, and so $L(P(v)) = 0$. Since $z(P(v)) = z(I(v)) = v$, it follows that $r(z(P(v))) = r(v) = L(P(v)) + ttl(v)$, and we are done.

For the induction step, suppose that for some $i = 0, 1, \ldots$, during the time interval $[\ell_i, \ell_{i+1})$, for every vertex $v \in V$, the equation (2) holds. Let $v_1, v_2, \ldots, v_k$, for some integer $k \geq 1$, be the vertices that change its labels at time $\ell_{i+1}$. Let $u_1, u_2, \ldots, u_k$ be the (not necessarily distinct) vertices such that the label $P(v_j)$ of $v_j$ is set to $P(u_j) + n$, for $j \in [k]$.

Let $P(x)$ and $ttl(x)$ (respectively, $P'(x)$ and $ttl'(x)$) denote the label and the $ttl$ of the vertex $x$ before (resp., after) the time moment $\ell_{i+1}$, respectively. It follows that $P'(v_j) = P(u_j)+n$, and $ttl'(v_j) = ttl(u_j) - 1$, for every $j \in [k]$. Hence $L(P'(v_j)) = L(P(u_j)) + 1$, and so $L(P'(v_j)) + ttl'(v_j) = L(P(u_j)) + ttl(u_j) = r(z(P(u_j)))$, for every $j \in [k]$. Moreover, as $P'(v_j) = P(u_j) + n$, it follows that $z(P'(v_j)) = z(P(u_j))$, and so $L(P'(v_j)) + ttl'(v_j) = r(z(P'(v_j)))$, for every $j \in [k]$, proving the lemma. ∎

By lines 3-4 of Algorithm 2, and by definition of the radii $r(v)$ of vertices $v$, for every vertex $v$, the variable $ttl(v)$ is always non-negative, and $r(v) \leq t - 1$. Hence, it follows that for every vertex $v$, $L(P(v))$ is always at most $t - 1$. We remark that there is no need for the line 4 to be executed as an atomic operation for arguing that the levels of vertex labels are never greater than $t-1$. If this line is not executed as an atomic operation, then after the first instruction of the line 4 and before the second instruction of this line it holds for a short period of time that $L(P(v)) + ttl(v) = r(z(P(v))) + 1$. However, at this point $ttl(v) \geq 1$, and so $L(P(v))$ is never greater than $r(z(P(v)))$, and we are done. We summarize this discussion by the following lemma.

**Lemma 5.3** *For every vertex $v$, $L(P(v))$ is always at most $t - 1$.*

Intuitively, if a vertex $v$ scans an edge $e$ adjacent to $v$ it means that the edge is treated by the algorithm exactly in the same way as it would have been treated by our streaming algorithm. Consequently, if we show that ultimately all the edges of the graph are scanned, we can then use the analysis of our streaming algorithm to derive the correctness of the distributed version of our algorithm.

**Lemma 5.4** *For every edge $e = (w, w') \in E$, the edge $e$ is scanned at some point of the execution of the algorithm.*

**Proof:** The edge $e$ is read $(2t-1)$ times by the vertex $w$, and $(2t-1)$ times by the vertex $w'$. (Once for each round $i \in \{2, 3, \ldots, 2t\}$ of the algorithm.) Recall that each time a level of a label grows, it grows at least by 1, and that initially all vertices have labels of level 0. Moreover, by Lemma 5.3, no label ever reaches a higher level than $t-1$. Consequently, it may happen that the level of $w$ grows on at most $t-1$ different rounds, and the same is true for $w'$. Since altogether the algorithm runs for $2t$ rounds, it follows that there is necessarily a round $j$, $j \in \{2, 3, \ldots, 2t\}$, on which neither the level of $w$ nor the level of $w'$ grows.

Observe that whenever a label of a vertex grows, the level of the label grows as well, and thus neither the label of $w$ nor the label of $w'$ grows on round $j$, respectively. Let $P(w) = P_j(w)$ and $P(w') = P_j(w')$ be the values of the labels of $w$ and $w'$ in the beginning of the round $j$, respectively. If $P(w) \succ P(w')$ then the algorithm run by $w'$ will discover this during the execution of the while loop on round $j$, because it received $P(w)$ in a message from $w$ sent on round $j-1$, and $P(w')$ does not change during the round $j$. Hence in this case the vertex $w'$ scans the edge $e$ on round $j$.

Symmetrically, if $P(w') \succ P(w)$ then the vertex $w$ scans the edge $e$ on round $j$. ∎

We next note that Lemma 3.2 holds for the distributed version of our algorithm as well.

**Lemma 5.5** *For every vertex $v \in V$, $|T(v)| \leq t - 1$.*

**Proof:** Fix a vertex $v \in V$. By exactly the same argument as in the proof of Lemma 3.2, $|T(v)| \leq L(P(v))$. The lemma now follows from Lemma 5.3. ∎

Hence the inequality (1) holds for the distributed algorithm as well.

For a label $n \cdot (t-1) + i$ of level $t-1$ to occur, it must hold that $r(u) = t-1$ for the vertex $u$ such that $I(u) = i$. The probability of this event is $p^{t-1}$. Hence Lemma 3.1 keeps holding too.

Fix a vertex $v \in V$. Exactly as in the analysis of the streaming algorithm, for every label $P$ of level $t-1$ that occurs during the algorithm, at most one edge $(u,v)$ with $P = P(u) \succ P(v)$ is inserted into $X(v)$. Hence, with high probability, at most $O(n^{1/t} \cdot (t \cdot \log n)^{1-1/t})$ edges $(u,v)$ with $P(u) \succ P(v)$, $L(P(u)) = t - 1$, are inserted into the set $X(v)$.

The sets $X^{(i)}(v)$, $i \in [(t-1)]$, are defined as in the Section 3.2, and $X^{(t-1)} = \emptyset$, by the same argument. We next show that the sets $X^{(i)}(v)$, $i \leq t-2$, also have small cardinality.

**Lemma 5.6** *For every vertex $v \in V$, and index $i \in [(t-2)]$, with high probability,*

$$|X^{(i)}(v)| = O\left(n^{1/t} \cdot \frac{\log^{1-1/t} n}{t^{1/t}}\right). \tag{3}$$

**Proof:** The value of $L(P(v))$ grows each time an edge $(u,v)$ with $P(u) \succ P(v)$, and such that $P(u)$ is a selected label, is scanned by the vertex $v$. Fix an index $i \in [(t-2)]$. During the time period that $L(P(v)) = i$, the value $|X^{(i)}(v)|$ is incremented every time an edge $(u,v)$ with $P(u) \succ P(v)$, such that $P(u)$ is *not* selected, $B(P(u)) \notin M(v)$, and $L(P(u)) \leq t-2$, is scanned by $v$.

Similarly to the proof of Lemma 3.4, we consider the sequence $\eta = (e_1 = (u_1, v), \ldots, e_k = (u_k, v))$ of edges that were scanned by $v$ during the time period that $L(P(v)) = i$, and such that $P(v)$ did not grow as a result of processing these edges. The sequences $\sigma = (P_1, \ldots, P_k)$, $P_j = P(u_j)$, $j \in [k]$, and $\sigma' = (P_{j_1}, P_{j_2}, \ldots, P_{j_\ell})$, for an integer $\ell$, $0 \leq \ell \leq k$, are defined as in the proof of Lemma 3.4. It is easy to verify that the equality $|X^{(i)}(v)| = \ell$ holds.

Since $L(P_{j_i}) < t-1$, for every $P_{j_i} \in \sigma'$, and since a label of level at most $t-2$ has a probability of $p = \left(\frac{t \cdot \log}{n}\right)^{1/t}$ to be selected independently of all other labels, it follows that the probability that $\ell$ or more not selected labels with distinct base values will appear in a row, with no selected labels in-between, is at most $(1-p)^\ell$. The inequality (3) follows now by exactly the same argument as in the proof of Lemma 3.4. ∎

Hence Corollaries 3.5 and 3.6 hold for the distributed version of our algorithm as well.

### 5.3 The Stretch Guarantee

In this section we argue that the subgraph constructed by the algorithm is indeed a $(2t-1)$-spanner of the input graph. We do it by showing that once the algorithm scans an edge $e = (w, w')$ (that is, either the vertex $w$ scans it or the vertex $w'$ does), the spanner guarantees a stretch of at most $(2t-1)$ for the edge.

Note that in our algorithm every vertex $v$ on every round $i$ reads all edges adjacent to $v$. For a fixed vertex $v$ and round $i$, let $(e_1, e_2, \ldots, e_{deg(v)})$ be the sequence of edges adjacent to $v$ read by $v$ on round $i$. (In other words, the vertex $v$ read first the edge $e_1$, then $e_2$, etc., and $e_{deg(v)}$ was the last edge read by $v$ on this round.) Let $P_{i,j}(v)$ be the value of the label $P(v)$ of the vertex $v$ on round $i$ before it reads the edge $e_j$. Let $T_{i,j}(v)$ denote the value of the set $T(v)$ at the same time moment.

For a pair of vertices $v, v' \in V$, and a pair of indices $i, i'$ of rounds, and a pair of indices $j, j'$, we write $\{v, i, j\} \preceq \{v', i', j'\}$ if the time moment on which the vertex $v$ reads its $j$th adjacent edge on round $i$ precedes or equal to the time moment on which the vertex $v'$ reads its $(j')$th adjacent edge on round $i'$. Let $T\{v, i, j\}$ denote the value of the set $T = \bigcup_{v \in V} T(v)$ at the time moment $\{v, i, j\}$.

**Lemma 5.7** *For a pair of vertices $v, v' \in V$, if there exist indices $i, j, i', j'$ such that $B(P_{i,j}(v)) = B(P_{i',j'}(v'))$, then the set $\bigcup_{v \in V} T(v)$ contains a path of length $L(P_{i,j}(v)) + L(P_{i',j'}(v'))$. Moreover, if $\{v, i, j\} \preceq \{v', i', j'\}$ then the set $T\{v', i', j'\}$ contains such a path.*

The proof of this lemma is identical to the proof of Lemma 3.7.

Let $H_i$ denote the set of edges present in the spanner (that is, in the set $\bigcup_{v \in V} Sp(v)$) at the beginning of the round $i$, $i = 1, 2, \ldots$.

**Lemma 5.8** *Consider an edge $e = (w, w')$ that was scanned by the algorithm on round $i$, $i \geq 2$. Then for every $i'$, $i' \geq i+1$, $H_{i'}$ spans the edge $e$.*

**Proof:** Suppose without loss of generality that the edge was scanned by the vertex $w$. Let $P(w')$ be the label of $w'$ that the vertex $w$ received from $w'$ just before the round $i$ started. Then, by definition, it follows that $w$ substituted $v = w$, $u = w'$, and discovered that $P(v) = P(w) \prec P(w') = P(u)$. If $P(w')$ is a selected label then the edge $e$ was inserted into the set $T(w)$, and we are done. Otherwise, if $P(w')$ is not a selected label.

Let $\hat{M}(w)$ denote the value of the set $M(w)$ at the moment that $w$ starts scanning the edge $e$. If $B(P(w')) \notin \hat{M}(w)$ then the edge $e = (w, w')$ was inserted into the set $X(w)$, and again we are done. Finally, if $B(P(w')) \in \hat{M}(w)$ the the assertion of the lemma follows using Lemma 5.7, and the proof argument of Lemma 3.8. ∎

Lemmas 5.4 and 5.8 imply that the algorithm constructs a $(2t-1)$-spanner of its input graph.

### 5.4 Local Processing

It is a common assumption that all local computations can be done in zero time in the distributed setting. Under this assumption one needs not to bother himself with the specific implementation of the data structure $M(v)$. However, if we do want an algorithm that is also efficient in terms of local computation, we can use the dynamic dictionary data structure of Dietzfilbinger et al. [20]. This dynamic dictionary guarantees an amortized expected time of $O(1)$ per edge, and an overall expected $O(deg(v))$ processing time-per-round, if $deg(v)$ is sufficiently large.

Recall that the data structure of Beame and Fich with the dynamization technique of Andersson and Thorup [18, 5] maintains a dynamic set with $q$ keys from an arbitrary universe using $O(\sigma(q)) = O\left(\sqrt{\frac{\log q}{\log \log q}}\right)$ time per update or membership query. Using this data structure we obtain a worst-case processing time-per-edge of $O(\sigma(deg(v)))$, and processing time-per-edge of $O\left(\sqrt{\frac{\log \log n + (\log n)/t}{\log(\log \log n + (\log n)/t)}}\right)$ with high probability. We denote the latter expression by

$$\chi(n,t) = \sqrt{\frac{\log \log n + (\log n)/t}{\log(\log \log n + (\log n)/t)}} \ .$$

For the first (respectively, second) bound we argue that $|M(v)| \leq deg(v)$ (resp., $|M(v)| = O(n^{1/t} \cdot (t \cdot \log n)^{1-1/t}))$ for every vertex $v \in V$. The second bound follows from Lemma 5.6, and since $|M(v)| \leq |X(v)|$ for every vertex $v$. For the first bound we need to guarantee that no edge is scanned more than once by the same vertex. One can do it if each vertex $v$ maintains a bit flag for each edge $e$ adjacent to $v$ indicating whether $v$ has already scanned $e$ or not. Note also that if we do allow a vertex to scan an edge more than once than still $|M(v)| \leq 2t \cdot deg(v)$, because $v$ can scan each of the adjacent to it edges at most once per round, and the algorithm runs for at most $2t$ rounds. Consequently, the worst-case processing time-per-edge will be only slightly higher, specifically, $O(\sigma(t \cdot deg(v)))$.

Finally, since $|M(v)| \leq deg(v)$, and $|Sp(v)| \leq deg(v)$, and since each edge and label can be represented with $O(\log n)$ bits, it follows that each vertex $v$ uses space of $O(deg(v) \cdot \log n)$. Since a vertex needs always to maintain at least one bit for each edge adjacent to it, this space requirement is *optimal up to a logarithmic factor*.

We need some additional notation. For a fixed graph $G = (V, E)$, a vertex $v \in V$, and a round $i$, let $Q(v, i)$ be the subset of edges adjacent to $v$ that are read by $v$ on round $i$, and let $\rho(v, i)$ denote the ordering of $Q(v, i)$ that determines the order in which the vertex $v$ reads the edges of the set $Q(v, i)$.

We summarize this section with the following theorem.

**Theorem 5.9** *Let $n, t$, $n \geq t \geq 1$, be positive integers. Consider an execution of our algorithm in the distributed synchronous static model on an input (unweighted undirected) $n$-vertex graph $G = (V, E)$ such that the graph and the orderings $\rho(v, i)$, for every vertex $v \in V$ and round $i \in [2t]$, are chosen by a non-adaptive adversary obliviously of the coin tosses of the algorithm. The algorithm constructs a $(2t-1)$-spanner of the input graph. The expected size of the spanner is $O(t \cdot n^{1+1/t})$ or the size of the spanner is $O((t \cdot \log n)^{1-1/t} \cdot n^{1+1/t})$ with high probability (depending on the choice of $p$, see Theorem 3.11). The running time of the algorithm is at most $2t$ rounds, its message complexity is $O(t \cdot |E|)$. Also, each vertex $v$ uses $O(deg(v) \cdot \log n)$ bits of space. Moreover, the local processing time-per-edge of each vertex $v$ running the algorithm on each round is expected $O(1)$, worst-case $O(\sigma(deg(v)))$, and $O(\chi(n, t))$ with high probability. The algorithm uses only messages of size $O(\log n)$.*

## 6 A Distributed Incremental Algorithm

We next show that the distributed static algorithm of Section 5 can be adapted in a very simple way to the distributed dynamic incremental environment, that is, to a network in which some edges may appear spontaneously, but no existing edge ever crashes.

## 6.1 The algorithm

The static algorithm of Section 5 can be seen in the following way. Each vertex $v$ maintains a local counter variable $ctr(v)$ which is initialized to 1. As long as the counter is no greater than $2t$, the algorithm invokes the Procedure *Round*.

To handle incremental topology updates (appearances of new edges) we introduce into the algorithm the following changes. First, it will now run forever, and invoke the Procedure *Round* as long as the variable $ctr(v)$ is no greater than $2t$. It will also increment the value of $ctr(v)$ after each invocation of the procedure. In addition, once a vertex $v$ detects a topology update (an appearance of a new edge adjacent to $v$) it will reset the value of $ctr(v)$ (set it to 1). The latter will make $v$ to invoke the Procedure *Round* for $2t$ additional times, regardless of the number of times $v$ invoked the procedure before the topology update was detected.

The formal description of the incremental algorithm follows.

---

**Algorithm 3** A distributed incremental synchronous algorithm for constructing a sparse $(2t-1)$-spanner. The pseudo-code is for a fixed vertex $v \in V$.

1: $ctr(v) \leftarrow 1$
2: On every round do (lines 3-9 in an infinite loop)
3: **if** some edge $e = (v, z)$ appeared on previous round **then**
4:     $ctr(v) \leftarrow 1$
5: **end if**
6: **if** $(ctr(v) \leq 2t)$ **then**
7:     Invoke Procedure *Round*
8:     $ctr(v) \leftarrow ctr(v) + 1$
9: **end if**

---

## 6.2 The analysis

In the context of dynamic distributed algorithms it is a common practice to consider the following scenario. Suppose that all the topology updates occur before a certain round $R$, and starting from the round $R$ the network becomes stable. At a certain later round $R'$, $R' \geq R$, the algorithm stabilizes and produces the correct output with respect to the last (stable) topology of the network. The number of rounds $R' - R$ that elapse between the last topology update and the first time when the output of the algorithm becomes correct is called the *quiescence time* of the algorithm. Particularly, this is the efficiency measure considered in [11].

Our result is, however, significantly stronger. Not only that we show that the quiescence time of our algorithm is at most $2t$, but rather that if an edge $e$ appears on certain round $R$, within $2t$ rounds after that event the spanner maintained by the algorithm will already adapt to guarantee the stretch of $(2t-1)$ for the edge $e$, even if many other (incremental) topology updates occurred in the meanwhile. Moreover, we show that the same statement holds true even if many edges showed up in the network at the same time as the edge $e$ did.

For an edge $e = (w, w')$, let the *treatment time of $e$* be the duration of the time period (in terms of the number of rounds for the synchronous environment, and in terms of the time units for the asynchronous one) that elapses between the moment that $e$ shows up in the network, and the moment that the spanner maintained by the algorithm starts to span the edge (that is, to provide a stretch guarantee of $2t - 1$ for it). We note that in our algorithm for the incremental environment edges are never deleted from the spanner, and so once it starts to span a certain edge, it will keep doing so forever. We will show that the worst-case treatment time of every edge in our algorithm is at most $2t$.

We start with arguing that every edge $e = (w, w')$ is scanned by the algorithm within $2t$ rounds after it shows up in the network. Suppose that $e$ appears on round $R$. Then at the beginning of the round $R+1$ both $w$ and $w'$ reset their counters $ctr(w)$ and $ctr(w')$, respectively, and read this edge for the $2t$ following rounds. (We assume here that both endpoints detect occurrence of the edge that connects them simultaneously.) Hence, by the same argument as in the proof of Lemma 5.4, the edge $e$ is scanned on round $R + 2t$ or earlier.

It is easy to see that Lemma 5.8 keeps holding, that is, if an edge is scanned by the algorithm on round $i$, the spanner starts to span the edge starting with the end of this round. Finally, assuming that the graph $G = (V, E)$, and the sequence of edge sets added to $G$ on different rounds is determined by a non-adaptive adversary, obliviously of the random coin tosses of the algorithm, the same argument as we used for proving Lemma 5.5 and Lemma 5.6 shows that the size of the spanner is $O((t \cdot \log n)^{1-1/t} \cdot n^{1/t})$ with high probability.

We summarize this discussion with the following theorem.

**Theorem 6.1** *For $n, t$, $n \geq t \geq 1$, and an adversary as in Theorem 5.9, Algorithm 3 maintains a dynamic $(2t - 1)$-spanner of an input n-vertex graph in a distributed incremental environment. This spanner contains $O((t \cdot \log n)^{1-1/t} \cdot n^{1+1/t})$ edges with high probability. The treatment time of an edge in the algorithm is at most $2t$. The space requirements, and the local processing time-per-edge of the algorithm are listed in Theorem 5.9. (The notation $deg(v)$ means here the maximum degree of the vertex $v$ at some point of the execution.)*

## 6.3 Extensions of the Incremental Distributed Algorithm

In this section we strengthen Theorem 6.1 further, and show that by introducing some minor changes to the algorithm one can guarantee that the treatment time decreases drastically if we restrict the sequence of topology updates to a certain convenient structure. In particular, we show that if the set of edges added on each round forms a partial matching, then the treatment time becomes *one single round*.

The modification that we introduce to the algorithm is based on the observation that once an edge $e = (v, v')$ is scanned, there is no need to scan it again, as long as the network experiences only incremental updates. This observation follows from Lemma 5.8. So it makes sense for a vertex $v$ to maintain an array of bit flags $Scanned\_Edges(v)$ with an entry for every edge adjacent to $v$. These flags are all initialized as 0, and once $v$ learns that $e = (v, v')$ was scanned it marks the corresponding entry of the array $Scanned\_Edges(v)$ by 1, and never reads this edge again. A vertex $v$ can learn that an edge $e = (v, v')$ was scanned in one of the two ways. Either it scans it itself, or it was scanned by the other endpoint $v'$, and $v$ receives a message from $v'$ indicating it. So on line 17 of the Procedure $Read\_Edge$ (see Algorithm 2), the vertex $v$ will send the message $(P(v), ttl(v))$ to each of its neighbors $u$ such that the edge $(v, u)$ is not scanned (to the best of the knowledge of $v$), and will send the message $SCANNED$ to each vertex $v'$ such that the edge $(v, v')$ was just scanned by $v$. The vertex $v'$ marks the corresponding entry of its $Scanned\_Edges(v')$ array upon receival of the message $SCANNED$. We will refer to this modified version of Algorithm 3 as Algorithm $Sync\_Incr$.

We assume that the network is *stable* before the first round, meaning that all edges already present in the network were scanned by the algorithm before the first round. Note that this assumption is not needed for Theorem 6.1 to hold.

In other words, we consider the scenario in which our distributed algorithm worked for $2t$ rounds in a static network, terminated and produced the spanner with the desired properties and the appropriate satellite data structures $P(v), M(v), Sp(v)$, for every vertex $v$. After this happened, the network kept running the algorithm. In other words, each vertex $v$ waits for updates, re-sets the counter $ctr(v)$ once it detects an update, and starts executing the algorithm again.

Let $F_1, F_2, \ldots$ be a sequence of edge sets, where $F_i$ is the set of edges that appear in the graph on round $i$, $i = 1, 2, \ldots$. Suppose that each set $F_i$ is a partial matching, that is, each vertex $v$ has degree at most 1 in each $F_i$. (Particularly, some sets $F_i$ may be empty.) Suppose also that the network is stable at the beginning of round 1. We show that in this case the treatment time of our modified algorithm is *one*, that is, for every $i = 1, 2, \ldots$, and $e \in F_i$, the spanner $H_{i+1}$ spans $e$.

This result is a consequence of the following lemma.

**Lemma 6.2** *For every $i = 1, 2, \ldots$, and edge $e \in F_i$, the edge $e = (w, w')$ is scanned on round $i$.*

**Proof:** The proof is by induction on $i = 1, 2, \ldots$. The induction claim is that before the round $i$ starts (but after the round $i - 1$ ends), the network is stable, and all the edges of $\bigcup_{j=1}^{i-1} F_j$ were already scanned, and, moreover, for each index $j = 1, 2, \ldots, i - 1$, and every edge $e \in F_j$, the edge $e$ was scanned on round $j$.

The induction base holds vacuously. For the induction step suppose that the statement holds for some integer $i$, $i \geq 1$, and consider an edge $e = (w, w') \in F_{i+1}$. Let $P(w)$ (respectively, $P(w')$) be the label of the vertex $w$ (resp., $w'$) at the beginning of the round $i + 1$. Suppose without loss of generality that $P(w) \prec P(w')$. By induction hypothesis and since $F_{i+1}$ is a partial matching, all the edges adjacent to $w$ but the edge $e$ were already scanned. Consequently, the vertex $w$ will start the round $i + 1$ with reading the edge $e$. Then it will discover that $P(w) \prec P(w')$, and scan the edge. ∎

By Lemma 5.8, once an edge is scanned, the spanner starts to span it. We summarize this extension of Theorem 6.1 with the following corollary.

**Corollary 6.3** *For $n, t$, $n \geq t \geq 1$, consider an execution of the Algorithm Sync_Incr on an input $n$-vertex graph $G$. Suppose that before the beginning of round 1 the network maintains a $(2t - 1)$-spanner $H$ constructed by the Algorithm Sync_Incr, and the network is stable. Let $F_1, F_2, \ldots$ be a sequence of edge sets, with $F_i$ being the set of edges that show up in the network on round $i$. Suppose that each $F_i$ is a partial matching.*

*Then the algorithm will keep maintaining a $(2t - 1)$-spanner $H$, and for every edge $e \in \bigcup F_i$, the treatment time of $e$ is 1. Moreover, if the graph, the ordering of edges, and the topology updates are determined by a non-adaptive adversary obliviously of the random coins of the algorithm, then with high probability the spanner contains $O((t \cdot \log n)^{1-1/t} \cdot n^{1+1/t})$ edges at every step of the algorithm.*

We next extend Corollary 6.3 to scenarios when the sets $F_i$ have small maximum degree. Let $\Delta_i$ be the maximum degree of a vertex in the graph $(V, F_i)$, and $R_i$ be the index of the round on which the edges of $F_i$ appear in the network. We show that if $R_{i+1} - R_i \geq 2\Delta_i - 1$ for every $i = 1, 2, \ldots$, then the treatment time of an edge $e \in F_i$ is at most $2\Delta_i - 1$, for every $i = 1, 2, \ldots$. We then argue that the bound of $2\Delta_i - 1$ cannot be further improved using our algorithm.

**Lemma 6.4** *Consider a sequence of update edge sets $F_1, F_2, \ldots$ that arrive on rounds $R_1, R_2, \ldots$, respectively, and assume that before the round $R_1$ starts the network is stable, and that $R_{i+1} - R_i \geq 2\Delta_i - 1$, for $i = 1, 2, \ldots$, where $\Delta_i$ be the maximum degree of a vertex in $F_i$. For $i = 1, 2, \ldots$, every edge $e \in F_i$ is scanned by the Algorithm Sync_Incr within $2\Delta_i - 1$ rounds after it appears in the network.*

**Proof:** Similarly to Lemma 6.2, we prove by induction on $i$, $i = 1, 2, \ldots$, that before the round $R_i$ starts the network is stable, and all the edges of the set $\bigcup_{j=1}^{i-1} F_j$ were already scanned, and moreover, each edge $e \in F_j$, $j \in [i-1]$, was scanned within $2\Delta_j - 1$ rounds after its appearance.

The induction base holds vacuously. Suppose the induction hypothesis, and consider an edge $e = (w, w') \in F_i$, appearing on round $R_i$. At each moment of the execution of the algorithm, let the *active* *degree* of an edge $e = (w, w')$, denoted $actdeg(e)$, be the number of the not yet scanned edges adjacent to either $w$ or $w'$. For a round $R$, let $actdeg_R(e)$ denote the active degree of the edge $e$ at the beginning of the round $R$.

There are at most $\Delta_i - 1$ edges $e' \in F_i$ adjacent to $w$, $e' \neq e$, at most $\Delta_i - 1$ edges $e' \in F_i$, $e' \neq e$, adjacent to $w'$, and the edge $e$ also contributes 1 to the active degree of $e$ on round $R_i$. Finally, by the induction hypothesis, the network is stable right before the round $R_i$ starts, and consequently, all edges adjacent to one of the endpoints of $e$ that do not belong to $F_i$ were already scanned at this stage. Hence $actdeg_{R_i}(e) \leq 2\Delta_i - 1$.

Consider a round $R$, $R_i \leq R \leq R_i + (2\Delta_i - 1)$, such that the edge $e$ is not yet scanned in the beginning of the round $R$. We argue that the active degree of $e$ necessarily decreases at least by 1 on each such a round.

Suppose without loss of generality that $P_R(w) \prec P_R(w')$, where $P_R(w)$ (respectively, $P_R(w')$) stands for the value of the label $P(w)$ (resp., $P(w')$) at the beginning of the round $R$. At the time moment $\alpha$ on which the vertex $w$ reads the edge $e = (w, w')$ on round $R$, if the value of $P(w)$ is still smaller (in terms of $\prec$) then $P_R(w')$ then the edge $e$ is scanned, and we are done. It can happen, however, that $P(w)$ grew during the time period starting with the beginning of the round $R$ and ending with the moment $\alpha$. But this could only happen if $w$ scanned another edge $e'$ adjacent to it, and then its active degree decreased by 1. This completes the proof. ∎

The desired extension of Corollary 6.3 follows now directly from Lemma 5.8.

**Corollary 6.5** *Under assumptions of Corollary 6.3, let $R_1, R_2, \ldots$ be a monotonely increasing sequence of rounds, and $F_1, F_2, \ldots$ be a sequence of edge sets with $F_i$ being the set of edges that appear on round $R_i$, $i = 1, 2, \ldots$. Let $\Delta_i$ be the maximum degree of a vertex in $F_i$, and suppose that $R_{i+1} - R_i \geq 2\Delta_i - 1$.*

*Then Algorithm Sync_Incr keeps maintaining a $(2t-1)$-spanner that satisfies the properties listed in Corollary 6.3, except for the treatment time of edges of $\bigcup F_i$. In this scenario for every $i = 1, 2, \ldots$, and an edge $e \in F_i$, the treatment time of $e$ is at most $2\Delta_i - 1$.*

We remark that Theorem 6.1 provides a uniform bound of $2t$ rounds on the treatment time of every edge appearing in the network in *any incremental scenario*, and the limitations of Corollaries 6.3 and 6.5 are geared to single out some cases where (the already very strong) uniform upper bound of $2t$ can be further improved.

Finally, we show that the bound of $2\Delta_i - 1$ cannot be improved using our algorithm as is. Consider the following directed graph $\hat{F}$. The vertex set of the graph consists of the "central vertex" $v$, the vertices $u(1), u(2), \ldots, u(\Delta)$, the vertices $z(1), z(2), \ldots, z(\Delta)$, and the vertices $w(1,1), w(1,2), \ldots, w(1, \Delta-2), w(2,1), \ldots, w(2, \Delta-2), \ldots, w(\Delta, 1), \ldots, w(\Delta, \Delta-2)$. The edge set contains the oriented edges $\{\langle u(i), v \rangle \mid i \in [\Delta]\}$, the edges $\{\langle z(i), w(i,j) \rangle \mid i \in [\Delta], j \in [\Delta-2]\}$, the edges $\{\langle u(i), w(i,j) \rangle \mid i \in [\Delta], j \in [\Delta-2]\}$, and, finally, $\{\langle z(i), u(i) \rangle \mid i \in [\Delta]\}$. See Figure 1.

We next describe an execution scenario of our algorithm. Suppose that the undirected underlying graph $F$ of $\hat{F}$ is the update edge set provided to the algorithm on round $R$. An edge is oriented from a vertex $x$ to a vertex $y$ if and only if $P(x) \prec P(y)$. Suppose that on round $R$ the labels of $V(\hat{F})$ satisfy the order relationship determined by the directed graph $\hat{F}$. The maximum degree of $F$ is $\Delta$.

On the first round (after detecting the edges of $F$) each vertex $u(i)$, $i \in [\Delta]$, scans the edge $(u(i), w(i, 1))$, and each vertex $z(i)$, $i \in [\Delta]$ scans the edge $(z(i), w(i, 1))$. (The vertex $v$ scans nothing as all its adjacent edges are incoming.)

On the second round each $u(i)$ scans the the edge $(u(i), w(i, 2))$, and each $z(i)$ scans $(z(i), w(i, 2))$, etc., for each round $j = 1, 2, \ldots, \Delta - 2$, on round $j$ the edges $(u(i), w(i, j))$ and $(z(i), w(i, j))$ for all $i \in [\Delta]$ are scanned. Hence after $\Delta - 2$ rounds we are left with the graph $F'$ in which only the edges $\{(z(i), u(i)), (u(i), v) \mid i \in [\Delta]\}$ are present.

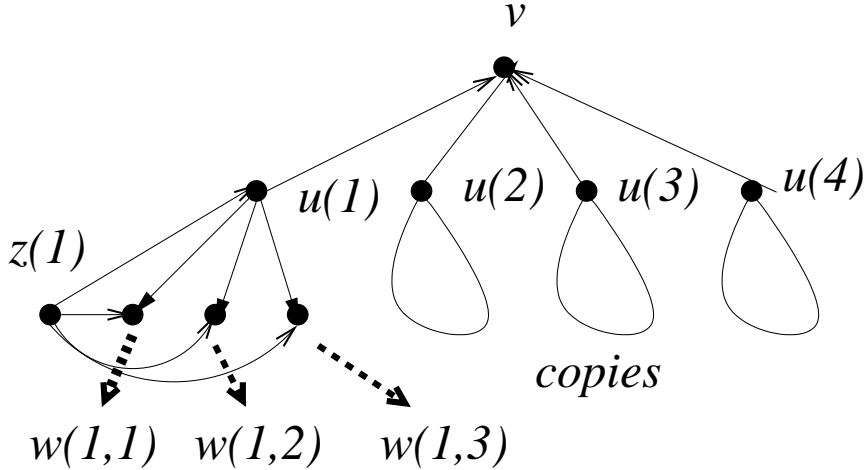

Figure 1: The graph $\hat{F}$ with $\Delta = 4$. The edges are depicted by thin solid edges. The dashed thick lines are used to associate the vertices $w(1,1)$, $w(1,2)$ and $w(1,3)$ with the points that represent these vertices in the figure. The three "sacks" depict three isomorphic copies of the graph induced by the vertices $z(1), u(1), w(1,1), w(1,2), w(1,3)$.

Moreover, suppose that on round $\Delta - 2$ each edge $(z(i), u(i))$ "switches orientation". This can happen if each $z(i)$, $i \in [\Delta]$, adapted the label of the "late" $w(i, \Delta - 2)$ (whose active degree has just become zero), and this label is greater than that of $u(i)$. The resulting edge set is $\{\langle u(i), z(i)\rangle, \langle u(i), v\rangle \mid i \in [\Delta]\}$.

On the $(\Delta-1)$st round every vertex $u(i)$ scans the edges $(u(i), z(i))$, $i \in [\Delta]$, and its labels grow enough for all the remaining edges $(u(i), v)$ to switch orientation. We then obtain the star $\{\langle v, u(i)\rangle \mid i \in [\Delta]\}$. Now $v$ takes its time scanning each of the $\Delta$ edges adjacent to it one by one. Consequently, in this scenario $2\Delta - 1$ rounds are required to the algorithm to scan all the edges of a set $F$ of maximum degree $\Delta$, and the analysis of Corollary 6.5 is tight.

# 7 Adapting the Algorithm to the Asynchronous Environment

In this section we adapt our algorithm to the asynchronous model of distributed computation.

Note that one can use a standard technique of $\alpha$-*synchronization* [7, 30] that converts any synchronous algorithm into an asynchronous one. However, this approach has several drawbacks. First, the upper bound on the running time of the algorithm would grow by a factor of 3, that is, become $6t$ instead of $2t$. While it is a common practice to ignore this constant factor of 3 in the running time, the running time of our algorithm is so small that this factor may well be significant. (Recall that $t$ is usually a constant, and it is always true that $t = O(\log n)$.)

Another drawback of $\alpha$-synchronization is that it also entails an increase in a message complexity of the algorithm. Specifically, a message is sent on each time unit through every edge of a network that runs an $\alpha$-synchronizer. Consequently, using $\alpha$-synchronization would increase the message complexity of our algorithm by a factor of 3 *for every input*, and there are many inputs on which the overhead is even larger. These are the scenarios on which most edges are scanned after being read only a constant number of times, but the algorithm still keeps running for $2t$ rounds in all, or almost all, vertices. The message complexity of the synchronous variant on these inputs is $O(|E|)$, while the message complexity of the asynchronous variant is $O(t \cdot |E|)$.

Finally, and perhaps, most importantly, $\alpha$-synchronization is a *static* synchronization technique, and it is not a-priori clear that it is applicable to a dynamic (though only incremantal) setting. In fact, Awerbuch and Sipser addressed this problem in [15], and have devised a dynamic analogue of $\alpha$-synchronization, but their synchronizer entails even greater overheads in time, message, and space complexities. Synchronization methods other than $\alpha$-synchronization require a significantly larger initialization time than the running time of our algorithm, and consequently, they are not applicable in this context (unless one is prepared to compromise the running time).

In view of these drawbacks it is advantageous to adapt our algorithm directly to the asynchronous environment, and we do so in this section.

## 7.1 The Asynchronous Algorithm

Each vertex $v$ marks each of the edges $e$ adjacent to $v$ by either *OLD* or *NEW* marks. These marks are stored in fields $v.mark(e)$, for every vertex $v$ and edge $e$ adjacent to $v$. When an edge $e = (w, w')$ appears in the network, both its endpoints mark it as *NEW* (set the fields $w.mark(e)$ and $w'.mark(e)$ to *NEW*) when they detect it, and each of them switches its mark to *OLD* after sending the first message through it.

Each vertex $v$ also maintains a field $v.status(e)$ for every edge $e$ adjacent to $v$, and this field is equal to either *SCANNED* or *NOTSCANNED*, depending on whether the edge was or was not scanned by the algorithm. All the statuses are initialized as *NOTSCANNED*.

To adapt Algorithm 3 to the asynchronous setting we invoke the Procedure *Async_Rnd* instead of the Procedure *Round* on line 7 of the algorithm. We next describe the Procedure *Async_Rnd*. First, the vertex $v$ initializes the fields $v.mark(e)$ and $v.status(e)$ for every edge $e$ adjacent to $v$ that just appeared in the network. Then for every edge $e = (v, u)$ adjacent to $v$ over which $v$ received the message *SCANNED*, $v$ changes the status of $e$ to *SCANNED*. Next, the vertex $v$ waits until it receives a message over each of the edges adjacent to $v$ that are marked as *OLD* and have status *NOTSCANNED*.

Then the vertex $v$ executes the while loop which is very similar to the while loop of Algorithm 2. The only difference is that every edge $(v, u)$ for which $P(u) \succ P(v)$ holds becomes scanned by $v$, and $v$ updates the status $v.status(e)$, and sends the message *SCANNED* to $u$ over $e$.

Finally, after the execution leaves the while loop, $v$ sends the message $(P(v), ttl(v))$ to each neighbor $u$ of $v$ such that $e = (v, u)$ has status *NOTSCANNED*, and sets the mark of each edge $e$ as above to *OLD* (if it was *NEW*). This means that the vertex $v$ will not start the while loop of the next round before getting a message over this edge.

The formal description of the Procedure *Async_Rnd* follows.

**Algorithm 4** The Procedure $Async\_Rnd$. The pseudo-code is for a fixed vertex $v \in V$.
1: **for** each edge $e = (v, u)$ that appeared in the network, and its appearance is now detected by $v$ **do**
2:   $v.status(e) \leftarrow NOTSCANNED$; $v.mark(e) \leftarrow NEW$
3: **end for**
4: **for** each edge $e$ such that the message $SCANNED$ arrived over $e$ **do**
5:   $v.status(e) \leftarrow SCANNED$
6: **end for**
7: Wait until you receive messages $(P(u), ttl(u))$ over all edges $e = (u, v)$ such that $v.status(e) = NOTSCANNED$ and $v.mark(e) = OLD$
8: **while** $\exists$ message $(P(u), ttl(u))$ with $P(u) \succ P(v)$ **do**
9:   lines 3-7 of the Procedure $Round$ (Algorithm 2)
10:   $v.status(e) \leftarrow SCANNED$
11:   Send the message $SCANNED$ to $u$
12: **end while**
13: Send to all your neighbors $u$ such that $v.status(e = (v, u)) = NOTSCANNED$ the message $(P(v), ttl(v))$
14: **for** each edge $e$ as above with $v.mark(e) = NEW$ **do**
15:   $v.mark(e) \leftarrow OLD$
16: **end for**

## 7.2 The Analysis

In this section we analyze the asynchronous incremental algorithm, Algorithm 4. We start with a few observations. First, the messages are not labeled by the current number of round of the sender, and, consequently, the version of the algorithm that we presented assumes that for a fixed edge $e = (v, u)$, all the messages sent by an endpoint $v$ over $e$ arrive to $u$ in the First-In-First-Out (henceforth, FIFO) manner. This assumption can be eliminated at the expense of making the analysis slightly more complex. (The main change in the algorithm will be attaching the current round of the sender to every message sent.) Since our argumentation here extends in a standard way to this more general case, we opt for presenting a somewhat simpler variant.

In this context note also that with this algorithm, a vertex $v$ never stores more than two messages received from the same neighbor $w$ of $v$. This is because the algorithm dictates a "ping-pong" of messages between $v$ and $w$. In other words, suppose that the vertex $v$ sends a message $\mathcal{M}_1$ to $w$, and receives a message $\mathcal{M}_2$ from $w$. It may happen that $w$ finishes processing $\mathcal{M}_1$, and sends $v$ a message $\mathcal{M}_3$, but for sending $v$ a yet additional message $\mathcal{M}_4$, the vertex $w$ has first to receive a new message $\mathcal{M}_1'$ from $v$. However, for $v$ to send $\mathcal{M}_1'$, $v$ must finish processing $\mathcal{M}_2$. Once $v$ finishes processing $\mathcal{M}_2$, $v$ discards it. See Figure 2. This property enables to abandon the assumption that messages arrive in the First-In-First-Out manner. Specifically, if this is not the case, it is sufficient to attach a parity bit to each message, and to make the receiver wait for the message with the correct parity.

Consider an edge $e = (v, u)$. Consider the pair of invocations of the Procedure $Async\_Rnd$ by the vertices $v$ and $u$ on which $v$ and $u$, respectively, detected the appearance of the edge $e$, and sent messages over $e$ on line 14. This pair of invocations is called the *first round with respect to the edge $e$*. Analogously, for some integer $j$, $j \geq 1$, consider the pair of invocations of the Procedure $Async\_Rnd$ by $v$ and $u$ on which $v$ and $u$, respectively, have sent their $j$th messages over $e$. This pair of invocations is called the *$j$th round with respect to $e$*.

Equipped with these observations and definitions, we can now use the definitions of the beginning of Section 5.2 of *reading* and *scanning* an edge on a certain round. It is easy to verify that the proofs of Lemmas 5.2-5.7 hold for Algorithm 4 as well (or, rather for a static version of Algorithm 4 that invokes

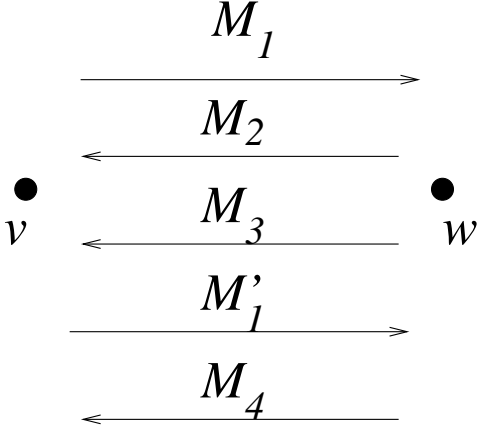

Figure 2: The figure illustrates the flow of messages between the vertices $v$ and $w$.

the Procedure $Async\_Rnd$ for $2t$ times).

The only change needed in the statement of Lemma 5.8 is that it has to refer now to the round $i$ *with respect to the edge $e$*. The proof of Lemma 5.8 holds for Algorithm 4. Note that $2t$ rounds with respect to an edge $e$ last at most $2t$ time units. Consequently, Theorems 5.9 and 6.1 extend to the asynchronous environment. (The running time becomes $2t$ time units. In the statement concerning the local processing time-per-edge one needs to refer to local clock cycles of a vertex rather than to rounds.)

To extend the results of Section 6.3 to the asynchronous setting one needs to re-cast the statements of these results. Let $F_1, F_2, \ldots$ be a sequence of edge sets that appear in the network at time moments $\alpha_1, \alpha_2, \ldots$, respectively, and assume that $\alpha_1 < \alpha_2 < \ldots$. Assume that the network is stable at time $\alpha_1$, and that for every index $j = 2, 3, \ldots$ the duration of the time period $[\alpha_{j-1}, \alpha_j)$, $\alpha_j - \alpha_{j-1}$, is greater than two time units, and that each $F_i$ is a partial matching. Then the treatment time of each edge $e \in F_i$ is at most *two time units*. (Note that there is a certain "price" for asynchrony here, in having the treatment time of two time units instead of one round in the synchronous case.)

To prove this statement, we re-formulate Lemma 6.2 as follows.

**Lemma 7.1** *For every $i = 1, 2, \ldots$, and edge $e \in F_i$, the edge $e = (u, w)$ is scanned at time $\alpha_i + 2$ or before.*

**Proof:** The induction claim is that at time $\alpha_i + 2$ the network is stable, all the edges of $\bigcup_{j=1}^{i} F_j$ were already scanned, and, moreover, for each index $j \in [i]$, and every edge $e \in F_j$, the edge $e$ was scanned no later than at time $\alpha_j + 2$.

For the induction base ($i = 1$) consider an edge $e = (w, w') \in F_1$. Let $\beta_1$ (respectively, $\beta_1'$) be the smallest time moment greater or equal to $\alpha_1$ such that $w$ (resp., $w'$) invokes the Procedure $Async\_Rnd$ at that time. The vertex $w$ (resp., $w'$) sends $P(w)$ (resp., $P(w')$) over the edge $e$ at time $\beta_1$ (resp., $\beta_1'$) on line 14 of this invocation of the Procedure $Async\_Rnd$. Let $\gamma_1$ (resp., $\gamma_1'$) be the time moment when $w$ receives and starts processing $P(w')$ (resp., $P(w)$) on line 7 of the Procedure $Async\_Rnd$.

Note that $w$ detects $e$ on its first round with respect to $e$, and it processes $e$ (enters the while loop with a message received over $e$) only on its second round with respect to $e$. The same statement is true for the vertex $w'$. In other words, $\alpha_1 \le \beta_1 \le \gamma_1$ and $\alpha_1 \le \beta_1' \le \gamma_1'$. See Figure 3 for an illustration.

It follows that $\beta_1, \beta_1' \le \alpha_1 + 1$, and $\gamma_1, \gamma_1' \le \alpha_1 + 2$. Since the network is stable at time $\alpha_1$, all edges adjacent to $w$ or $w'$ are scanned at time $\alpha_1$. Moreover, $F_1$ is a partial matching, and so $e$ is the only new edge appearing in the network during the time period $[\alpha_1, \alpha_1 + 2]$, adjacent to either $w$ or $w'$. It follows that the label of $w$, $P(w)$, did not change during the time period $[\alpha_1, \gamma_1)$. Analogously, $P(w')$ did not change during the time period $[\alpha_1, \gamma'_1)$.

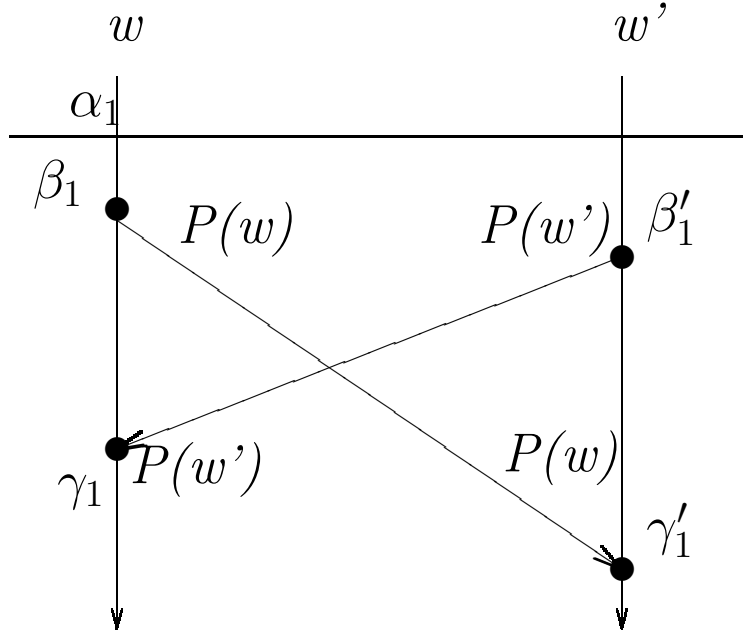

Figure 3: The time scale of the events in the proof. The events occurring with the vertex $w$ (respectively, $w'$) are depicted in the left (resp., right) vertical line, the horizontal line represents the time moment $\alpha_1$, and the diagonal lines represent to flow of the messages between $w$ and $w'$. The time axis is depicted by vertical lines, and it is oriented from top to bottom.

Hence if $P(w) \prec P(w')$ then $w$ scans $e$ at time $\gamma_1 \leq \alpha_1 + 2$, and otherwise $w'$ scans $e$ at time $\gamma'_1 \leq \alpha_1 + 2$. In either case we are done.

The induction step is proven in an analogous way, by combining this argument with that of proof of Lemma 6.2. ∎

Hence, Corollary 6.3 extends to the asynchronous setting, provided that the update edge sets $F_1, F_2, \ldots$ arrive with more than two time units between each pair of consequent sets. Corollary 6.5 extends in an analogous way. The extended result is summarized in the following corollary.

**Corollary 7.2** *Consider a sequence of update edge sets that arrive at time moments $\alpha_1, \alpha_2, \ldots$, $\alpha_1 < \alpha_2 < \ldots$, and assume that the network is stable at time $\alpha_1$, and that $\alpha_{i+1} - \alpha_i > 2\Delta_i = 2\Delta(F_i)$, for all $i = 1, 2, \ldots$. For $i = 1, 2, \ldots$, every edge $e \in F_i$ is scanned no later than at time $\alpha_i + 2\Delta_i$.*

# 8 Adapting the Algorithm to the Distributed Semi-Decremental Setting

In the following two sections we present a fully dynamic version of our distributed algorithm, and analyze it.

## 8.1 Preliminaries

Each vertex maintains explicit variables $X(v)$ and $T(v)$, $X(v) \cup T(v) = Sp(v)$. The edges of $X(v)$ (respectively, $T(v)$) will be called the $X$-edges (resp., $T$-edges) adjacent to $v$. A crash of an edge $e = (u, w)$ is called a *soft* crash if $e \notin (T(u) \cup T(w))$, and is called *hard* crash otherwise. The *semi-decremental scenario* is the scenario that assumes that no hard crashes occur. In this section we analyze this scenario, and we extend the analysis to the fully dynamic setting in Section 9.

During an execution of the algorithm, each vertex $v$ acquires labels once in awhile, and loses labels once in awhile (as a result of edge crashes). A label $P$ of $v$ is called *active* at a certain time moment $\alpha$ if $v$ was labeled by $P$ at some earlier time during the execution, and the label $P$ was not lost since then. Note that a vertex is allowed to acquire a new label without losing his previous label, and his previous label remains active unless it was explicitly lost as a result of certain edge crashes. We will soon elaborate on it.

Each vertex $v$ maintains a table $A(v)$ that keeps track of all active labels of $v$. The number of active labels for a given vertex never exceeds the parameter $t$. In the semi-decremental setting labels are never lost by vertices, and thus, each vertex is labeled by at most $t$ labels during the execution. Note that the table $A(v)$ has a different functionality than the table $M(v)$. For every base value $B$ of a label $P$ appearing in $M(v)$, the algorithm maintains a queue $M(v)[B]$ of edges $(v, z)$ with $P(z) \succ P(v) = P$, $B(P(z)) = B$. (The notations $P(z)$ and $P(v)$ refer to the labels of $z$ and $v$, respectively, at the time of scanning the edge $(v, z)$ by $v$.) The edges $(v, z)$ of this kind form a set denoted $D(v)$, and are called *dropped* or *D-edges*. In the static and incremental variants of the algorithm such edges were plainly discarded. However, if edges are allowed to crash, we may need these edges as a *replacement* for edges that are currently $X$-edges, in case that the latter edges happen to crash.

Fix a vertex $v$. For every edge $e = (v, w)$ adjacent to $v$, the vertex $v$ maintains a number of local variables. Particularly, the variable $v.mark(e)$ is equal to either $NEW$ or $OLD$ (see Section 7.1), and the variable $v.status(e)$ is equal to either $SCANNED$, $NOTSCANNED$ or $CRASHED$. (This variable was equal to $SCANNED$ or $NOTSCANNED$ so far because we did not consider edge crashes until now.) There is also a variable $v.scan\_status(e)$, called the *scan-status* of $e$, that accepts values $T$, $X$ or $D$, depending on whether the edge was scanned as a $T$-edge, $X$-edge, or $D$-edge, respectively.

There is also a variable $v.label(e)$ maintaining the label of the edge $e$ at the time it was scanned for the last time. (The label of $e$ is the label of one of the endpoints of $e$ at time that $e$ was scanned.) The label of the other endpoint of $e$ at the time of its scanning is stored at another variable $v.sec\_label(e)$. This variable is called the *second label* of the edge $e$. There is also a variable $v.own(e)$ that accepts values $SELF$ and $PEER$, depending on whether the edge $e = (v, w)$ was scanned by $v$ or $w$. (If $v$ scanned it then $v.own(e) = SELF$, and $w.own(e) = PEER$, and if $w$ scanned it then vice versa.) We also use the notation $v.X$, $v.T$, $v.D$, $v.Sp$, $v.A$, $v.M$ for denoting $X(v)$, $T(v)$, $D(v)$, $Sp(v)$, $A(v)$, and $M(v)$, respectively.

Fix a vertex $v$. We say that the edge $e = (v, u)$ adjacent to $v$ is *self-scanned* (respectively, *peer-scanned*) if $v.status(e) = SCANNED$ and $v.own(e) = SELF$ (resp., $v.own(e) = PEER$). Observe that the set $v.T$ (respectively, $v.X$; $v.D$) is the set of *self-scanned* $T$-edges (resp., $X$-edges; $D$-edges) adjacent to $v$.

We remark that the size of the data structures maintained by a vertex $v$ is still almost as small as it is for the incremental algorithm. Specifically, it is $O((deg(v) + t) \cdot \log n)$. (Though the constant hidden by $O$-notation is larger here than for the incremental algorithm). In particular, unlike many previous dynamic distributed algorithms (such as [15, 11]), our algorithm maintains *no explicit history* of previous communication.

## 8.2 The Algorithm

We next describe the version of our algorithm that is robust against soft crashes. The algorithm is identical to Algorithm 3 except that it invokes the Procedure *Dyn_Rnd* instead of the Procedure *Async_Rnd*.

We next describe the Procedure *Dyn_Rnd*. Let $v$ denote the vertex that executes the procedure. The procedure uses the same data structures as in Procedure *Async_Rnd*, and, in addition, it maintains the table $A(v) = v.A$ of *active labels* of the vertex $v$. This table is initialized as empty. As the execution proceeds, each label acquired by the vertex $v$ is recorded in the table $A(v)$. Later on crashes of certain edges may cause the vertex $v$ to remove some of the labels from the table $A(v)$. This cannot, however, happen in the semi-decremental setting, and thus we defer the discussion of this issue to Section 9.

The procedure starts with a loop that initializes the status and mark variables for all the edges that just appeared in the network. Next, the procedure goes over all the edges $e$ such that $v$ has just received the message *SCANNED* over them, and sets their status to *SCANNED*. However, unlike the Procedure *Async_Rnd*, in the Procedure *Dyn_Rnd* each message *SCANNED* will also convey some additional information, such as the scan-status of $e$, the label of $e$, and the second label of $e$ (that is, the label of the other endpoint of $e$ at the time it was scanned).

Next, the procedure enters the *crash-loop* (Procedure *CrashLoop*). The crash-loop consists of two for-loops. In the first for-loop all edges $e$ adjacent to $v$ that crashed on the previous round of $v$ are processed. If the edge $e$ was not yet scanned, or scanned by its other endpoint, then $v$ needs to do nothing beyond recording this crash, that is, setting $v.status(e)$ to *CRASHED*. Otherwise, if the status of $e$ is $D$ (the edge was dropped) then, in addition to marking the edge as *CRASHED*, the algorithm also updates the data structure $M(v)$ accordingly. We remark that even if a queue $v.M[B]$ becomes empty as a result of removing $e$, still the value $B$ should stay in the table $v.M$ because there is an $X$-edge adjacent to $v$ connecting $v$ to the cluster $B$. (The *cluster $B$* is the set of vertices that have an active label of base $B$.)

If the status of $e$ is $X$ ($e$ is an $X$-edge), then the procedure fetches from $M(v)$ a dropped edge that can replace $e$. This is done by the Procedure *XReplace*, described later in the sequel. Finally, if $e$ is a $T$-edge then the procedure updates the set $T(v)$, and invokes the Procedure *Crash*. Note, however, that in the semi-decremental scenario the execution cannot possibly reach this stage, because in this scenario $T$-edges never crash. Hence we defer the description of this procedure to Section 9.

Finally, the second for-loop of the Procedure *CrashLoop* iterates over all edges $e$ over which the vertex $v$ received the message *CRASH* on its previous round. However, messages *CRASH* are sent only from the Procedure *Crash*, and the Procedure *Crash* can be invoked only as a result of a crash of a $T$-edge (that is, a hard crash). Since in the semi-decremental setting hard crashes never occur, it follows that in this setting the execution never enters the second for-loop of the Procedure *CrashLoop*.

Next, the Procedure *Dyn_Rnd* waits until $v$ receives messages $(P(u), ttl(u))$ over all not scanned edges $e = (u, v)$, and processes them in a way similar to lines 8-13 of the Procedure *Async_Rnd* (Algorithm 4). One difference is that in the Procedure *Async_Rnd*, as well as in the Procedure *Round*, the edges $e = (u, v)$ such that $ttl(u) = 0$ (meaning that the label of the vertex $u$ is not selected) and $B(P(u)) \in M(v)$ are discarded, and in the Procedure *Dyn_Rnd* they are stored in an appropriate queue in $M(v)$. Another difference is that once a vertex $v$ acquires a new label $P(v)$ (this can happen only on line 11), this label is recorded in the table $A(v)$ of active labels of $v$.

Finally, the procedure sends the messages $(P(v), ttl(v))$ over all non-scanned edges adjacent to $v$, and updates the marks of edges over which the first message was sent (exactly like in the Procedure *Async_Rnd*).

The formal description of the Procedure *Dyn_Rnd* follows. Figure 4 illustrates the rules that are used by the procedure to determine the values of different variables for $T$-, $X$- and $D$-edges.

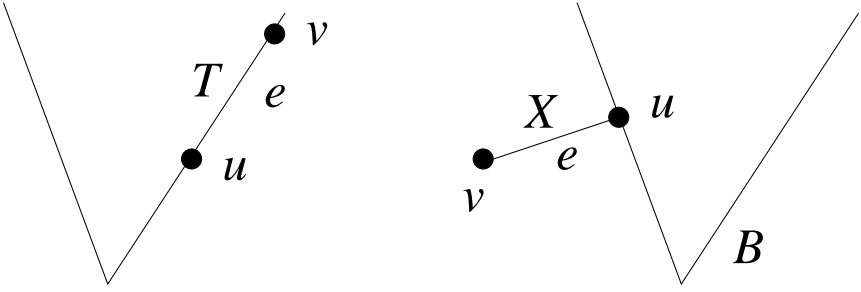

Figure 4: In the left-hand figure $e = (u, v)$ is a $T$-edge scanned by $v$. In this case $P(v)$ is set as $P(u) + n$, $v.label(e)$ is set to $P(v)$, $v.sec\_label(e)$ is set to $P(u)$, and $v.own(e)$ is set to $SELF$. In the right-hand figure $e = (u, v)$ is an $X$- or $D$-edge scanned by $v$. In this case the label of $v$ does not change, $v.label(e)$ is set to $P(u)$, $v.sec\_label(e)$ is set to $P(v)$, and $v.own(e)$ is set to $SELF$.

---

**Algorithm 5** The Procedure $Dyn\_Rnd$. The pseudo-code is for a fixed vertex $v \in V$.

1: **(Showup-loop)** lines 1-3 of the Procedure $Async\_Rnd$ (Algorithm 4)
2: Wait until some message is received over all edges $e = (u, v)$ such that $((v.status(e) = NOTSCANNED)$ and $(v.mark(e) = OLD))$
3: **for** each edge $e$ such that the message $SCANNED(scan\_status, P, P')$ arrived over $e$ **do**
4:    $v.status(e) \leftarrow SCANNED$; $v.scan\_status(e) \leftarrow scan\_status$; $v.label(e) \leftarrow P$; $v.sec\_label(e) \leftarrow P'$; $v.own(e) \leftarrow PEER$
5: **end for**
6: Invoke Procedure $CrashLoop$
7: Iterate over messages $(P(u), ttl(u))$ received over edges $e = (v, u)$ that satisfy the condition of line 2 in an arbitrary order and do
8: **while (Scan-loop)** $\exists$ message $(P(u), ttl(u))$ with $P(u) \succ P(v)$ **do**
9:    $v.status(e) \leftarrow SCANNED$; $v.own(e) \leftarrow SELF$
10:    **if** $ttl(u) > 0$ **then**
11:       $P(v) \leftarrow P(u) + n$; $ttl(v) \leftarrow ttl(u) - 1$; $v.A \leftarrow v.A \cup \{P(v)\}$
12:       $v.label(e) \leftarrow P(u) + n$; $v.T \leftarrow v.T \cup \{e\}$; $v.scan\_status(e) \leftarrow T$; $v.sec\_label(e) \leftarrow P(u)$
13:       Send the message $SCANNED(T, P(v), P(u))$ to $u$ over $e$
14:    **else**
15:       $v.label(e) \leftarrow P(u)$; $B \leftarrow B(P(u))$; $v.sec\_label(e) \leftarrow P(v)$
16:       **if** $B \notin v.M$ **then**
17:          $v.M \leftarrow v.M \cup \{B\}$; $v.X \leftarrow v.X \cup \{e\}$; $v.scan\_status(e) \leftarrow X$
18:          Send the message $SCANNED(X, P(u), P(v))$ to $u$ over $e$
19:       **else**
20:          $v.M[B] \leftarrow v.M[B] \cup \{e\}$; $v.scan\_status(e) \leftarrow D$; $v.D \leftarrow v.D \cup \{e\}$
21:          Send the message $SCANNED(D, P(u), P(v))$
22:       **end if**
23:    **end if**
24: **end while**
25: lines 14-17 of the Procedure $Async\_Rnd$

---

The formal description of the Procedure $CrashLoop$ follows.

**Algorithm 6** The Procedure *CrashLoop*.
1: **for** all edges $e$ adjacent to $v$ that crashed on the previous round of the vertex $v$ **do**
2:     *CrashItern(CRASHED)*
3: **end for**
4: **for** all edges $e$ adjacent to $v$ over which the message $CRASH$ arrived on the previous round of the vertex $v$ **do**
5:     *CrashItern(NOTSCANNED)*
6: **end for**

The Procedure *CrashItern* accepts as input one parameter *crash_status*, which may be equal to either $CRASHED$ or $NOTSCD$. This parameter determines the crash status of edges processed by the Procedure *CrashItern*. The pseudo-code of the Procedure *CrashItern* follows.

**Algorithm 7** The Procedure *CrashItern(crash_status)*.
1: **if** $(v.status(e) = NOTSCANNED)$ or $(v.own(e) = PEER)$ **then**
2:     $v.status(e) \leftarrow crash\_status$
3: **else**
4:     $v.status(e) \leftarrow crash\_status$; $P \leftarrow v.label(e)$; $B \leftarrow B(P)$
5:     **if** $v.scan\_status(e) = D$ **then**
6:         $v.M[B] \leftarrow v.M[B] \setminus \{e\}$; $v.D \leftarrow v.D \setminus \{e\}$
7:     **else if** $v.scan\_status(e) = X$ **then**
8:         *XReplace(P, e)*
9:     **else**
10:        $v.T \leftarrow v.T \setminus \{e\}$
11:        Invoke *Crash(P)*
12:     **end if**
13: **end if**

The Procedure *XReplace* accepts as input the label $P$ and the edge $e$. The procedure is invoked when the $X$-edge $e$ crashes, and the procedure makes an attempt to find a replacement for it. This replacement can come only from the set of dropped edges with base value $B$, stored in the queue $v.M[B]$. If this queue is not empty, the procedure fetches an edge $e'$ from this queue, removes it from the queue, and inserts it into the set $v.X$ of the self-scanned $X$-edges adjacent to $v$. If the queue is empty the procedure realizes that this edge has no replacement, and updates the set $v.M$ to reflect the fact that $v$ has no longer an edge connecting it to the cluster $B$. When the next edge of this kind will be scanned, the algorithm will not drop it, but rather use it as an $X$-edge. Finally, the crashed edge $e$ is removed from the set $v.X$ in either case.

There is one delicate point in this description. Some of the dropped edges from $v.M[B]$ may themself crash at the same time as $e$ crashes, and the algorithm still did not have a chance to update their status. To handle this difficulty, the procedure employs a while loop that fetches an edge until it finds an edge that did not crash, or discovers that the queue is empty. This loop is henceforth called *fetch-loop*.

The formal description of the Procedure *XReplace* follows.

**Algorithm 8** The Procedure *XReplace(P, e)*.

1: $B \leftarrow B(P)$; $fetched \leftarrow FALSE$
2: **while** (fetch-loop) $(v.M[B] \neq \emptyset)$ and $(fetched = FALSE)$ **do**
3:     Fetch an edge $e' = (v, u')$ from the queue $v.M[B]$
4:     **if** $e'$ crashed on the previous round of the vertex $v$ **then**
5:         $v.M[B] \leftarrow v.M[B] \setminus \{e'\}$; $v.D \leftarrow v.D \setminus \{e'\}$
6:     **else**
7:         $fetched \leftarrow TRUE$;
8:     **end if**
9: **end while**
10: **if** $fetched = TRUE$ **then**
11:     $v.X \leftarrow v.X \setminus \{e\} \cup \{e'\}$; $v.scan\_status(e') \leftarrow X$
12:     $v.M[B] \leftarrow v.M[B] \setminus \{e'\}$; $v.D \leftarrow v.D \setminus \{e'\}$
13:     Send the message $SCANNED(X, P, v.sec\_label(e'))$ to $u'$ over $e'$
14: **else**
15:     $v.M \leftarrow v.M \setminus \{B\}$; $v.X \leftarrow v.X \setminus \{e\}$
16: **end if**

We remark that the algorithm does not really need to maintain the set $v.D$ explicitly.

## 8.3 The analysis

In this section we show that the algorithm described in Section 8.2 maintains a sparse spanner in a semi-decremental setting.

### 8.3.1 The Preliminaries

We start with presenting some notation. Fix a vertex $v$, and a round $R$ of $v$. (If the setting is synchronous, this is a global round $R$. If it is asynchronous, this is the $R$th clock cycle of $v$.) We refer to the values of different variables local to $v$ at the beginning of round $R$ using the bracket-notation $\{R\}$. Particularly, $v.T\{R\}$ is the value of the variable $v.T$ at the beginning of round $R$ of $v$. We also use the notation $v.T\{R, END\}$ to denote the value of the variable $v.T$ at the end of round $R$ of $v$. Note that $v.T\{R, END\} = v.T\{R+1\}$.

Let $\ell$, $\ell \geq 0$, be the number of edges adjacent to $v$ that are processed by $v$ in the crash-loop of round $R$ of $v$. Let $k$, $k \geq 0$, be the number of edges adjacent to $v$ that $v$ scans on its round $R$. For an index $i$, $1 \leq i \leq \ell + k$, let $\{v, R, i\}$ denote the time moment on which the vertex $v$ starts processing the $i$th edge adjacent to it on its round $R$, and $v.T\{R, i\}$ denote the value of the variable $v.T$ at time $\{v, R, i\}$. Note that the time moment $\{v, R, 1\}$ is the time moment when the vertex $v$ starts the crash-loop of its round $R$, and it is different from the moment when the round $R$ of $v$ starts.

If $1 \leq i \leq \ell$ then the $i$th edge is processed inside crash-loop (that is, by the Procedure *CrashLoop*). Otherwise, it is processed in the while loop, on lines 8-24 of the Procedure *Dyn_Rnd*. This loop will be henceforth referred to as the *scan-loop*. In the asynchronous setting, for a time moment $\alpha$, we use the notation $v.T\{\alpha\}$ to denote the value of the variable $v.T$ at time $\alpha$. (This notation, like the notation $v.T\{R, i\}$, will be used for other variables than $v.T$ as well.)

One more local variable that we will refer to is the variable $v.E$ that stores the set of edges incident to $v$ whose status is either *SCANNED* or *NOTSCANNED*, but not *CRASHED*. This variable needs not to be maintained explicitly, but will be useful in the analysis.

The following two properties of the semi-decremental setting will be very helpful for the analysis. Recall that $Sp = \bigcup_{v \in V} v.Sp$, and $T = \bigcup_v v.T$.

**Observation 8.1** *If an edge $e$ belongs to the set $Sp$, the only way for $e$ to stop belonging to $Sp$ is to crash. Moreover, the only way for a scanned edge $e$ to stop being scanned is to crash.*

**Proof:** The set $Sp$ consists of $T$-edges and $X$-edges. $T$-edges never crash in the semi-decremental setting. An edge $e$ is removed from $v.X$ only on lines 11 and 15 of the Procedure *XReplace* (Algorithm 8). However, in both these cases the edge $e$ is an input parameter of the Procedure *XReplace*. Hence the edge $e$ either crashed, or the message *CRASH* was received by $v$ over $e$. Messages *CRASH* are sent only as a result of a hard crash, and hard crashes are impossible in the semi-decremental setting. ∎

The next observation states that Lemma 5.7 still holds.

**Observation 8.2** *For a pair of vertices $v, v' \in V$, if there exist indices $R, R', i, i'$ such that $B(v.P\{R, i\}) = B(v'.P\{R', i'\})$, and $(v, R, i) \preceq (v', R', i')$ (using the notation of Section 5.3), then the set $T\{v', R', i'\}$ contains a path of length at most $L(v.P\{R, i\}) + L(v'.P\{R', i'\})$ connecting the vertices $v$ and $v'$.*

To verify this observation note that the $T$-edges are not allowed to crash in the semi-decremental setting. In the asynchronous setting the more convenient form of this observation is the following one.

**Observation 8.3** *For a pair of vertices $v, v' \in V$, if there exist time moments $\alpha, \alpha'$, $\alpha \leq \alpha'$, such that $B(v.P\{\alpha\}) = B(v'.P\{\alpha'\})$, then $T\{\alpha'\}$ contains a path of length at most $L(v.P\{\alpha\}) + L(v'.P\{\alpha'\})$ connecting the vertices $v$ and $v'$.*

### 8.3.2 Validity of Local Data Structures

In this section we show that Algorithm 5 maintains correct data structures. The next lemma focuses on the validity of the sets $v.M[B]$.

**Lemma 8.4** *For a vertex $v$ and a round $R$ of the vertex $v$, let $\ell$, $\ell \geq 0$ (respectively, $k$, $k \geq 0$) be the number of edges adjacent to $v$ processed in the crash-loop (resp., scan-loop) of the round $R$ of the vertex $v$. Let $i$, $i \in [\ell + k]$, be an index.*

1. *If a value $B$ belongs to $v.M\{R, i\}$ then there exists an edge $e' = (v, u')$ such that*

    (a) *$v.status(e')\{R, i\} = SCANNED$, $v.scan\_status(e')\{R, i\} = X$, Also, let $P = v.label(e')$. Then $B(P) = B$. In addition, $e' \in v.X\{R, i\}$, $v.own(e')\{R, i\} = SELF$.*

    (b) *The vertex $u'$ was labeled by a label $P'$ of base $B$ at some time $\alpha$, $\alpha' \leq \beta$, where $\beta$ is the time moment $\{v, R, i\}$.*

2. *If an edge $e'' = (v, u'') \in v.M[B]\{R, i\}$ then $v.scan\_status(e'')\{R, i\} = D$, $e'' \in v.D\{R, i\}$, $B(v.label(e'')\{R, i\}) = B$, $v.own(e'')\{R, i\} = SELF$, and $v.status(e'')\{R, i\} = SCANNED$. Moreover, the paragraph 1b holds for the vertex $u''$.*

**Proof:** The proof is by induction on time moments $\{v, R, i\}$. The induction base is the time $\{v, 1, 1\}$. This is the time when the vertex $v$ invokes the Procedure *Dyn_Rnd* for the first time. At this point the table $v.M$ is empty, and thus the lemma holds vacuously.

Next, we prove the induction step for Part 1. We assume the statement for a moment $\{v, R, i\}$, $R \geq 1$, $i \geq 1$ are integers, and prove it for the "next" time moment. The next time moment is either $\{v, R, i+1\}$ if the vertex $v$ processes $i + 1$ or more edges on its round $R$, or $\{v, R+1, 1\}$ otherwise. Our analysis for these two cases is identical, and thus, to simplify the notation we assume that $\{V, R, i+1\}$ is the next time moment after $\{v, R, i\}$.

Consider a value $B$, $B \in v.M\{R, i+1\}$. The analysis splits into two cases depending on whether $B \notin v.M\{R, i\}$ or not.

1. ($B \notin v.M\{R, i\}$)
   Consequently, the value $B$ joins $v.M$ when $v$ processes the $i$th edge $\tilde{e} = (v, \tilde{u})$ adjacent to $v$ on its round $R$. Since no base value is ever inserted into the set $v.M$ during the crash-loop, it follows that the edge $\tilde{e}$ is processed in the scan-loop (lines 8-24, Algorithm 5). Moreover, the only step when a base value can join the set $v.M$ is line 17 of Algorithm 5. In this case, however, the edge $\tilde{e}$ satisfies all the requirements from the edge $e'$ in the lemma, and we are done. (Particularly, the time moment $\beta$ is equal to $\alpha = \{v, R, i+1\}$.)

2. ($B \in v.M\{R, i\}$)
   By the induction hypothesis, there exists an edge $e' = (v, u')$ that satisfies the requirements of the lemma at time $\{v, R, i\}$. The discussion now splits into two subcases, depending on whether the $i$th processed edge $\tilde{e} = (v, \tilde{u})$ is processed in the scan-loop or in the crash-loop.

   (a) (The edge $\tilde{e}$ is processed in the scan-loop.)
       No edge is ever removed from $v.X$, and no edge $\hat{e}$ ever changes its status from *SCANNED* to *NOTSCANNED* or *CRASHED*, and no scanned edge ever changes its fields $v.own(\hat{e})$ or $v.label(\hat{e})$ during the scan-loop. Consequently, the edge $e'$ satisfies the requirements of the lemma at time $\{v, R, i+1\}$ as well.

   (b) (The edge $\tilde{e}$ is processed in the crash-loop.)
       Recall that $B \in v.M\{R, i\}$ and $B \in v.M\{R, i+1\}$. If $\tilde{e} \neq e'$ then the edge $e'$ satisfies the requirements of the lemma at time $\{v, R, i+1\}$ as well, and we are done.
       If $\tilde{e} = e'$ then the execution invokes the Procedure *XReplace* with parameters $(P, e')$, with $P = v.label(e')\{R, i\}$. Moreover, $B(P) = B$. If $v.M[B]\{R, i\} = \emptyset$ then the execution of the Procedure *XReplace* does not enter the fetch-loop, and proceeds directly to line 15 of Algorithm 8. It follows that in this case $B \notin v.M\{R, i+1\}$, contradiction. Hence there exists some edge $e'' = (v, u'') \in v.M[B]\{R, i\}$ that is fetched in the fetch-loop of the Procedure *XReplace* (lines 2-9, Algorithm 8) of this invocation. By the induction hypothesis of Part 2 of the lemma, since $e'' \in v.M[B]\{R, i\}$, the edge $e''$ satisfies all the desired properties except that its scan-status is $D$ instead of $X$, and it belongs to $v.D$ and not to $v.X$. However, these two things are corrected on line 11 of this invocation of the Procedure *XReplace*, and we are done proving Part 1.

Next, we prove the induction step for Part 2. Consider an edge $e'' = (v, u'') \in v.M[B]\{R, i+1\}$. Analogously to the proof of Part 1, we split the discussion into two cases, depending on whether the condition $e'' \notin v.M[B]\{R, i\}$ does or does not hold.

1. ($e'' \notin v.M[B]\{R, i\}$)
   Consequently, the edge $e''$ joins the set $v.M[B]$ when $v$ processes the $i$th edge $\tilde{e} = (v, \tilde{u})$ adjacent to $v$ on the round $R$ of $v$. Analogously to the proof of Part 1, in this case the edge $\tilde{e}$ is necessarily processed in the scan-loop. Moreover, the edge $e''$ can be added to $v.M[B]$ only on line 20 of Algorithm 5 (Procedure *Dyn_Rnd*), and also, necessarily $e'' = \tilde{e}$. In this case the different fields of $\tilde{e}$ (that is, $v.status(\tilde{e})$, $v.scan\_status(\tilde{e})$, $v.own(\tilde{e})$, $v.label(\tilde{e})$, etc.) were set correctly on lines 9, 15 and 20 of the procedure.

   Moreover, the label $B$ belongs to the set $v.M\{R, i\}$, since otherwise the execution would not reach line 20). Also, note that $B$ is the base value of the label of the endpoint $u''$ of $e''$ at time $\{v, R, i\}$, proving the lemma.

2. ($e'' \in v.M[B]\{R,i\}$)

   By the induction hypothesis, the edge $e'' = (v, u'')$ satisfies the requirements of Part 2 at time $\{v, R, i\}$. The discussion splits into two subcases, depending on whether the $i$th processed edge $\tilde{e} = (v, \tilde{u})$ is processed in the scan-loop or in the crash-loop.

   (a) (The edge $\tilde{e}$ is processed in the scan-loop.)
       Exactly the same argument as in Subcase 2a of the proof of Part 1 applies.

   (b) (The edge $\tilde{e}$ is processed in the crash-loop.)
       If $\tilde{e} \neq e''$ then the edge $e''$ satisfies the requirements of the lemma at time $\{v, R, i+1\}$, and we are done.

       If $\tilde{e} = e''$ then since by the induction hypothesis $v.scan\_status(e'')\{R,i\} = D$ it follows that the execution of the Procedure *CrashItern* reaches line 6 and removes the edge $e''$ from the set $v.M[B]$. This, however, contradicts the assumption of the lemma that $e'' \in v.M[B]\{R, i+1\}$, and we are done.

∎

The next lemma argues that the edge records stored by the vertices running the algorithm are maintained correctly.

**Lemma 8.5** *For $v, R, k, \ell$ and $i$ as in Lemma 8.4, and an edge $e = (v, u)$ adjacent to $v$ that satisfies that $v.status(e)\{R, i\} = SCANNED$ and $v.own(e)\{R, i\} = SELF$, the following statements hold.*

1. *The set $Sp\{v, R, i\}$ spans the edge $e$.*

2. *If $v.scan\_status(e)\{R, i\} = T$ then $e \in v.T\{R, i\}$.*

3. *If $v.scan\_status(e)\{R, i\} = X$ then $e \in v.X\{R, i\}$.*

4. *If $v.scan\_status(e)\{R, i\} = D$ then the following statement holds. Let $\beta$ be the time moment $\{v, R, i\}$. Then there exists an edge $e' = (v, u') \in v.X\{R, i\} = v.X\{\beta\}$, and a pair of time moments $\alpha, \alpha'$ such that $\alpha, \alpha' \le \beta$ such that $B(u.P\{\alpha\}) = B(u'.P\{\alpha'\}) = B(v.label(e')\{R,i\}) = B(v.label(e')\{\beta\})$. Let $B = B(u.P\{\alpha\})$. Moreover, $e \in v.M[B]\{R,i\}$, and $B \in v.M\{R,i\}$.*

**Remark:** Intuitively, the statement 4 means that for every edge $e \in v.D$, there exists an edge $e' \in v.X$ such that the edge $e$ can serve as a replacement for the edge $e'$.

**Proof:** Like the proof of Lemma 8.4, the proof of this lemma is by induction on time moments $\{v, R, i\}$. The induction base is the moment $\{v, 1, 1\}$, and it holds vacuously.

For the induction step we assume the statement for a moment $\{v, R, i\}$, $R \ge 1$, $i \ge 1$, are integers, and prove it for the "next" time moment. Like in the proof of Lemma 8.4, to simplify the notation we assume that the next time moment is $\{v, R, i+1\}$.

Consider an edge $e = (v, u)$ with $v.status(e)\{R, i+1\} = SCANNED$, $v.own(e)\{R, i+1\} = SELF$. The analysis splits into two cases depending on whether the vertex $v$ has set the value of the variable $v.status(e)$ to $SCANNED$ while processing the $i$th edge on round $R$, or earlier.

1. (The vertex $v$ sets the value of the variable $v.status(e)$ to $SCANNED$ while processing the $i$th edge on round $R$.)

   By inspection of the algorithm it is easy to verify that in this case the $i$th edge processed by $v$ on round $R$ must have been the edge $e$, and moreover, this processing occurred in the scan-loop (and not in the crash-loop). The reason for the latter is that in the crash-loop no unscanned edge ever becomes scanned, but rather already scanned edges may change their scan-status.

Consider the iteration of the scan-loop on which the message $(P(u), ttl(u))$ was processed. If $ttl(u) > 0$ (meaning that the label of $u$ is a selected label) or if $B(P(u)) \notin v.M$ then the edge $e$ was inserted to $v.Sp$ on lines 12 or 17 of Algorithm 5, respectively, and in both cases we are done.

Consider the case that $ttl(u) = 0$ and $B(P(u)) \in v.M\{R,i\}$. In this case the scan-status of the edge $e$ was set to $D$ on line 20. By Lemma 8.4 there exists an edge $e' = (v, u')$, $e' \in v.X\{R, i\}$, and such that $u'$ was labeled by a label $P'$ of base $B = B(P(u))$ at the time when the vertex $v$ scanned the edge $e'$. Then, by Observation 8.2, there exists a path $\Pi$ of length $L(P(u)) + L(P') \leq 2t - 2$ connecting between $u$ and $u'$ in $T\{v, R, i\}$. Since $e' = (v, u') \in v.X\{R, i\}$, it follows that there exists a path of length at most $2t - 1$ connecting $v$ and $u$ in $Sp\{v, R, i\}$. Since no crashes of $T$-edges occur, all edges of the path $\Pi$ survive till the moment $\{v, R, i+1\}$. The edge $e' = (v, u')$ belongs to $v.X\{R, i\}$, and is not removed from this set when the vertex $v$ processes its $i$th edge on round $R$. The reason for the latter is that the $i$th processed edge is the the edge $e = (v, u)$, and it is processed in the scan-loop. Hence $e' \in v.X\{R, i+1\}$, and we are done proving that the edge $e$ is spanned by the edge set $Sp\{v, R, i+1\}$. The statements 2, 3, and 4 follow directly from the induction hypothesis, and from the observation that the $i$th processed edge is processed in the scan-loop.

2. (The vertex $v$ sets the value $v.status(e)$ to $SCANNED$ before it processes the $i$th edge on round $R$.) The discussion splits into two subcases depending on whether the $i$th edge $\tilde{e}$, $\tilde{e} \neq e$, was processed in the scan-loop or in the crash-loop. In the former case by the induction hypothesis we are done. Hence we next consider the case that the edge was processed in the crash-loop.

If $v.scan\_status(e)\{R, i+1\} = T$ then since $\tilde{e} \neq e$, the scan-status of the edge $e$ was $T$ at time $\{v, R, i\}$ as well (that is, $v.scan\_status(e)\{R, i\} = T$), and thus, by the induction hypothesis, $e \in v.T\{R, i\}$. Since $\tilde{e} \neq e$, it holds that $e \in v.T\{R, i+1\}$, and we are done (proving the statement 2 of the lemma).

If $v.scan\_status(e)\{R, i+1\} = X$ then by the same argument $e \in v.X\{R, i+1\}$, proving the statement 3. Note that since $v.T\{R, i+1\} \cup v.X\{R, i+1\} = v.Sp\{R, i+1\} \subseteq Sp\{v, R, i+1\}$, it follows that $Sp\{v, R, i+1\}$ spans the edge $e$ in both cases, as required in the statement 1.

Consider the case
$$v.scan\_status(e)\{R, i+1\} = D \ . \tag{4}$$

Recall that the $i$th edge $\tilde{e} = (v, \tilde{u})$ processed by $v$ on its round $R$, is processed in the crash-loop, and $\tilde{e} \neq e$. By (4), it follows that $v.scan\_status(e)\{R, i\}$ is equal to either $D$ or $X$.

If $v.scan\_status(e)\{R, i\} = X$ then by the induction hypothesis, $e \in v.X\{R, i\}$. However, the only way for the edge $e$ to stop belonging to the set $v.X$ is to crash. Since $e \neq \tilde{e}$, it follows that $e \in v.X\{R, i+1\}$, and $v.scan\_status(e)\{R, i+1\} = X$, contradicting (4).

Hence $v.scan\_status(e)\{R, i\} = D$. By the induction hypothesis there exists an edge $e' = (v, u') \in v.X\{R, i\}$ that satisfies the statement 4 of the lemma. If $e' \neq \tilde{e}$ then $e' \in v.X\{R, i+1\}$, and we are done. Finally, if $e' = \tilde{e}$ then the processing of the edge $e' = (v, u')$ in the crash-loop resulted in an invocation of the Procedure $XReplace$ on line 8 of the Procedure $CrashItern$. This procedure was invoked with parameters $(P, e')$, with $P = v.label(e')\{R, i\}$. By the induction hypothesis, $B(P) = B(u.P\{\alpha\}) = B(u'.P\{\alpha'\})$, where $\alpha$ and $\alpha'$ are the time moments whose existence is guaranteed by the induction hypothesis of the statement 4 of the lemma.

The Procedure $XReplace$ starts with the fetch-loop (lines 2-9 of Algorithm 8). Observe that by the induction hypothesis, $e \in v.M[B]\{R, i\}$, and $v.status(e)\{R, i\} = SCANNED$. Hence the fetch-loop necessarily succeeds in fetching an edge $\check{e} = (v, \check{u})$. If $\check{e} = e$ then the edge $e$ joins $v.X = v.X\{R, i+1\}$ on line 11, and $v.scan\_status(e)\{R, i+1\} = X$. This, however, contradicts (4). Thus $\check{e} \neq e$. In this case, however, the edge $\check{e}$ joins $v.X\{R, i+1\}$, and it is easy to verify that $\check{e}$ is the edge $e'$ whose existence we need to prove to establish the statement 4. ∎

We will mostly use the statement 1 of Lemma 8.5, that is, that edge scanned at a time $\{v.R, i\}$ is actually spanned by $Sp\{v, R, i\}$. Note, however, that when crashes are allowed, some edges belonging to $v.Sp\{R, i\}$ may not be present in the network at time $\{v, R, i\}$. On the other hand, we next show that all edges of $v.Sp\{R, i\}$ were present in the network at most one time unit before the time $\{v, R, i\}$.

**Lemma 8.6** *For a vertex $v$, a round $R$, $R \geq 2$ of $v$, and indices $i$, $\ell$ and $k$ as in Lemma 8.5, the following statement holds. If $i \geq \ell + 1$ then $v.Sp\{R, i\} \subseteq v.E\{R, 1\}$. Otherwise (if $i \leq \ell$), $v.Sp\{R, i\} \subseteq v.E\{R-1, 1\}$.*

**Remark:** The set $v.E\{R, 1\}$ contains all edges whose appearance was detected during the showup-loop of round $R$ at vertex $v$ (line 1 of Algorithm 5).

**Remark:** Intuitively, as long as the crash-loop of round $R$ is not over, the vertex $v$ "does not know" which edges crashed on round $R - 1$, and thus it may use some edges that were present in the network in the beginning of the round $R - 1$ but crashed in the meanwhile as a part of the spanner it maintains.

**Proof:** For a fixed time moment $\{v, R, i\}$, let $v.Old\{R, i\}$ denote the set of edges $e$ adjacent to $v$ that satisfy $v.mark(e)\{R, i\} = OLD$ and $v.status(e)\{R, i\} = SCANNED$. Note that $v.Sp\{R, i\} \subseteq v.Old\{R, i\}$. We prove by induction on time moments $\{v, R, i\}$ that $v.Old\{R, i\} \subseteq v.E\{R, 1\}$ if $i \geq \ell + 1$, and $v.Old\{R, i\} \subseteq v.E\{R-1, 1\}$ otherwise.

The induction base is the moment $\{v, 2, 1\}$. Note that edges are marked as $OLD$ for the first time on line 25 of the first invocation (round 1) of the Procedure $Dyn\_Rnd$. However, for an edge $e$ to be marked as $OLD$ on line 25 of round 1 of $v$, it must have appearred in the showup loop of round 1, and so $e \in v.Old\{2, 1\}$ implies $e \in v.E\{1, 1\}$.

We next prove the induction step. The discussion splits into a number of cases depending on the value of $i$.

1. $(i = \ell)$
   In this case the induction hypothesis is $v.Old\{R, \ell\} \subseteq v.E\{R-1, 1\}$. The $i$th processed edge $\tilde{e} = (v, \tilde{u})$ is the last edge processed in the crash-loop of round $R$. We need to argue that $v.Old\{R, \ell + 1\} \subseteq v.E\{R, 1\}$.

   Consider an edge $e = (v, u) \in v.Old\{R, \ell + 1\}$. Since no edges join the set $v.Old$ during the crash-loop, it follows that $e \in v.Old\{R, \ell\}$, and thus, by the induction hypothesis, $e \in v.E\{R-1, 1\}$. If the edge $e$ did not crash between the time $\{v, R-1, 1\}$ and $\{v, R, 1\}$ then we are done. If the edge $e$ did crash on round $R - 1$ of the vertex $v$ then it was detected in the crash-loop of the round $R$, and the variable $v.status(e)$ was set to $CRASHED$ on line 4 of the Procedure $CrashItern$. Hence $e \notin v.Old\{R, \ell + 1\}$, contradiction. (Observe that the time moment $\{v, R, \ell + 1\}$ is the moment that follows the end of the crash-loop and precedes the beginning of the scan-loop of round $R$ of the vertex $v$.)

2. $((i < \ell)$ or $(\ell < i < \ell + k))$
   Edges are never marked as $OLD$ during either the crash-loop or the scan-loop. Hence in both these cases $v.Old\{R, i+1\} = v.Old\{R, i\}$, and if $i \geq \ell + 1$ then $v.Old\{R, i\} \subseteq v.E\{R, 1\}$. Otherwise, if $i \leq \ell$ then $v.Old\{R, i\} \subseteq v.E\{R-1, 1\}$. The statement of the lemma follows.

3. $(i = \ell + k)$
   This is the case when the induction hypothesis is $v.Old\{R, \ell + k\} \subseteq v.E\{R, 1\}$, and we need to argue that $v.Old\{R+1, 1\} \subseteq v.E\{R, 1\}$.

Between the time $\{v, R, \ell + k\}$ and $\{v, R + 1, 1\}$ the vertex $v$ processes the last edge of round $R$ of $v$ (this is done in the scan-loop), and then performs line 25 of the invocation of the Procedure $Dyn\_Rnd$ of the round $R$ of the vertex $v$, and lines 1-4 of the invocation of this procedure on round $R + 1$ of the vertex $v$. The processing of the last $((\ell + k)$th) edge adjacent to $v$ in the scan-loop of round $R$ does not add any edges to the edges of the set $v.Old$. The line 25 of the Procedure $Dyn\_Rnd$, however, does mark some edges as $OLD$. On the other hand, all the latter edges were necessarily discovered on line 1 of the $R$th invocation of the Procedure $Dyn\_Rnd$ local at the vertex $v$. Thus, these edges belong to $v.E\{R, 1\}$. Finally, the lines 1-5 of the invocation of the Procedure $Dyn\_Rnd$ on round $R + 1$ local at $v$ mark no edges as $OLD$, and thus we are done.

∎

### 8.3.3 Quiescence Time

In this section we show that Algorithm 5 has small quiescence time.

We start with arguing that in the setting that allows only soft crashes and no incremental updates, the quiescence time is just one round (or one time unit if the setting is asynchronous).

**Lemma 8.7** *Consider a synchronous network. Suppose that for some pair of rounds $R$, $R'$, $R \leq R'$, the network is stable at the beginning of round $R$. (See the definition in Section 6.3.) Suppose also that (arbitrarily many) soft crashes occur between the beginning of round $R$ and the end of round $R'$. Suppose also that no incremental topology updates occur, and that no topology updates occur starting with the beginning of round $R' + 1$. Then at the beginning of round $R' + 2$ and later the algorithm maintains a $(2t - 1)$-spanner of the network.*

**Proof:** For a round $\tilde{R}$, we will use the notation $Sp\{\tilde{R}\}$ and $E\{\tilde{R}\}$ to denote the sets of edges $Sp$ and $E$, respectively, at the beginning of round $\tilde{R}$.

Since the network is stable at the beginning of round $R$, it follows that all edges present in the network at this stage are scanned. Moreover, as we have seen, in the semi-decremental setting the only way for an edge to stop being scanned is to crash, and thus all edges that are still present in the network in the beginning of round $R' + 2$ are scanned as well. Consequently, by Lemma 8.5, all these edges are spanned by $Sp\{R' + 2\}$. Moreover, by Lemma 8.6, $Sp\{R' + 2\} \subseteq E\{R' + 1\}$. Finally, by an assumption of the lemma, $E\{R' + 1\} \subseteq E\{R' + 2\}$ (because no edge crashes on round $R' + 1$; in fact, in this case $E\{R + 1\} = E\{R' + 2\}$), and so $Sp\{R' + 2\} \subseteq E\{R' + 2\}$. This proves that the set $Sp\{R' + 2\}$ spans all edges of the network, and is contained in $E\{R' + 2\}$.

To argue that after the beginning of round $R' + 2$ the algorithm keeps maintaining a $(2t - 1)$-spanner, we note that as no edges crash after the beginning of round $R' + 2$, edges are not removed from the sets $Sp$ and $E$. Since the network is stable at this point, and no topology updates occur, no edges are added into the set $Sp$ from this point on. ∎

To extend Lemma 8.7 to the asynchronous setting we first re-state Lemma 8.6 in the following more convenient form.

**Lemma 8.8** *For a time moment $\delta$, and a vertex $v$, there exists $\eta$, $\delta \geq \eta \geq \delta - 1$, such that $v.Sp\{\delta\} \subseteq v.E\{\eta\}$.*

**Proof:** Let $R$ be the round of the vertex $v$ such that the moment $\delta$ is between the beginning of the round $R$ of $v$ and the beginning of the round $R + 1$ of $v$. The proof splits into a number of cases depending on the relation between the time moment $δ$ and the time period during which the vertex $v$ is on its round $R$.

1. ($\delta$ is before the execution of the Procedure $Dyn\_Rnd$ (local at $v$) ends the crash-loop of round $R$.)
   Note that since the set $v.Sp$ does not change before the crash-loop starts (that is, does not change on lines 1-5 of the Procedure $Dyn\_Rnd$), it follows that between the time that the Procedure $Dyn\_Rnd$ begins and the time it invokes the Procedure $CrashLoop$ the set $v.Sp$ is equal to $v.Sp\{R, 1\}$.

   By Lemma 8.6, $v.Sp\{\delta\} \subseteq v.E\{R-1, 1\}$. Let $\eta$ denote the time moment $\{v, R-1, 1\}$. At most one time unit elapses between the time moments $\{v, R-1, 1\}$ and $\{v, R, 1\}$. The entire crash-loop involves only local computation, and thus, requires zero "distributed time". Consequently, $\eta \geq \delta - 1$, and $v.Sp\{\delta\} \subseteq v.E\{\eta\}$,

2. ($\delta$ is after the execution of the Procedure $Dyn\_Rnd$ (local at $v$) ends the crash-loop of round $R$.)
   By Lemma 8.6, $v.Sp\{\delta\} \subseteq v.E\{R, 1\}$. Let $\eta$ be the time moment $\{v, R, 1\}$. Since at most one time unit elapses between the beginning of round $R$ and the time moment $\delta$, it follows that $\eta \geq \delta - 1$, and so $v.Sp\{\delta\} \subseteq v.E\{\eta\}$, proving the lemma.

∎

Observe that in Lemma 8.8 we assume that local computation requires zero time. Alternatively, one can assume that there is a small $\epsilon > 0$, such that all local computations done by a vertex in a fixed round require at most $\epsilon$-fraction of a time unit. In this case $\eta$ in Lemma 8.8 satisfies a slightly weaker inequality $\delta \leq \eta \leq \delta - 1 - \epsilon$.

We next extend Lemma 8.7 to the asynchronous setting.

**Lemma 8.9** *Consider the asynchronous setting. Let $\alpha$ and $\alpha'$ be two time moments, $\alpha < \alpha'$. Suppose that the network is stable at time $\alpha$, and that soft crashes occur between the time $\alpha$ and $\alpha'$, no topology updates occur starting from time $\alpha'$, and no other (incremental) topology updates occur. Fix $\beta = \alpha' + 1$. Then the algorithm maintains a correct $(2t-1)$-spanner starting from time $\beta$.*

**Proof:** All edges present in the network at time $\alpha$ are scanned. Also, all edges that are still present in the network at time $\beta$ are scanned as well. (The argument here is analogous to the proof argument of Lemma 8.7.) Consider an edge $e = (v, u)$, and suppose without loss of generality that it is scanned by $v$. Then by Lemma 8.5, the edge $e$ is spanned by $Sp\{\beta\}$. Moreover, by Lemma 8.5 (Parts 2-4) and by Observation 8.3, there is a path $\Pi$ that connects $v$ and $u$ in $Sp\{\beta\}$, has length at most $2t - 1$, and contains at most one $X$-edge $e' = (v, u')$. (This edge $e'$ is necessarily adjacent to the vertex $v$.)

Moreover, by Lemma 8.8, there exists a time moment $\gamma$, $\gamma \geq \beta - 1$, such that $v.Sp\{\beta\} \subseteq v.E\{\gamma\}$. Hence $\gamma \geq \alpha'$, and so no edge crashes between the time $\gamma$ and $\beta$. Hence $v.Sp\{\beta\} \subseteq v.E\{\gamma\} \subseteq v.E\{\beta\}$. All other edges of the path $\Pi$ are $T$-edges, and so they belong to $v.E\{\beta\}$ as well. ∎

Hence we have shown that in the (pure) semi-decremental setting (when only soft crashes are allowed, and incremental updates are not) the quiescence time is *one time unit*.

We next extend Lemma 8.9 to a scenario when soft crashes occur in more than one single burst.

**Lemma 8.10** *Let $\alpha_1, \alpha'_1, \alpha_2, \alpha'_2, \ldots$ be a sequence of time moments such that $\alpha_1 < \alpha'_1 < \alpha_2 < \alpha'_2 < \ldots$, and $\alpha_{i+1} \geq \alpha'_i + 1$ for every index $i = 1, 2, \ldots$. Suppose that the network is stable at time $\alpha_1$, and that soft crashes occur during the time intervals $[\alpha_1, \alpha'_1], [\alpha_2, \alpha'_2], \ldots$ only, and no other topology updates occur. Then the algorithm maintains a correct $(2t-1)$-spanner during the time intervals $(\alpha'_1, \alpha_2), (\alpha'_2, \alpha_3), \ldots$.*

The proof of this lemma is by a straightforward induction that uses Lemma 8.9 to prove the induction step.

Lemma 8.10 shows that, amazingly, the algorithm takes care of the bursts of semi-decremental topology updates *on the fly*, no matter how complex is the structure of the update sets. This is in contrast to our results for the incremental topology updates (Corollaries 6.3, 6.5), where some limitations on the update sets do exist.

Moreover, by inspection of the proof of Lemma 8.9 it is easy to verify that incremental updates can be introduced, and the analysis extends to this more general scenario. This statement is formalized in the following lemma.

**Lemma 8.11** *Consider the scenario of Lemma 8.10 with the only change that incremental updates may appear at any time. Let $E_0$ be the set edges present at time $\alpha_1$. For $i = 1, 2, \ldots$, let $E_i$ be the subset of $E_0$ that contains the edges that are present in the network during the time period $(\alpha'_i, \alpha_{i+1})$. (Note that edges do not crash during these periods.) Then for every index $i = 1, 2, \ldots$, during the period $(\alpha'_i, \alpha_{i+1})$ the algorithm maintains a correct $(2t-1)$-spanner for the subset $E_i$.*

We next argue that the quiescence time of our algorithm in the semi-decremental scenario is at most $2t$ time units, no matter what is the structure of the incremental and decremental topology updates. (As long as the latter are soft crashes.)

We start with proving this for the synchronous setting.

**Lemma 8.12** *The quiescence time in the synchronous semi-decremental setting (that allows both incremental and soft decremental updates) is at most $2 \cdot t$ rounds.*

**Proof:** Suppose all topology updates occur on round $R$ or before, for some round $R \geq 1$. Consider an edge $e = (v, u)$ present in the network at the beginning of round $R+1$. By the proof argument of Lemma 5.4 (used also in Section 6.2 to analyze the incremental algorithm), the edge $e$ is scanned at the end of round $R + 2t$ or earlier.

Suppose without loss of generality that $e$ is scanned by the vertex $v$, i.e., $v.own(e)\{R + 2t, END\} = SELF$. Then by Lemma 8.5, the edge $e$ is spanned by the set $Sp\{v, R + 2t, END\}$. Consider the path $\Pi$ of length at most $2t - 1$ connecting between the endpoints $v$ and $u$ of the edge $e$ in $Sp\{v, R + 2t, END\}$. By Lemma 8.5 and Observation 8.3, this path contains at most one $X$-edge $e' = (v, u')$, and moreover, this edge (if exists) is adjacent to $v$. By Lemma 8.6, $v.Sp\{R+2t, END\} \subseteq v.E\{R+2t, 1\}$. It follows that $e' \in v.E\{R + 2t, 1\}$. Since no edge crashes on round $R + 2t$, it follows that $e' \in v.E\{R + 2t, END\}$.

$T$-edges never crash in this setting, and thus all other edges of $\Pi$ belong to $E\{v, R + 2t, END\}$ (as well as to $Sp\{v, R + 2t, END\}$). Hence the edge $e$ is spanned by $Sp\{v, R + 2t, END\}$, and all the edges of the path spanning it belong to $E\{v, R + 2t, END\}$. (Note that since the setting is synchronous, $E\{v, R + 2t, END\} = E\{R + 2t, END\}$, and analogously, $Sp\{v, R + 2t, END\} = Sp\{R + 2t, END\}$.) ∎

The argument extends to the asynchronous setting.

**Lemma 8.13** *The quiescence time in the asynchronous semi-decremental setting is at most $2 \cdot t$ time units.*

**Proof:** Suppose that no topology update occurs after time $\alpha$, for some $\alpha > 0$. Consider an edge $e = (v, u)$ present in the network at time $\alpha$. By the proof argument of Lemma 5.4, the edge $e$ is scanned at time $\beta$, for $\beta$ such that $\alpha \leq \beta \leq \alpha + 2t$. Suppose without loss of generality that $e$ is scanned by the vertex $v$. By Lemma 8.5, the edge $e$ is spanned at time $\beta$. Moreover, since no edge crashes after $\beta$ (as $\alpha \leq \beta$), it follows that the edge $e$ is spanned at time $\alpha + 2t$ as well.

Consider the path $\Pi$ as in the proof of Lemma 8.8 in $Sp\{\alpha + 2t\}$. By Lemma 8.8, there exists $\eta$, $\alpha + 2t \geq \eta \geq \alpha + 2t - 1$ such that $v.Sp\{\alpha + 2t\} \subseteq v.E\{\eta\}$. Since no edge crashes between the time $\eta$ and $\alpha + 2t$, it follows that $e' \in v.Sp\{\alpha + 2t\} \subseteq v.E\{\alpha + 2t\}$. All $T$-edges of $\Pi \subseteq Sp\{\alpha + 2t\}$ belong to $E\{\alpha + 2t\}$ as well, because they never crash. Hence the path $\Pi$ is contained in $Sp\{\alpha + 2t\}$ and in $E\{\alpha + 2t\}$, and it spans the edge $e$. ∎

Finally, we strengthen Lemma 8.13, and show that if an edge $e = (v, u)$ is present in the network for $2t$ time units, and at least one single time unit elapsed since the time the last soft crash occurred, then the edge $e$ is spanned by the spanner maintained by the algorithm.

**Lemma 8.14** *Let $\alpha$ be the time when the last (soft) decremental update occurs, and $\delta$ be the time when the edge $e = (v, u)$ appears in the network. Suppose that $e$ never crashes after the time $\delta$. Let $\zeta = \max\{\delta + 2t, \alpha + 1\}$. Then the spanner maintained by the algorithm spans the edge $e$ starting with the time $\zeta$.*

**Remark:** Interestingly, the lemma places no restriction whatsoever on incremental updates. As many of these updates as one wishes may occur at any time, and the statement of the lemma keeps holding.

**Proof:** The crucial observation for this proof is that even when soft crashes are allowed, the proof argument of Lemma 5.4 is applicable for an edge $e = (v, u)$ that *does not crash itself*. Consequently, the edge $e$ is scanned at time $\zeta$ because $\zeta \geq \delta + 2t$. Suppose without loss of generality that the edge $e$ is scanned by $v$. By Lemma 8.5, the edge $e$ is spanned at time $\delta + 2t$, and its spanning path $\Pi \subseteq Sp\{\zeta\}$ contains at most one $X$-edge $e'$, and $e'$ (if exists) is adjacent to $v$.

By Lemma 8.8, there exists $\eta$, $\zeta \geq \eta \geq \zeta - 1 \geq \alpha$, such that $v.Sp\{\zeta\} \subseteq v.E\{\eta\}$. The rest of the argument is identical to that of the proof of Lemma 8.13. (It uses the assumption that no edges crash during the time interval $[\eta, \zeta]$).  ∎

### 8.3.4 The Size of the Spanner

In this section we argue that the spanner maintained by the algorithm is sparse.

Fix a vertex $v$. By inspection it is easy to verify that vertices never lose labels in this algorithm (in the semi-decremental setting). In other words, each label $P$ that the vertex $v$ ever acquired is stored in the table $A(v) = v.A$, and is never removed from there. Hence the proof argument of Section 5.2 applies, and particularly, $|T(v)| \leq t - 1$, and for every index $i$, $i \in [(t-1)]$, the table $A(v)$ contains at most one label of level $i$. Exactly the same arguments as the argument preceding Lemma 5.6 and the proof argument of Lemma 5.6 show that the number of edge inserted into $X(v)$ *in the scan-loops* on all different invocations of the Procedure $Dyn\_Rnd$ is $O(\log^{1-1/t} n \cdot t^{1-1/t} \cdot n^{1/t})$ with high probability.

Moreover, whenever an edge $e'$ is inserted into the set $X(v)$ outside of the scan-loops, this can happen only on line 11 of the Procedure *XReplace* (Algorithm 8). However, on this line some other edge $e$ is necessarily removed from $X(v)$, and thus, these operations never increase the size of $X(v)$.

The desired upper bound on the size of the spanner follows.

**Corollary 8.15** *The algorithm maintains a $(2t - 1)$-spanner of size $O(t^{1-1/t} \cdot \log^{1-1/t} n \cdot n^{1+1/t})$ in the semi-decremental synchronous and asynchronous settings. The guarantee on the size of the spanner holds with high probability.*

### 8.3.5 Local Space and Time Requirements

In this section we analyze the local space and time requirements for a fixed vertex $v$ that executes Algorithm 5.

**Lemma 8.16** *The vertex $v$ uses $O(deg(v))$ words of storage, each with $O(\log n)$ bits.*

**Proof:** Consider an edge $e = (v, u)$ adjacent to the vertex $v$. If $e$ is either not scanned or if it is scanned by $u$, then $e$ contributes $O(1)$ words to the total number of words stored in the data structures of the vertex $v$. Suppose that the edge $e = (v, u)$ is scanned by $v$. Let $P = v.label(e)$, and $B = B(P)$. Regardless of the scan-status of $e$, $O(1)$ words are stored in the different fields $v.status(e)$, $v.scan\_status(e)$, $v.label(e)$, $v.sec\_label(e)$, etc.

We also need to take into account the words stored in the tables $v.A$ and $v.M$. If $e$ is a $T$-edge then by lines 11-12 of the Procedure $Dyn\_Rnd$ the label $P$ is stored in the table $v.A$. We charge this label to the edge $e$. If $e$ is an $X$-edge then by line 17 of the Procedure $Dyn\_Rnd$ the value $B$ is stored in the table $v.M$. We charge this value to the edge $e$. Finally, if $e$ is a $D$-edge then by line 20 of the Procedure $Dyn\_Rnd$, the edge $e$ is stored in the table $v.M[B]$. This storage space is charged to the edge $e$.

It is easy to verify that all storage space used by the algorithm is charged to one of the edges adjacent to $v$, and also, there are $O(1)$ words charged to every edge $e$ adjacent to $v$. The lemma follows. ∎

We next argue that the time that the vertex $v$ needs to process locally a fixed edge $e = (v, u)$ adjacent to $v$ on a fixed invocation of the Procedure $Dyn\_Rnd$ is small. This processing time will be henceforth referred as the *local processing time-per-edge (per round)*.

If an edge $e$ shows up in the network and its appearance is detected by the vertex $v$ on line 1 of the round $R$ of $v$ (in a fixed invocation of the Procedure $Dyn\_Rnd$ local at $v$), then on the same round the edge is processed in time $O(1)$. Then the edge is processed on the next $\ell$ rounds of the vertex $v$ until it is scanned, for some $\ell \leq 2t - 1$. Consider the round on which the vertex $v$ scans the edge $e$. (It can happen that $e$ is scanned by the other endpoint $u$. In this case the vertex $v$ devotes $O(1)$ time-per-round for processing $e$ for at most $2t$ rounds, until it receives a message from the other endpoint $u$ informing $v$ that the edge $e$ was scanned.)

If $e = (v, u)$ is $T$-scanned then its processing (lines 11-13 of the Procedure $Dyn\_Rnd$) requires $O(1)$ time. If $e$ is $X$- or $D$-scanned then the execution reaches line 16 and tests whether $B = B(P(u)) \in v.M$. As we have seen in Section 5.4, $|v.M| \leq deg(v)$, and also, with high probability, $|v.M| = O(n^{1/t} \cdot (t \cdot \log n)^{1-1/t})$. Hence, using the data structure of Beame and Fich [18] with the dynamization technique of Andersson and Thorup [5] this test can be carried out in worst-case time $O(\sigma(deg(v))) = O\left(\sqrt{\frac{\log deg(v)}{\log\log deg(v)}}\right)$, and with high probability, in time $O(\chi(n, t)) = O\left(\sqrt{\frac{\log\log n + (\log n)/t}{\log(\log\log n + (\log n)/t)}}\right)$. All other operations of the execution require only a constant time.

Consider a set $F$ of edges adjacent to the vertex $v$ that crash, and their crashes are processed by $v$ in the crash-loop of a certain round $R$. We also assume that all these crashes are soft crashes. All not scanned edges $e \in F$ are processed on lines 1-2 of the Procedure $CrashItern$, and their processing requires $O(1)$ time. The same is true for edges $e = (v, u)$ that are scanned by $u$.

For each edge $e = (v, u)$ scanned by $v$, its scan-status is either $D$ or $X$. In the former case it is processed on lines 5-6 of the Procedure $CrashItern$. This processing requires accessing the queue $v.M[B]$, and this entails going through the data structure of Beame and Fich [18, 5]. Hence the worst-case processing time in this case is $O(\sigma(deg(v)))$, and with high probability, the processing time is $O(\chi(n, t))$.

Hence we are left with the case that $e$ is an $X$-edge. In this case the Procedure $XReplace$ is invoked on line 8 of the Procedure $CrashItern$. This procedure fetches edges $e' = (v, u')$ from the queue $v.M[B]$, $B = B(v.label(e))$, until it finds an edge not in $F$, or until the queue $v.M[B]$ becomes empty. For such a fetched edge $e' \in F$, we charge the current processing of it (that occurs as a part of processing $e$) to the edge $e'$. Note that since the edge $e'$ is removed from both $v.M[B]$ right after this processing, each edge $e'$ can be processed this way (as a part of processing another edge $e$) at most once. This processing entails accessing the queue $v.M[B]$, and thus requires the worst-case time $O(\sigma(deg(v)))$, and time $O(\chi(n, t))$ with high probability.

Finally, once a satisfactory edge $e'$ is fetched, the processing of the edge $e$ requires additional $O(1)$ time. In other words, the edge set $F$ of crashing edges adjacent to $v$ is processed by $v$ locally within worst-case time $O(|F| \cdot \sigma(deg(v)))$, that is, in $O(\sigma(deg(v)))$ time-per-edge. We summarize with the following corollary.

**Corollary 8.17**  1. Each edge $e = (v, u)$ that shows up in the network is processed by its endpoints $v$ and $u$ for at most $2t$ rounds, and on each round each endpoint spends $O(1)$ time to this edge, except for one round on which one of the endpoints spends $O(\sigma(deg(v)))$ time worst-case, and $O(\chi(n,t))$ time with high probability, processing this edge.

2. Each edge $e = (v, u)$ that crashes (assuming $e$ is not a $T$-edge) is processed by both $v$ and $u$ for one round, and each endpoint spends at most $O(\sigma(deg(v)))$ time worst-case, and $O(\chi(n,t))$ time with high probability, processing this edge.

Finally, we turn our attention back to the dynamic *centralized* model (see Section 4), and argue that in that model each soft crash can be processed in expected $O(1)$ time, and with high probability, in $O(\sigma(h))$ time, where $h = \max\{deg(e), \log n\}$.

For this end the tables $v.M = M(v)$ are maintained in the way described in Section 3.4, using the hash table $h : [n] \to [N]$, $N = \Theta((t \cdot \log n)^{1-1/t} \cdot n^{1/t})$. As we have seen in Section 3.4, for each vertex $v$ and base value $B$, the expected size of $v.M[B]$ is $O(1)$, and thus the lookup and deletion operations that manipulate the queues $v.M[B]$ in procedures *Dyn_Rnd*, *CrashLoop* and *XReplace* can be carried out in expected time $O(1)$. Moreover, with high probability, the size of $v.M[B]$ is $O\left(\frac{\log n}{\log \log n}\right)$, and it is always no greater than $deg(v)$. Hence, with high probability, these lookup and deletion operations can be carried out in time $O(\sigma(h))$, where $h = \max\{deg(v), \log n\}$. For an edge $e = (v, u)$ that experiences a soft crash, by definition, the degree of its endpoint $v$ that processes the crash is smaller or equal to $deg(e)$, and this $h \leq \max\{deg(e), \log n\}$, as required.

This fills in the gap that was left in our analysis of the dynamic centralized spanner-maintaining algorithm of Section 4.

## 9 Adapting the Algorithm to the Distributed Fully Dynamic Setting

Our algorithm for maintaining a sparse $(2t-1)$-spanner in the semi-decremental setting (Algorithm 5) can be extended to maintain a sparse spanner in the fully dynamic setting. For this end we need to specify the pseudo-code of the Procedure *Crash*. Recall that this procedure cannot be possibly invoked in the semi-decremental setting, because in this setting an execution of the algorithm never reaches line 11 of the Procedure *CrashItern* (Algorithm 7). This is, however, not the case in the fully dynamic setting.

### 9.1 The Preliminaries

We first introduce some terminology. An edge $e = (v, u)$ is said to *crash really* on a round $R$ of the vertex $v$ if the edge disappears from the network, and $v$ detects it while executing the crash-loop of its round $R$. Particularly, in this case the edge cannot be used for communication until it re-appears. Note that in this case the vertex $v$ sets the status of the edge $e$ to $CRASHED$, and moreover, this is the only case in which the vertex $v$ does so. We say that the edge $e = (v, u)$ *crashes virtually* on a round $R$ of the vertex $v$ if it switches the status of the edge $e$ from $SCANNED$ to $NOTSCANNED$ while executing the crash-loop of round $R$ of $v$. Also, an edge $e$ is said to *super-crash* on a round $R$ of $v$ if $e$ crashes either really or virtually on round $R$ of $v$.

We also distinguish the time moments when the edge $e$ crashes *physically* and crashes *really*. The moment when the edge $e$ disappears from the network is the moment when the edge crashes *physically*. The moment when $e$ crashes *really* is the moment when the owner of $e$ detects it, and sets its status to $CRASHED$. One of the assumptions of the computational model is that at most one time unit passes between these two events.

Before describing the Procedure *Crash* we prove the following lemma that characterizes scenarios in which the procedure may be invoked.

**Lemma 9.1** *For a vertex $v$ to invoke the Procedure Crash with a parameter $P$ on a round $R$ of $v$, there must exist a self-scanned $T$-edge $e = (v, u)$ that super-crashes on round $R$ of $v$. Moreover, $v.label(e) = P$, and $L(P) \geq 1$.*

**Proof:** The Procedure *Crash* is always invoked from the Procedure *CrashItern*. The latter procedure accepts as input the parameter *crash_status* that may be equal to either $CRASHED$ or $NOTSCD$. If the invocation of the Procedure *CrashItern* that invoked the Procedure *Crash* has *crash_status* $= CRASHED$, it means that a $T$-edge $e = (v, u)$ adjacent to $v$ crashed really.

Otherwise, if this invocation has *crash_status* $= NOTSCD$ then the vertex $v$ received the message $CRASH$ over a $T$-edge $e = (v, u)$ adjacent to $v$, and so in this case the edge $e$ crashed virtually. In both cases it holds that $(v.status(e) = SCANNED)$ and $(v.own(e) = SELF)$. Note also that in both cases the Procedure *Crash* is invoked with a parameter $P$ equal to $v.label(e)$. Observe that for a self-scanned $T$-edge $e = (v, u)$, the label $v.label(e)$ of the edge $e$ has level greater or equal to 1. Consequently, $L(P) \geq 1$.
∎

We introduce some additional terminology. Suppose that at some point during the execution of the algorithm a vertex $v$ scans an edge $e = (v, u)$, and sets its label $P = v.P$ to $u.P + n$. In this case we say that the vertex $v$ *acquires* the label $P$ from the vertex $u$ (through the edge $e$).

For a base value $B$, the *cluster $B$* is the set of all vertices $v$ that have an active label $P$ of base $B$. In other words, for a vertex $v$ to belong to a cluster $B$ there must exist a label $P \in v.A$ such that $B(P) = B$. A vertex $x$ such that $I(x) = B$ is called the *root of the cluster $B$*.

We say that a vertex $u$ is a *parent* of a vertex $v$ in a cluster $B$ if both vertices belong to the cluster $B$, and the vertex $v$ acquired its label $P_v$ of base $B$ from the vertex $u$. In this scenario the vertex $v$ is said to be a child of $u$ in the cluster $B$. See Figure 5 for an illustration.

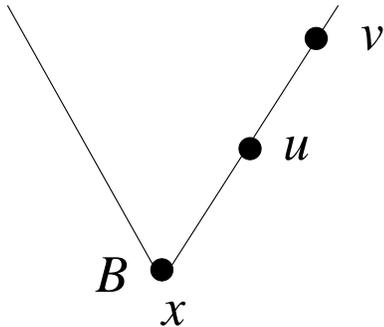

Figure 5: The vertex $u$ is the parent of $v$ in the cluster $B$. The label $P_v$ of base $B$ of the vertex $v$ is related to the label $P_u$ of the base $B$ of the vertex $u$ by the equation $P_v = P_u + n$. The vertex $x$ is the root of the cluster $B$.

Suppose that a vertex $v$ acquired a label $P$ from the vertex $u$ at some point of an execution of the algorithm. At some later stage of the execution it may happen that either the edge $e$ or some other edge that belongs to the path connecting the root of the cluster $B(P)$ with the vertex $v$ crashes. In this case our algorithm ensures that the vertex $v$ will realize that it no longer belongs to the cluster $B$, and update its data structures accordingly. Particularly, if this happens, the vertex $v$ will remove the label $P$ from the table $v.A$ of active labels. We will refer to this event by saying that "the vertex $v$ *loses* the label $P$".

Interestingly, in Algorithm 5 for a fixed vertex $v$ and a cluster $B$, it may happen that the vertex $v$ has more than one parent in the cluster $B$. For example, consider a triangle with vertices $v$, $u$, $x$. Suppose that $I(x) = 3$, $I(u) = 2$, $I(v) = 1$, and that the labels 3 and $n + 3 = 6$ are the only selected labels. (See Section 3.1 for the definition of a *selected* label.) When the edge $(x, v)$ is scanned, $v$ acquires the label 6 from $x$, and when the edge $(x, u)$ is scanned, $u$ acquires the label 6 from $x$. When the edge $(v, u)$ is scanned, the algorithm discovers that $P(v) = P(u) = 6$, but $I(u) > I(v)$, and so $P(u) \succ P(v)$. Since 6 is a selected label, it follows that the vertex $v$ acquires now the label 9 from $u$. Altogether, $v$ ends up having three active labels 1, 6, and 9, and labels 6 and 9 are both of base 3. Consequently, the vertex $v$ has two parents in the cluster 3, specifically, $x$ and $u$. See Figure 6.

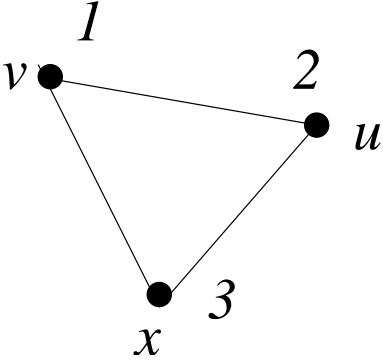

Figure 6: An example of a graph and an execution of the original algorithm in which the vertex $v$ has two parents $x$ and $u$ in the cluster $B = 3$.

We next modify the algorithm slightly to prevent this anomaly, and to guarantee that each vertex $v$ will have at most one parent in each cluster $B$ it belongs to. We remark that this anomaly does not need to be corrected for our results for the semi-decremental setting to hold. In fact, it is likely that the algorithm with this anomaly is good enough for the fully dynamic setting as well. However, the version of the algorithm that corrects this anomaly is more convenient for analysis.

Let $B(v.A) = \{B \mid \exists P \in v.A \text{ s.t. } B = B(P)\}$ be the set of base values of active labels of $v$.

We modify the lines 10-23 of Algorithm 5, and replace them by the following pseudo-code. All references to line numbers in the pseudo-code below refer to Algorithm 5.

1: **if** $((ttl(u) > 0)$ and $(B(P(u)) \notin B(v.A)))$ **then**
2:    execute lines 11-13 (of Algorithm 5)
3: **else**
4:    execute line 15
5:    **if** $(B(P(u)) \notin (v.M \cup B(v.A)))$ **then**
6:      execute lines 17-18
7:    **else**
8:      execute lines 20-21
9:    **end if**
10: **end if**

This modification of the algorithm has the following effect. Whenever a vertex $v$ scans an edge $e = (v, u)$ with the other endpoint $u$ belonging to one of the clusters to which $v$ belongs, the vertex $v$ records the edge $e$ as dropped. The rational of this modification is two-fold. First, it ensures that each vertex has at most one parent in each cluster it belongs to. Second, if either of the endpoints of $e$ will lose the label $P(u)$ in future, the fact that the edge $e$ is recorded as a dropped edge will make the vertex $v$ to re-scan $e$. Note that the edge $e = (v, u)$ as above can indeed be dropped "painlessly", because its endpoints belong to the same cluster, and thus, by Observation 8.3, the maintained spanner already contains a path of length at most $2t - 2$ between them.

This modification affects slightly, however, the size analysis of the constructed spanner. The change is in the argument that provides an upper bound on $|X^{(i)}(v)|$, for a fixed vertex $v$ and index $i$, $0 \le i \le t-2$. (See the proof of Lemma 3.4.) The subsequence $\sigma' = (P_{j_1}, P_{j_2}, \ldots, P_{j_\ell})$, for some $\ell \ge 0$, is now defined as the subsequence of the sequence $\sigma$ that contains only labels $P_{j_q} = P(u_{j_q})$ of level at most $t-2$ such that no other labels with the same base appear in $\sigma$ with an index small than $j_q$, *and such that $B(P_{j_q}) \notin v.A$*. Equipped with this modification, the proof argument of Lemma 3.4 becomes applicable, and the upper bounds of Corollaries 3.5 and 3.6 on the size of maintained spanner hold true for the modified algorithm as well.

To summarize, the properties of the original algorithm that were proved in previous sections are satisfied by the modified algorithm too, and, in addition, it guarantees the following property.

**Lemma 9.2** *For a base value $B$, the modified algorithm guarantees that the set $\tau_B = \{(v, u) \mid v.label(e) = P, B(P) = B\}$ of edges labeled by labels of base $B$ forms a spanning tree of the cluster $B$.*

**Proof:** This set is acyclic because each vertex in the cluster $B$ has at most one parent in $B$. By definition, it spans all the vertices in the cluster $B$. Let $x$ be the root of $B$. Given a vertex $v$ in $B$, let $P_v$ be the label of base $B$ in $v.A$. By a straightforward induction on the level $L(P_v)$ of $P_v$ it is easy to see that $\tau_B$ contains a path connecting the root $x$ with the vertex $v$. ∎

Finally, we introduce to the algorithm one more modification. Specifically, this modification makes sure that each vertex $v$ sends *some message* to each of its neighbors on every round of $v$. In other words, after finishing executing the lines 1-25 of Algorithm 5, the vertex $v$ will check for every edge $e = (v, u)$ adjacent to $v$ that satisfies $v.status(e) = SCANNED$ or $v.status(e) = NOTSCANNED$ whether $v$ has sent some message to $u$ over $e$ on the current round or it has not. In the latter case $v$ will send the message *VOID*. In addition, on each round the vertex $v$ will await receiving the messages from all its neighbors before processing. The messages *VOID* are discarded right after being received; they are used to maintain a certain level of synchronization between neighboring vertices.

## 9.2 The Procedures *Crash* and *Update_Label*

We next describe the Procedure *Crash*. The procedure accepts as input one parameter $P$ that satisfies $L(P) \ge 1$ (by Lemma 9.1). This parameter is the label of the edge $e$ adjacent to $v$ that super-crashed, and whose super-crash triggered the invocation of the Procedure *Crash*. The Procedure *Crash* starts with invoking the Procedure *Update_Label* with the same parameter $P$. The Procedure *Update_Label* returns a label $\check{P}$ that becomes the new label $P(v)$ of the vertex $v$. The new label $\check{P}$ is necessarily smaller than the label $P$, and moreover, $L(\check{P}) < L(P)$. In addition, $\check{P}$ is always taken from the set $v.A$ of active labels of $v$. Observe that since $L(P) \ge 1$, and since each vertex $z$ always has the label $I(z)$ of level 0 in its set $z.A$ of active labels, such a label $\check{P}$ always exists.

The Procedure *Update_Label* updates also the table $v.A$ of active labels of $v$. Specifically, the vertex $v$ loses all its active labels that are greater than $\check{P}$. We will describe the Procedure *Update_Label* in full detail later in the sequel. After returning from the invocation of the Procedure *Update_Label*, the Procedure *Crash* enters the for-loop. This for-loop updates the sets $v.T$ and $v.X$, as well as the statuses of some edges $e'$ adjacent to $v$. In addition, this for-loop propagates the message *CRASH* over some of the edges adajcent to $v$.

More specifically, the execution enters the for-loop with each edge $e' = (v, z)$ adjacent to $v$ that satisfies the following two conditions. First, the edge $e'$ is necessarily scanned, that is, $v.status(e') = SCANNED$.

Second, the edge $e'$ satisfies one of the following three conditions. Note that these conditions are mutually exclusive.

**Condition I:** The edge $e'$ is a $T$-edge and its label is (strictly) greater than the input parameter $P$ of the Procedure *Crash*.

**Condition II:** The edge $e' = (v, z)$ is either an $X$- or a $D$-edge, the owner of $e'$ is $z$, and its label is greater or equal to $P$.

**Condition III:** The edge $e' = (v, z)$ is either an $X$- or a $D$-edge, its owner is $v$, and its *second* label is greater or equal to $\check{P}$. (Recall that $\check{P}$ is the label returned by the Procedure *Update_Label*.)

If the execution enters the loop with an edge $e' = (v, z)$, it always sets the status of $e'$ (the variable $v.status(e')$) to $NOTSCANNED$, and sends the message $CRASH$ to $z$ over $e'$. In addition, if Condition I holds then the edge $e'$ is removed from the set $v.T$. If Condition II holds then no additional operations are performed. If Condition III holds then if the edge $e'$ is an $X$-edge, then the edge $e'$ is removed from the set $v.X$. Otherwise, if Condition III holds and the edge $e'$ is a $D$-edge, then $e'$ is removed from the set $v.M[B']$, where $B'$ is the cluster to which the vertex $z$ belonged at the time that the vertex $v$ scanned the edge $e'$. In addition, if the set $v.M[B']$ becomes empty as a result of the removal of the edge $e'$, then the value $B'$ is removed from the set $v.M$. The pseudo-code of the Procedure *Crash* follows.

---

**Algorithm 9** The Procedure $Crash(P)$.

1:  $\check{P} \leftarrow Update\_Label(P)$
2:  **for** every edge $e' = (v, z)$ such that $((v.status(e') = SCANNED)$ **and**
    I $(((v.scan\_status(e') = T)$ and $(v.label(e') > P))$ **or**
    II $((v.scan\_status(e') = X$ or $D)$ and $(v.label(e') \geq P)$ and $(v.own(e') = PEER))$ **or**
    III $((v.scan\_status(e') = X$ or $D)$ and $(v.sec\_label(e') \geq \check{P})$ and $(v.own(e') = SELF))))$ **do**
3:    send the message $CRASH$ to $z$ over $e'$
4:    $v.status(e') \leftarrow NOTSCANNED$
5:    **if** (Condition I) **then**
6:      $v.T \leftarrow v.T \setminus \{e'\}$
7:    **else if** (Condition III) **then**
8:      **if** $(v.scan\_status(e') = X)$ **then**
9:        $v.X \leftarrow v.X \setminus \{e'\}$
10:      **else**
11:        $B' \leftarrow B(v.label(e'))$; $v.M[B'] \leftarrow v.M[B'] \setminus \{e'\}$; $v.D \leftarrow v.D \setminus \{e'\}$
12:        **if** $(v.M[B'] = \emptyset)$ **then**
13:          $v.M \leftarrow v.M \setminus \{B'\}$
14:        **end if**
15:      **end if**
16:    **end if**
17: **end for**

---

To provide some intuition for the algorithm, we next explain which edges $e' = (v, z)$ satisfy one of the three conditions (Conditions I, II, and III), and why does the algorithm undertake certain actions for each of these edges.

**Condition I:** There are two types of edges $e' = (v, z)$, $v.label(e') > P$, that satisfy Condition I. Let $P' = v.label(e')$.

**First type:** In edges of the first type the vertex $z$ is the parent of the vertex $v$ in the cluster $B' = B(P')$,

$B' \neq B(P)$. See Figure 7.

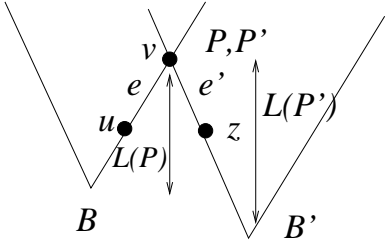

Figure 7: The edge $e'$ is an edge of the first type satisfying Condition I. Both edges $e$ and $e'$ are $T$-edges owned by $v$.

Edges of this type are characterized by $v.status(e') = SCANNED$, $v.own(e') = SELF$, $v.scan\_status(e') = T$, and $v.label(e') = P' > P$. Observe that as $P$ and $P'$ are both active labels of $v$ at this point of the execution, and $P' > P$, it follows that $L(P') > L(P)$. (Recall that a vertex may have at most one active label of any given level.)

In this case when the edge $e'$ was scanned, the vertex $v$ was already labeled by $P$. (In other words, $P$ belonged to the set $v.A$ at the time when $v$ scanned the edge $e'$. This follows since $v$ never acquires a label $P''$ that is smaller than one of its active labels.) Let $P(v)$ denote the label of $v$ at that time. It follows that $P(v) \geq P$. However, as a result of the current invocation of the Procedure $Crash$, the vertex $v$ loses the label $P$ and all its other active labels greater than $P$. Consequently, the edge $e'$ needs to be re-scanned.

To implement this intuition, the algorithm removes the edge $e'$ from the set $v.T$, switches the status of $e'$ to $NOTSCANNED$, and sends the message $CRASH$ to $z$ over $e'$. The latter two operations are done for all edges $e'$ for which the execution enters the loop (lines 3-4 of Algorithm 9). We will call these operations the *standard* operations.

**Second type:** In the edges of the second type the vertex $z$ is a child of the vertex $v$ either in the cluster $B = B(P)$, or in a cluster $B' = B(P')$, $P' > P$. See Figure 8.

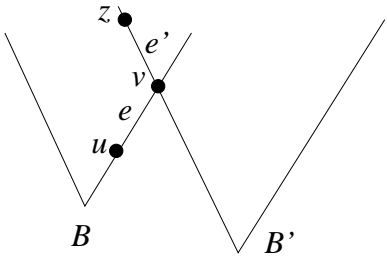

Figure 8: The edge $e'$ is an edge of the second type satisfying Condition I. Both edges $e$ and $e'$ are $T$-edges. The edge $e$ is owned by $v$, and the edge $e'$ is owned by $z$.

Edges of this type are characterized by $v.status(e') = SCANNED$, $v.scan\_status(e') = T$, $v.own(e') = PEER$, $v.label(e') = P' > P$. If $B(P') = B(P) = B$ then $z$ is a child of $v$, and $v$ is a child of $u$ in the cluster $B$, and thus $L(P')$ (respectively, $L(P)$) is the distance between the root $r$ of the cluster $B$ and the vertex $z$ (resp., $v$). Hence in this case $L(P') = L(P) + 1 > L(P)$.

If $B(P') \neq B(P)$ then the argument that shows that $L(P') > L(P)$ is slightly more elaborate. We first introduce the following notation. For a vertex $x$ and a cluster $\beta$ such that $x \in \beta$, let $P_\beta(x)$ denote the active label of the vertex $x$ that has base $\beta$.

Consider the labels $P_B(v)$ and $P_{B'}(v)$, $B = B(P)$, $B' = B(P')$. We note that $P_B(v) = P$, and $P' = P_{B'}(v) + n$. Both labels $P_B(v)$ and $P_{B'}(v)$ belong to the set $v.A$ of active lables of $v$ at the time when the vertex $v$ processes the super-crash of the edge $e$.

**Claim 9.3** $P_{B'}(v) > P_B(v) = P$.

**Proof:** Since $B \neq B'$, it follows that $P_{B'}(v) \neq P_B(v)$. Suppose for contradiction that $P_{B'}(v) < P_B(v)$. Since $P_{B'}(v)$ and $P_B(v)$ belong to the set $v.A$ simultaneously at some point fo the execution, it follows that $L(P_{B'}(v)) < L(P_B(v))$, and so $P_{B'}(v) \leq P_B(v) - n$. However, $P' = P_{B'}(v) + n$, and thus $P' \leq P_B(v) = P$. This is a contradiction to Condition I. ∎

Hence $L(P') = L(P_{B'}(v)) + 1 > L(P)$ in this case too. Since at the time when the edge $e'$ was scanned, the label of the vertex $v$ was $P_{B'}(v)$, it follows from Claim 9.3 that the label of $v$ at that time was greater or equal to $P$. (It was equal to $P$ if and only if $B = B'$.) Since now the vertex $v$ loses the label $P$, and all its other labels that are greater or equal to $P$, it follows that, in particular, it loses the label $P_{B'}(v)$. Hence the edge $e'$ needs to be re-scanned. To this end the algorithm undertakes the same operations as for edges of the first type that satisfy Condition I.

**Condition II:** There are two types of edges $e' = (v, z)$ that satisfy Condition II.

**First type:** Edges of the first type connect a neighbor $z$ of $v$ to one of the clusters $B'$ to which the vertex $v$ belongs. Moreover, for each edge $e'$ of this type, the label $P'$ of the edge $e'$ (i.e., $P' = v.label(e')$) satisfies $P' \geq P$. Also, the edge $e'$ is either an $X$- or a $D$-edge, and it is scanned by $z$. See Figure 9.

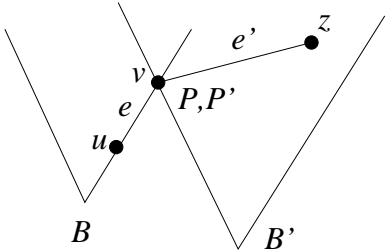

Figure 9: The edge $e'$ is an edge of the first type satisfying Condition II. The edge $e$ is a $T$-edge, and it is experiencing a super-crash. The edge $e'$ is an $X$- or a $D$-edge, owned by $z$, and its label $P' = v.label(e')$ is greater or equal than the label $P$ of $v$.

The edge $e'$ was scanned by the vertex $z$, and $z$ discovered that its current label is smaller than the label $P'$ of the vertex $v$. Currently, as a result of a super-crash of the edge $e$, the vertex $v$ loses both labels $P$ and $P'$. Hence the edge $e'$ has to be re-scanned. For this end the vertex $v$ sends to $z$ over the edge $e'$ the message $CRASH$, and sets the status of $e'$ to $NOTSCANNED$ (that is, performs the two standard operations).

We note also that the labels $P$ and $P'$ both belong to $v.A$ at the time when the edge $e$ super-crashes. Since $P' \geq P$ then either $P' = P$ or $L(P') > L(P)$.

**Second type:** Edges of the second type that satisfy Condition II are $D$-edges $e' = (v, z)$ that are scanned by $z$, and such that when the vertex $z$ scanned $e'$ it discovered that $B(P(v)) \in z.A$. As a result the edge $e'$ was dropped. Let $P(v)$ (respectively, $P(z)$) be the label of $v$ (resp., $z$) at the time when the vertex $z$ scanned the edge $e'$. Note that necessarily $P(z) \prec P(v)$. See Figure 10.

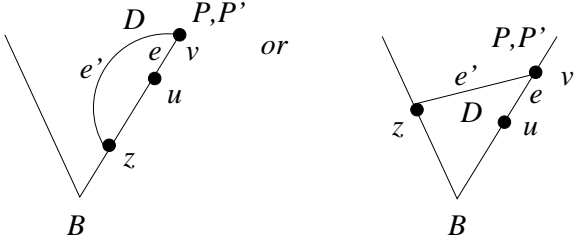

Figure 10: The two possible configurations of edges $e' = (v, z)$ of the second type that satisfy Condition II. In both configurations these are dropped edges connecting two endpoints $v$ and $z$ belonging to the same cluster $B$. In the first configuration $z$ is an ancestor of $v$ in $B$, and in the second configuration the edge $e'$ connects two different branches of the spanning tree of $B$, but $z$ is no farther from the root of $B$ than $v$.

Note that $v.label(e') = P(v) = P'$. Since $P(v) \geq P$, and because the vertex $v$ loses the label $P(v)$ as a result of a super-crash of the edge $e$, it follows that the edge $e'$ has to be re-scanned. To accomplish it, the two standard operations are undertaken in this case too.

The same argument as the one we used for the edges of the first type that satisfy Condition II shows that either $P' = P$ or $L(P') > L(P)$.

**Condition III:** Like for Conditions I and II, edges $e' = (v, z)$ that satisfy Condition III can be of two types.

**First type:** Edges $e'$ of the first type connect the vertex $v$ with a cluster $B_z$. These edges are either $X$- or $D$-edges. Also, edges $e' = (v, z)$ of this type were scanned by the vertex $v$ while the label of $v$ was $\check{P}$ or greater. Currently, as a result of the super-crash of the edge $e$, the label of $v$ decreases and becomes $\check{P}$. Consequently, the algorithm needs to re-scan edges of this type. The label of $v$ at the time of scanning the edge $e'$ is recorded in the variable $v.sec\_label(e')$. See Figure 11.

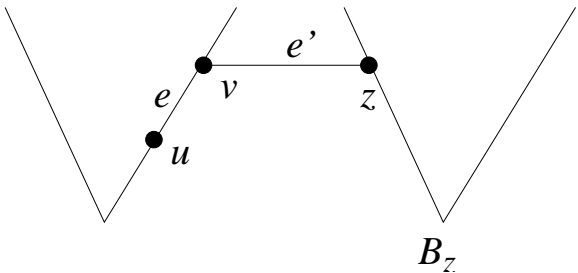

Figure 11: The edge $e'$ is an edge of the first type satisfying Condition III. It is either an $X$- or a $D$-edge owned by $v$. Its label is the label $P_z = P_{B_z}(z)$ of the vertex $z$ in the cluster $B_z$, and its second label is the label of $v$ at the time that $v$ scans the edge $e'$.

**Remark:** It may look counter-intuitive that the vertex $v$ needs to re-scan edges $e'$ that it scanned while having the label $\check{P}$, even though $v$ does not lose the label $\check{P}$ as a result of the super-crash of $e$. (The vertex $v$ loses all labels strictly greater than $\check{P}$.) The algorithm nevertheless re-scans these edges for the following reason. If an edge $e'$ were to stay in the set $v.X$, the cardinality of $v.X$ may have exceeded the desired bound of $O((t \cdot \log n)^{1-1/t} \cdot n^{1/t})$. Indeed, recall that $v.X = X(v) \supseteq \bigcup_{i=0}^{t-2} X^{(i)}(v)$. (See the notation of Section 3.2 and Lemma 3.4.) Let $i_0 = L(\check{P})$, and consider the set $X^{(i_0)}(v)$. Consider the sequence $\eta$ of all edges $\eta = (e_1, e_2, \ldots, e_k)$, for some integer $k \geq 0$, that arrived during the time period that the condition $L(P(v)) = i_0$ was satisfied, and such that the label $P(v)$ did not grow as a result of processing these edges. Note that it is not necessarily true that $X^{(i_0)}(v)$ contains only edges from the sequence $\eta$, and thus the proof argument of Lemma 3.4 would not carry over for our fully dynamic algorithm, unless edges $e'$ with second label $\check{P}$ are re-scanned.

We continue the analysis of edges of the first type satisfying Condition III. For an edge $e'$ of this type, its scan-status is either $X$ or $D$, its owner is $v$, and its second label is greater or equal to $\check{P}$. The label $v.label(e')$ is set to the label of $z$ at the time when the vertex $v$ scanned the edge $e'$. We denote this label by $P_z$.

For edges of this type the two standard operations are executed. In addition, if $e'$ is an $X$-edge then the edge is removed from the set $v.X$. Otherwise, the edge $e'$ is a $D$-edge, and in this case it is removed from the set $v.M[B_z]$, $B_z = B(v.label(e')) = B(P_z)$. Moreover, if as a result of removing the edge $e'$ from the set $v.M[B_z]$ the latter set becomes empty, then the value $B_z$ is removed from the set $v.M$.

**Second type:** The edges $e' = (v, z)$ of the second type that satisfy Condition III are $D$-edges that connect two vertices of the same cluster $B$. See Figures 12 and 13 for an illustration.

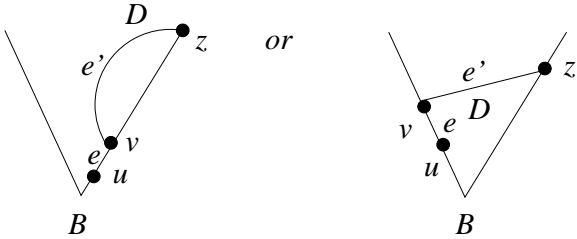

Figure 12: Two possible configurations of edges $e' = (v, z)$ of the second type that satisfy Condition III. In both configurations the edge $e'$ is a $D$-edge owned by $v$.

An edge $e' = (v, z)$ of this kind satisfies $v.label(e') = P(z)$, $v.sec\_label(e') = P(v)$, where $P(v)$ (respectively, $P(z)$) is the label of $v$ (resp., $z$) at the time that $v$ scans the edge $e'$. In addition, $v.sec\_label(e') \geq \check{P}$, and $v.own(e') = SELF$.

In this case the algorithm proceeds exactly in the same way as it does for edges of the first type that satisfy Condition III.

We remark also that in contrast to edges that satisfy Conditions I and II, for edges $e'$ that satisfy Condition III it is not necessarily true that $L(P') \geq L(P)$.

We next describe the Procedure *Update_Label*. Recall that this procedure is invoked from the Procedure *Crash*, and it accepts as input the input parameter $P$ of the Procedure *Crash*. The parameter $P$ is necessarily an active label of the host vertex $v$, i.e, $P \in v.A$, and moreover, $L(P) \geq 1$. (The host vertex of an algorithm is the vertex that runs the algorithm).

In the Procedure *Update_Label* the vertex $v$ loses the label $P$, as well as all its other active labels that are greater than $P$. It then adapts the largest label $\check{P}$ still left in $v.A$ as its current label $P(v)$. This label is returned by the procedure. Observe that each vertex $v$ always has the label $I(v)$ of level 0 present in $v.A$, and thus the set $v.A$ is never empty.

The pseudo-code of the Procedure *Update_Label* is presented below. The word *VOID* is used to denote the default value; if a value of a variable is equal to *VOID* it means that the variable contains no meaningful data.

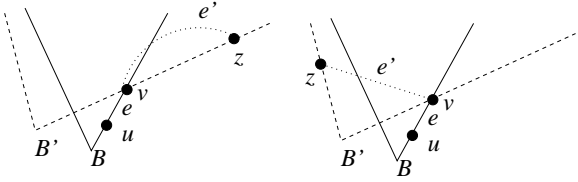

Figure 13: Two more possible configurations of edges $e' = (v, z)$ of the second type that satisfy Condition III. The cluster $B$ (respectively, $B'$) is depicted by solid (dashed) lines, and the edge $e'$ is depicted by dotted lines.
Like in Figure 12, the edge $e'$ is a $D$-edge owned by $v$, but in contrast to Figure 12, the edge $e' = (v, z)$ connects two endpoints from a cluster $B' \ne B$. Note that it may happen that $P > v.sec\_label(e') \ge \check{P}$.

**Algorithm 10** The Procedure $Update\_Label(P)$.
1: **for** every label $P' \in v.A$, $P' \ge P$ **do**
2:     $v.A[L(P')] \leftarrow VOID$
3: **end for**
4: $v.P \leftarrow \max\{P^* \mid P^* \in v.A\}$
5: $\text{return}(v.P)$

This completes the description of the fully dynamic algorithm for maintaining sparse spanners. We conclude this section with a number of remarks.

**Remark:** The algorithm can be simplified so that each $X$- or $D$-edge adjacent to a $T$-edge that super-crashes is re-scanned. This change does not affect the worst-case efficiency of our algorithm. However, it seems that in many cases this simplified version of the algorithm will re-scan much more edges than needed (or, rather, than the algorithm presented above re-scans).

**Remark:** Note that in the algorithm presented above a vertex $v$ may invoke the Procedure *Crash* many times on the same round. This may cause the vertex $v$ to send more than one message $CRASH$ through a fixed edge $e$ adjacent to $v$ on a fixed round. We observe, however, that messages $CRASH$ do not carry with them any additional information, and thus all these multiple messages can be replaced by just one single message $CRASH$ sent through each edge $e$ as above. A simple way to implement this is to maintain a variable $P_{\min}$ initialized as $\infty$. Each time $v$ wishes to invoke the Procedure *Crash* with a parameter $P$ it compares $P$ with $P_{\min}$, and sets $P_{\min}$ to be equal to the smaller value among these two values. At the very end of the crash-loop the vertex $v$ invokes the Procedure *Crash* with the parameter $P_{\min}$, unless the latter is equal to $\infty$. For convenience, we will ignore this issue in the analysis, and analyze the (simpler) version of the algorithm that is given by the pseudo-code.

**Remark:** Interestingly, a vertex $v$ may receive a message $(P(u), ttl(u))$ from a neighbor $u$ of $v$ over an edge $e = (v, u)$ on a round $R$ of $v$ on which it sends the message $CRASH$ to $u$ over $e$. We observe that this message $(P(u), ttl(u))$ was sent by $u$ before $u$ knew of the virtual crash of the edge $e$, and consequently, this message may contain irrelevant information. (Particularly, if $u$ will lose the label $P(u)$ on the next round as a result of the message $CRASH$ just sent by $v$.)

However, in our algorithm the vertex $v$ processes the message $(P(u), ttl(u))$ as usual, and does not discard it. As a result, it may happen that the record of the edge $e$ local at $v$ will contain not up-to-date information. The approach that the algorithm employs to handle this anomaly is based on the observation that when $u$ receives the message $CRASH$ from $v$, it realizes that the edge $e$ is scanned incorrectly, and informs $v$ of this fact.

**Remark:** A vertex $v$ may send the message $CRASH$ to its neighbor $u$ over the edge $e = (v, u)$ on a round $R$, and also send the message $(P(u), ttl(u))$ to $u$ over $e$ on the same round. Observe that once $v$ sends the message $CRASH$ over $e$ it sets its status to $NOTSCANNED$ (lines 3-4 of the Procedure *Crash*). Then on line 25 of the Procedure *Dyn_Rnd* (Algorithm 5; see also Algorithm 4) $v$ sends the message $(P(v), ttl(v))$ to $u$. Obviously, sending these two messages over the same edge on the same round causes no congestion problem, because the message $CRASH$ requires only a constant number of bits. These two messages will be sent as a single message, to make sure that they arrive at their destination simultaneously even in the asynchronous setting.

Similarly, a vertex $v$ may send the message $CRASH$ and a message $SCANNED$ to the same neighbor $u$ of $v$ on the same round $R$. This may happen if $v$ receives a message $(P(u), ttl(u))$ from $u$ on round $R$, and as a result, $v$ processes the edge $e = (v, u)$ in the scan-loop.

Note, however, that messages $SCANNED$ and $(P(v), ttl(v))$ are never sent over the same edge $e$ on the same round. To justify this claim we note that for $v$ to send a message $SCANNED$ over $e$, it must enter the scan-loop with $e$. However, if $v$ enters the scan-loop with $e$, when it ends the scan-loop the variable $v.status(e)$ is equal to $SCANNED$. However, the vertex $v$ sends the message $(P(v), ttl(v))$ only over those edges $e$ that satisfy that $v.status(e) = NOTSCANNED$ at the time when the scan-loop ends. (See line 25 of Algorithm 5, and line 14 of Algorithm 4.)

We next describe the behavior of a vertex $v$ that receives two messages over the same edge $e = (v, u)$ on the same round $R$ of $v$. Suppose first that $v$ receives the message $CRASH$ and the message $(P(u), ttl(u))$ on round $R$ of $v$. If $v.status(e) = SCANNED$ then, by lines 2 and 7 of Algorithm 5, the algorithm ignores the message $(P(u), ttl(u))$. If $v.status(e) = NOTSCANNED$ then, by lines 1-2 of Algorithm 7, the algorithm ignores the message $CRASH$. Specifically, it does not invoke the Procedure *Crash*, but rather sets $v.status(e)$ to $NOTSCANNED$ again, and proceeds to the scan-loop to process the message $(P(u), ttl(u))$.

The more complicated case is when the vertex $v$ receives the message $CRASH$ and the message $SCANNED(scan\_status, P, P')$ on the same round. As we have seen, in this case the vertex $u$ produced first the message $CRASH$ during its crash-loop, and later in the scan-loop, it produced the message $SCANNED$. In this case, for consistency, it is desirable that the algorithm local at the vertex $v$ will first process the message $CRASH$, and only then process the message $SCANNED$. For this end we need to change slightly the pseudo-code of Algorithm 5. Specifically, the for-loop of lines 3-5 will process only messages $SCANNED$ that arrived *alone*, that is, with no message $CRASH$ attached. In addition, right after the crash-loop the algorithm will execute the for-loop of lines 3-5 again, but this time only for edges $e$ over which both messages $CRASH$ and $SCANNED$ were received.

An alternative approach is to discard each message $(P(u), ttl(u))$ or $SCANNED$ that arrives in conjunction with the message $CRASH$, and to process only the message $CRASH$. It will be apparent from the analysis of Section 9.3 that the quiescence time of the algorithm is bounded by $3t$ even if some of the messages $(P(u), ttl(u))$ or $SCANNED$ fail to arrive. It is, however, crucial for all messages $CRASH$ that are sent to arrive, and also, it is important that the algorithm will not get stuck waiting for one of the missing messages $(P(u), ttl(u))$ or $SCANNED$. The latter does not happen if the "missing" message arrived in conjunction with the message $CRASH$, and was discarded.

## 9.3 Analysis

### 9.3.1 The Quiescence Time

In this section we analyze the quiescence time of the fully dynamic algorithm presented in the previous section.

The following lemma shows that a vertex cannot self-scan the same edge for two rounds in a row.

**Lemma 9.4** *For a vertex $v$, and an edge $e = (v, u)$ adjacent to $v$, and a round $R$ of $v$, the vertex $v$ cannot possibly self-scan $e$ on rounds $R$ and $R+1$ of $v$.*

**Proof:** If $v$ self-scans the edge $e = (v, u)$ on round $R$, it holds that $v.status(e)\{R, END\} = v.status(e)\{R+1\} = SCANNED$. For $v$ to self-scan the edge $e$ on round $R+1$ too, it must hold that $v.status(e)\{R+1\} = NOTSCANNED$, and thus the lemma follows. ∎

We next prove a technical lemma that asserts that the data structures of neighboring vertices are consistent to a certain extent. To formulate the lemma, the following notation is needed. For a vertex $v$ and a round $R$ of $v$, the time moment $\{v, R, CRASH\}$ is the time moment when the vertex $v$ starts the crash-loop of its round $R$. The notation $v.T\{R, CRASH\}$ stands for the value of the variable $v.T$ at time $\{v, R, CRASH\}$.

**Lemma 9.5** *Consider the synchronous setting, and an edge $e = (v, u)$. Suppose that $v.status(e)\{R, CRASH\} = SCANNED$, $v.scan\_status(e)\{R, CRASH\} = T$, $v.own(e)\{R, CRASH\} = SELF$, and that the vertex $v$ receives the message $CRASH$ from the vertex $u$ over $e$ on round $R$. Then $u.status(e)\{R-1, CRASH\} = SCANNED$, $u.scan\_status(e)\{R-1, CRASH\} = T$, $u.own(e)\{R-1, CRASH\} = PEER$, and $u.label(e)\{R-1, CRASH\} = v.label(e)\{R, CRASH\}$.*

**Proof:** Let $R'$ denote the maximum number of round, $R' \leq R$, on which the vertex $v$ self-scanned the edge $e$. We note that the vertex $v$ could not scan the edge $e$ on round $R$ before the crash-loop starts, because $v$ receives the message $CRASH$ over $e$ on round $R$. Hence, even if the vertex $v$ receives the message $CRASH$ along with a message $SCANNED$ over the edge $e$ on round $R$, nevertheless, $v$ enters the crash-loop before processing the message $SCANNED$.

Hence $R' \leq R - 1$. We next argue that $R' < R - 1$. Suppose for contradiction that $R' = R - 1$. In this case it follows that the vertex $v$ executes the following operations.

1. It receives the message $CRASH$ on round $R$ from $u$ over the edge $e$.

2. It self-scans the edge $e$ on round $R - 1$.

3. It receives a message $(P(u), ttl(u))$ on round $R - 1$ from $u$ over the edge $e$.

To justify the statement 3, we note that $v$ needs to receive a message $(P(u), ttl(u))$ on round $R - 1$ to self-scan the edge $e$ as a $T$-edge on round $R - 1$.

Also, it follows that the vertex $u$ executes the following operations.

1. It sends the message $CRASH$ to $v$ on round $R - 1$.

2. It sends the message $(P(u), ttl(u))$ to $v$ on round $R-2$. (This follows from the statement 3 regarding the vertex $v$.)

Also, since the vertex $u$ executes the operation 1, the vertex $u$ must enter the for-loop of the Procedure *Crash* with $e$ on round $R - 1$. Hence, $u.status(e)\{R - 1, CRASH\} = SCANNED$.

Since a message $(P(u), ttl(u))$ is sent by $u$ on line 25 of Algorithm 5, it follows that for the vertex $u$ to execute operation 2 it must satisfy $u.status(e)\{R-2, END\} = NOTSCANNED$. Since $u.status(e)\{R-2, END\} = u.status(e)\{R-1\}$, it follows that $u.status(e)\{R-1\} = NOTSCANNED$. However, $u.status(e)\{R-1, CRASH\} = SCANNED$. Note that no edge $e$ is ever *self-scanned* by a vertex $u$ between the beginning of a round and the beginning of the crash-loop of the same round. Hence, $u.own(e)\{R-1, CRASH\} = PEER$, i.e., the vertex $u$ peer-scanned the edge $e$ on round $R-1$. Hence the vertex $v$ self-scanned $e$ on round $R-2$, and sent a message $SCANNED$ to $u$ on round $R-2$. By the operation 2 of the vertex $v$, the vertex $v$ also self-scans $e$ on round $R-1$. By Lemma 9.4, this is a contradiction.

Hence $R' \le R - 2$. The vertex $v$ self-scans the edge $e$ on round $R'$, and sends the message $SCANNED(T, P(v), P(u))$ to $u$ on round $R'$. Hence, the vertex $u$ processes this message on lines 3-5 of Algorithm 5 of round $R' + 1 \le R - 1$, and it follows that $u.status(e)\{R'+1, CRASH\} = SCANNED$, $u.scan\_status\{R'+1, CRASH\} = T$, $u.own(e)\{R'+1, CRASH\} = PEER$, and $u.label(e)\{R'+1, CRASH\} = v.label(e)\{R, CRASH\}$. (For the last equality note that by the maximality of $R'$, $v.label(e)\{R', CRASH\} = v.label(e)\{R, CRASH\}$).

To complete the proof we argue that the record of the edge $e$ local at the vertex $u$ does not change between the time moments $\{u, R'+1, CRASH\}$ and $\{u, R-1, CRASH\}$. First, observe that if $R' = R-2$ then we are done. Hence we are left with the case that $R' \le R - 3$. However, since $e$ is a $T$-edge, for its record to change it must super-crash on a round $\tilde{R}$, $R' + 1 \le \tilde{R} < R - 1$. (Note that if $e$ super-crashes on round $R - 1$, it does so during the crash-loop, and so its super-crash has no effect on the value of $u.label(e)\{R-1, CRASH\}$.) If the edge $e$ super-crashes on round $\tilde{R}$, the vertex $v$ detects this super-crash on round $\hat{R}$, $\hat{R} \in \{\tilde{R}, \tilde{R}+1\}$. However, since by an assumption of the lemma, at time moment $\{v, R, CRASH\}$ the record of the edge $e$ local at $v$ indicates that the edge is scanned, it follows that $v$ scans the edge $e$ on a round $\check{R}$, $\hat{R} \le \check{R} \le R$. However, $R \ge \check{R} \ge \hat{R} \ge \tilde{R} \ge R' + 1$, i.e., $R \ge \hat{R} > R'$, contradicting the maximality of $R'$. ■

We next extend Lemma 9.5 to the asynchronous setting.

**Lemma 9.6** *Consider an asynchronous network, and an edge $e = (v, u)$. Suppose that at a certain time moment $\alpha_v$ it holds that $v.status(e)\{\alpha_v\} = SCANNED$, $v.scan\_status(e)\{\alpha_v\} = T$, $v.own(e)\{\alpha_v\} = SELF$, and that the vertex $v$ receives the message $CRASH$ from $u$ over $e$, and starts processing it at time $\alpha_v$. Let $\alpha_u$ be the time moment when the other endpoint $u$ of $e$ has sent the message $CRASH$ over $e$. Then $u.status(e)\{\alpha_u\} = SCANNED$, $u.scan\_status(e)\{\alpha_u\} = T$, $u.own(e)\{\alpha_u\} = PEER$, and $u.label(e)\{\alpha_u\} = v.label(e)\{\alpha_v\}$.*

**Proof:** Let $\beta$ be the maximum value such that $\beta \le \alpha_v$ and the vertex $v$ scanned the edge $e$ at time $\beta$. Let $\gamma$ be the maximum value such that $\gamma \le \alpha_v$ and the edge $e$ appears in the graph at time $\gamma$. If $\gamma > \beta$ then $v$ re-scanned the edge $e$ at time $\delta$, for some $\delta$, $\beta < \gamma \le \delta \le \alpha_v$, contradicting the maximality of $\beta$. Hence $\gamma \le \beta \le \alpha_v$.

Let $R_v$ (respectively, $R_u$) be the first round of $v$ (resp., $u$) that starts after time $\gamma$. The vertex $v$ (resp., $u$) sets the status of the edge $e$ to $NOTSCANNED$ on round $R_v$ (resp., $R_u$), and sends its first message to $u$ (resp., $v$) on its round $R_v$ (resp., $R_u$). The vertex $v$ (resp., $u$) receives the first message from $u$ (resp., $v$) on its round $R_v + 1$ (resp., $R_u + 1$). Note that by definition of $\gamma$, the edge $e$ does not crash really during the time interval $(\gamma, \alpha_v]$. Hence the vertices $v$ and $u$ correspond continuously during this time interval.

Let $\xi$ be the time moment on which the edge $e$ crashes really for the first time after the time moment $\gamma$. (It may happen that $\xi = \infty$ if $e$ never crashes really after time $\gamma$.) For a positive integer $j \ge 1$ we say that the vertex $v$ (resp., $u$) sends (or receives) a message to (or from) the vertex $u$ (resp., $v$) *on round $j$ with respect to the edge $e = (v, u)$ (after $\gamma$)*, if it does so on its round $R_v + j$ (resp., $R_u + j$), and it happens before the time moment $\xi$.

In this scenario, for a round $j$ with respect to the edge $e$, we denote by $\{v, e, j, CRASH\}$ the time moment on which the vertex $v$ is on round $j$ with respect the edge $e$, and the execution of the Procedure $Dyn\_Rnd$ local at $v$ is at the beginning of the crash-loop.

Let $j$, $j \geq 2$, denote the number of round with respect to the edge $e$ on which the vertex $v$ receives the message $CRASH$ from $u$ at time $\alpha_v$. In other words, using the notation that we have just defined, the time moment $\{v, e, j, CRASH\}$ is equal to the time moment $\alpha_v$. Following the proof argument of Lemma 9.5 we conclude that $u.status(e)\{u, e, j-1, CRASH\} = SCANNED$, $u.scan\_status(e)\{u, e, j-1, CRASH\} = T$, $u.own(e)\{u, e, j-1, CRASH\} = PEER$, and $u.label(e)\{u, e, j-1, CRASH\} = v.label(e)\{v, e, j, CRASH\}$. The time moment $\{u, e, j-1, CRASH\}$ is equal to $\alpha_u$, completing the proof. ∎

Note that in Lemma 9.6, $\alpha_v - 1 \leq \alpha_u \leq \alpha_v$.

We next use Lemma 9.6 to analyze the quiescence time of our algorithm.

**Lemma 9.7** *Let $\alpha$, $\alpha \geq 0$, be a time moment, and $e = (v, w)$ be an edge owned by $v$ that super-crashes at time $\alpha$. Let $L(e)$ denote the level of the label of $e$ local at $v$ at time $\alpha$, i.e., $L(e) = L(v.label(e)\{\alpha\})$. Then there exists an edge $\check{e}$ that crashes really at time $\beta$, $\beta \geq \alpha - (t-1)$. Moreover, if at time $\alpha$ the edge $e$ is a T-edge, then the same statement applies with $\beta \geq \alpha - (L(e) - 1)$.*

**Proof:** If the edge $e$ crashes really at time $\alpha$ then we are done. Hence we are left with the case that $e$ crashes virtually at time $\alpha$, that is, the endpoint $v$ switches the value of the variable $v.status(e)$ from $SCANNED$ to $NOTSCANNED$ at time $\alpha$.

Let $R = v.R\{\alpha\}$ denote the round on which the vertex $v$ is at time $\alpha$. Denote $\tilde{e}_1 = e = (z_0, z_1)$, $z_0 = u$, $z_1 = v$, $\alpha_1 = \alpha$. For the edge $e$ to crash virtually on round $R$ of $v$ there must exists a T-edge $\tilde{e}_2 = (z_1, z_2)$, $z_1 = v$, adjacent to $z_1$ and owned by $z_1$ that super-crashes on round $R$ of $z_1 = v$. If the edge $\tilde{e}_2$ crashes really on round $R$ of $z_1$ then we are done. Otherwise, $\tilde{e}_2$ crashes virtually on round $R$ of $z_1$. In other words, the vertex $z_2$ sends the message $CRASH$ to $z_1$ over the edge $\tilde{e}_2$, and $z_1$ receives this message at time $\alpha = \alpha_1$. Note that it may be the case that $e = \tilde{e}_1 = \tilde{e}_2$, and then the message $CRASH$ was received by the vertex $z_1$ over the edge $e$ itself, and $z_2 = z_0$. In any case, however, the message $CRASH$ was sent by $z_2$ to $z_1$ over $\tilde{e}_2$ at time $\alpha_2 \geq \alpha_1 - 1$.

First, we consider the case that $\tilde{e}_1 \neq \tilde{e}_2$ (and, consequently, $z_0 \neq z_2$). In this case $z_1.status(\tilde{e}_2)\{\alpha_1\} = SCANNED$, $z_1.scan\_status(\tilde{e}_2)\{\alpha_1\} = T$, $z_1.own(\tilde{e}_2)\{\alpha_1\} = SELF$. Let $\tilde{P}_2 = z_1.label(\tilde{e}_2)\{\alpha_1\}$. By Lemma 9.6, $z_2.status(\tilde{e}_2)\{\alpha_2\} = SCANNED$, $z_2.scan\_status(\tilde{e}_2)\{\alpha_2\} = T$, $z_2.own(\tilde{e}_2)\{\alpha_2\} = PEER$, and $z_2.label(\tilde{e}_2)\{\alpha_2\} = \tilde{P}_2$.

In addition, for the vertex $z_2$ to send the message $CRASH$ over a T-edge $\tilde{e}_2$ at time $\alpha_2$, it must have invoked the Procedure *Crash* with a parameter $P'_2$ on round $R_2 = z_2.R\{\alpha_2\}$. Moreover, since $z_2.scan\_status(\tilde{e}_2)\{\alpha_2\} = T$, by Condition I of the Procedure *Crash* it holds that $P'_2 < \tilde{P}_2$. Furthermore, by the argument that follows the description of Procedure *Crash* in Section 9.2, $L(P'_2) < L(\tilde{P}_2)$. (See the subsection devoted to Condition I.)

Since the vertex $z_2$ invokes the Procedure *Crash* with the parameter $P'_2$ at time $\alpha_2$, it follows that there exists an edge $\tilde{e}_3 = (z_2, z_3)$ that super-crashes at time $\alpha_2$. Moreover, $z_2.status(\tilde{e}_3)\{\alpha_2\} = SCANNED$, $z_2.scan\_status(\tilde{e}_3)\{\alpha_2\} = T$, $z_2.own(\tilde{e}_3)\{\alpha_2\} = SELF$, and $z_2.label(\tilde{e}_3)\{\alpha_2\} = P'_2$. We denote $\tilde{P}_3 = z_2.label(\tilde{e}_3)\{\alpha_2\}$, and more generally, $\tilde{P}_{j+1} = z_j.label(\tilde{e}_{j+1})\{\alpha_j\}$, for $j = 0, 1, \ldots, \ell-1$, where $\ell$ is a positive integer to be specified in the next paragraph.

By employing this argument inductively we conclude the following statements.

1. The vertex $z_1$ invokes the Procedure *Crash* at time $\alpha = \alpha_1$, and the edge $\tilde{e}_2 = (z_1, z_2)$ super-crashes at time $\alpha_1$.

2. There is a sequence of edges, called a *crash-path* of the edge $e = \tilde{e}_1$, $\sigma = (\tilde{e}_2, \tilde{e}_3, \ldots, \tilde{e}_\ell)$, $\ell \geq 2$, $\tilde{e}_j = (z_{j-1}, z_j)$ for $j = 2, 3, \ldots, \ell$, and a sequence of values $\alpha_1, \alpha_2, \ldots, \alpha_{\ell-1}$, $\alpha_1 > \alpha_2 > \cdots >$

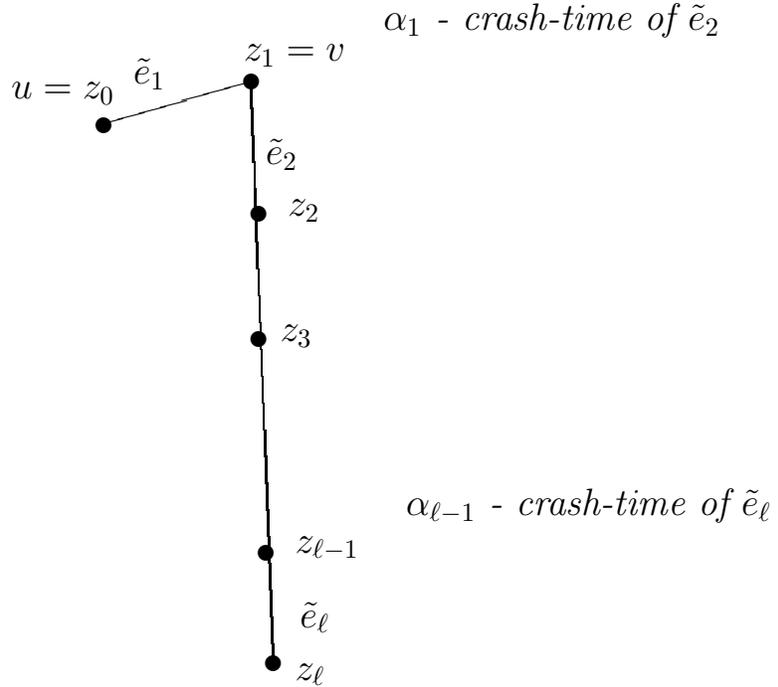

Figure 14: The crash-path of the edge $e = \tilde{e}_1$. The edge $\tilde{e}_\ell$ crashes really.

$\alpha_{\ell-1}$, $\alpha_{j-1} > \alpha_j \geq \alpha_{j-1} - 1$ for $j = 2, 3, \ldots, \ell$, such that $z_{j-1}.status(\tilde{e}_j)\{\alpha_{j-1}\} = SCANNED$, $z_{j-1}.scan\_status(\tilde{e}_j)\{\alpha_{j-1}\} = T$, $z_{j-1}.own(\tilde{e}_j)\{\alpha_{j-1}\} = SELF$, and each edge $\tilde{e}_j$ super-crashes at time $\alpha_{j-1}$, $j = 2, 3, \ldots, \ell$.

3. The edge $\tilde{e}_\ell$ crashes *really* at time $\alpha_{\ell-1}$.

See Figure 14 for an illustration.

It also holds that $L(\tilde{P}_2) > L(\tilde{P}_3) > \cdots > L(\tilde{P}_\ell)$, and $L(\tilde{P}_2) = L(z_1.label(\tilde{e}_2)\{\alpha_1\}) \leq t - 1$. Also, since $\tilde{P}_\ell = z_{\ell-1}.label(\tilde{e}_\ell)\{\alpha_{\ell-1}\}$, and $z_{\ell-1}$ is the owner of the $T$-edge $\tilde{e}_\ell$, it follows that $L(\tilde{P}_\ell) \geq 1$. Hence the sequence $(\tilde{P}_2, \tilde{P}_3, \ldots, \tilde{P}_\ell)$ is of length at most $t - 1$, and $\ell \leq t$. The edge $\check{e} = \tilde{e}_\ell$ crashes really at time $\beta = \alpha_{\ell-1} \geq \alpha - (\ell - 2) \geq \alpha - (t - 2)$.

Consider now the complementary case that $e = \tilde{e}_1 = \tilde{e}_2$. In this case the endpoint $z_2$ of $e$ sent the message $CRASH$ at time $\alpha_2 \geq \alpha_1 - 1$, and $z_2$ did it as a result of an invocation of the Procedure *Crash*, local to $z_2$ at time $\alpha_2$. Note that in this case the edge $\tilde{e}_2$ is not necessarily a $T$-edge at time $\alpha_2$. This is in contrast to the first case when the edge $\tilde{e}_2$ is always a $T$-edge.

For the vertex $z_2$ to invoke the Procedure *Crash* at time $\alpha_2$, there must exist an edge $\tilde{e}_3 = (z_2, z_3)$ adjacent to $z_2$ that super-crashes at time $\alpha_2$, owned by $z_2$, and $\tilde{e}_3$ is a $T$-edge at time $\alpha_2$. From this point on the same argument that we used in the first case applies for the $T$-edge $\tilde{e}_3$. The $T$-edge $\tilde{e}_3$ super-crashes at time $\alpha_2 \geq \alpha_1 - 1$, and consequently, the edge $\tilde{e}_\ell = \check{e}$ crashes really at time $\beta = \alpha_{\ell-1} \geq \alpha - (\ell - 2) \geq \alpha - (t - 1)$.

Moreover, suppose now that, in either of the two cases considered above, the edge $\tilde{e}_1$ is a $T$-edge. Then $\beta = \alpha_{\ell-1} \geq \alpha - (\ell - 2) \geq \alpha - (L(\tilde{P}_2) - 1)$. If $\tilde{e}_1 = \tilde{e}_2 = e$ then $\tilde{P}_2 = \tilde{P}_1$, and we are done. If $\tilde{e}_1 \neq \tilde{e}_2$ then by a previous argument, $L(\tilde{P}_1) > L(\tilde{P}_2)$, and $\beta \geq \alpha - (L(\tilde{P}_1) - 1)$. ∎

In the context of the last proof we remark that there may be more than one crash-path for a given super-crash of an edge $e$.

We need some additional terminology. Fix a time moment $\alpha$, and consider a vertex $w$ that belongs to a cluster $B$. Let $x \in V$ be the root of the cluster $B$. Then the *acquire-path* $\Pi_B(w)$ of the vertex $w$ in the cluster $B$ is the unique path connecting the vertices $x$ and $w$ in the spanning tree of the cluster $B$. At a time moment $\alpha$, the *time-stamp* of the acquire-path $\Pi_B(w)$ is the largest time moment no greater than $\alpha$ on which the vertex $w$ scanned the edge $e$ that is adjacent to $w$ and belongs to the path, and as a result the vertex $w$ joined the cluster $B$.

Recall also that for an edge set $H$ and an edge $e = (v, u)$, we say that $H$ *spans* $e$ if $dist_H(v, u) \le 2t - 1$. (See Section 5.3.)

**Lemma 9.8** *In an asynchronous network, for some time moment $\gamma$, suppose that no edge crashes really at time $\gamma$ or later. Suppose that for a time moment $\alpha$, $\alpha \ge \gamma + (t-1)$, it holds that $v.status(e)\{\alpha\} = SCANNED$, and $v.own(e)\{\alpha\} = SELF$. Then the following two statements hold.*

1. *The edge $e$ is spanned by $Sp\{\alpha\}$.*

2. *For any time moment $\delta$, $\delta > \alpha$, the edge $e$ is spanned by $Sp\{\delta\}$.*

**Proof:** Consider an edge $e = (v, w)$ that satisfies the assumption of the lemma. We start with proving statement 1.

If $e \in Sp\{\alpha\}$ then we are done. Otherwise, $v.scan\_status(e)\{\alpha\} = D$. Let $e' = (v, w') \in v.X\{\alpha\}$ be the edge that satisfies that $B(v.label(e)\{\alpha\}) = B(v.label(e')\{\alpha\})$. Let $P = v.label(e)\{\alpha\}$, $P' = v.label(e')\{\alpha\}$, and $B = B(P) = B(P')$. Such an edge exists by Lemma 8.5(4).

Let $\Pi_B(w) = (x = w_0, w_1, \dots, w_h = w)$, (for some $h \ge 0$, and $x$ such that $I(x) = B$), be the acquire-path of $w$ with respect to $B$ with the largest time-stamp smaller or equal than $\alpha$. Let $e_i = (w_i, w_{i+1})$ for $i = 0, 1, \dots, h-1$. The path $\Pi_B(w')$ is defined analogously, $\Pi_B(w') = (x = w'_0, w'_1, \dots, w'_g = w')$, for some $g \ge 0$, and $e'_i = (w'_i, w'_{i+1})$, for $i = 0, 1, \dots, g-1$.

Observe that all edges of the set $\{e\} \cup \Pi_B(w) \cup \Pi_B(w')$ were all inserted into the set $Sp$ on or before time $\alpha$, and thus one of them super-crashed no later than at time $\alpha$. Moreover, $e' \in Sp\{\alpha\}$, and thus the edge that super-crashed belongs to the set $\Pi_B(w) \cup \Pi_B(w')$.

For some index $i = 0, 1, \dots, h-1$, if the edge $e_i \in \Pi_B(v)$ super-crashes at time $\xi = \xi_i$, then $e_{i+1}$ super-crashes at time $\xi_{i+1} \le \xi_i + 1$, and the edge $e_{h-1}$ super-crashes at time $\xi_{h-1} \le \xi_i + (h-1-i) = \xi + (h-1-i)$. Furthermore, the edge $e$ super-crashes at time $\xi^* \le \xi + (h-i)$. See Figure 15 for an illustration.

Since $v.status(e)\{\alpha\} = SCANNED$ it follows that $\alpha < \xi^* = \xi + (h-i)$. Consequently, the edge $e_i$ super-crashes at time $\xi$, $\xi > \alpha - (h-i)$. The edge $e_i$ is a $T$-edge, and the level of $e_i$ at time it super-crashes is $L(e_i) = i + 1$. Hence, by Lemma 9.7, there exists an edge $\check{e}$ that crashes really at time $\eta$,

$$\eta \ge \xi - (L(e_i) - 1) = \xi - i > \alpha - h$$
$$\ge \alpha - (t-1) \ge \gamma \,.$$

However, no edge crashes really after time $\gamma$, contradiction. If an edge $e'_j \in \Pi_B(w')$, $j = 0, 1, \dots, g-1$, super-crashes, a contradiction is derived by the same argument (using that $v.status(e)\{\alpha\} = SCANNED$).

We next prove the statement 2 of the lemma. Consider an edge $e = (v, w)$ that satisfies the assumption of the lemma, and consider a time moment $\delta$, $\delta \ge \alpha \ge \gamma + (t-1)$. Note that the edge $e$ is owned by $v$ at time $\alpha$. The proof splits into a number of cases according to the value of $v.scan\_status(e)\{\alpha\}$.

In the first case $v.scan\_status(e)\{\alpha\}$ is equal to either $T$ or $X$. Hence, $e \in Sp\{\alpha\}$. The only way for $e$ to stop belonging to $Sp$ is to super-crash at time $\eta$, $\alpha < \eta \le \delta$. However, by Lemma 9.7, there exists an edge $\check{e}$ that crashes really at time $\eta - (t-1)$ or later. Since $\eta - (t-1) > \gamma$, this is a contradiction.

In the second case $v.scan\_status(e)\{\alpha\}$ is equal to $D$. By the statement 1 of this lemma, the edge $e$ is spanned by $Sp\{\alpha\}$. Assuming that $e$ is not spanned by $Sp\{\delta\}$, there must exist some edge $\tilde{e}$ in the network that belongs to the path connecting the endpoints of $e$ of length at most $2t-1$ in $Sp\{\alpha\}$, such that this edge ($\tilde{e}$) super-crashes at time $\eta$, $\alpha < \eta \leq \delta$. By Lemma 9.7, there exists an edge $\check{e}$ that crashes really at time $\eta - (t-1)$ or later. Since $\eta - (t-1) > \alpha - (t-1) = \gamma$, this is a contradiction. ∎

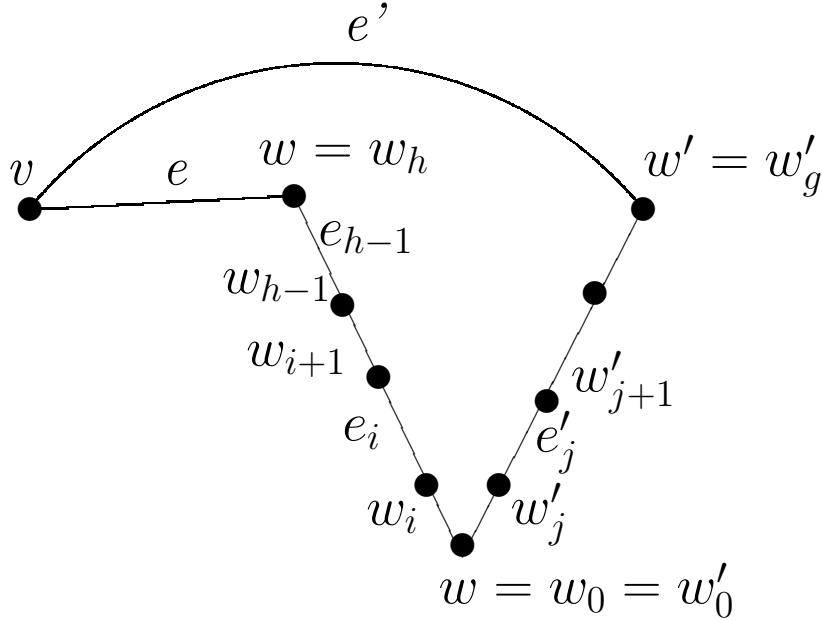

Figure 15: The paths $\Pi_B(w)$ and $\Pi_B(w')$.

We recall that the moment an edge $e$ crashes really is the moment when its owner $v$ switches its status from *SCANNED* to *CRASHED*. By one of the assumptions of the model, the time period that elapses between the moment that the edge $e$ crashes physically and crashes really is at most one time unit. (See the beginning of Section 9.1.)

Hence, the next lemma follows directly from Lemma 9.8

**Lemma 9.9** *In an asynchronous network, for some time moment $\gamma$, suppose that no edge crashes really at time $\gamma$ or later. Suppose that for a time moment $\alpha$, $\alpha \geq \gamma + t$, it holds that $v.status(e)\{\alpha\} = SCANNED$, $v.own(e)\{\alpha\} = SELF$. Then statements 1 and 2 of Lemma 9.8 hold.*

Analogously, the next lemma follows directly from Lemma 9.7.

**Lemma 9.10** *In an asynchronous network, for a time moment $\gamma$, suppose that no edge crashes physically on or after time $\gamma$. Then no edge super-crashes on or after time $\gamma + t$.*

The next lemma shows that the quiescence time of our algorithm is at most $3t$. For a time moment $\alpha$, recall that $E\{\alpha\}$ is the set of edges present in the network at this time.

**Lemma 9.11** *In an asynchronous network, for a time moment $\gamma$, suppose that no edge crashes physically at time $\gamma$ or later. Then for every time moment $\alpha$, $\alpha \geq \gamma + 3t$, every edge $e \in E\{\alpha\}$ is spanned by $Sp\{\alpha\}$.*

**Proof:** Fix a time moment $\alpha$, $\alpha \geq \gamma + 3t$, and consider an edge $e = (v, w) \in E\{\alpha\}$. Let $\delta$ be the maximum time moment such that $\delta \leq \alpha$ and the edge $e$ super-crashes at time $\delta$.

If $\delta \leq \gamma + 1$ then by the proof argument of Lemma 5.4, at time $\delta + 2t \leq \gamma + 2t + 1$ the edge $e$ is scanned. Hence, by Lemma 9.9, the edge $e$ is spanned starting with the time $\max\{\delta + 2t, \gamma + t\}$. Since $\max\{\delta + 2t, \gamma + t\} \leq \gamma + 2t + 1 \leq \alpha$, we are done.

Hence we are left with the case that $\delta > \gamma + 1$. In this case the edge $e$ crashes virtually at time $\delta$. (Otherwise, if $e$ crashes really at time $\delta$ then it crashes physically at time greater or equal to $\delta - 1 > \gamma$, contradiction.) Suppose without loss of generality that the endpoint $v$ is the owner of the edge $e$ at time $\delta$. It follows that the vertex $v$ receives the message $CRASH$ from $u$ at time $\delta > \gamma + 1$. By the proof argument of Lemma 5.4 and since the edge $e$ does not super-crash during the time interval $(\delta, \alpha)$, it follows that at time $\delta + 2t$ the edge $e$ is scanned. Thus, at time $\zeta = \max\{\delta + 2t, \gamma + t\}$ and later the edge $e$ is spanned. Moreover, by Lemma 9.10, $\delta \leq \gamma + t$, and thus $\zeta = \max\{\delta + 2t, \gamma + t\} \leq \gamma + 3t$. ∎

### 9.3.2 The Size of the Maintained Spanner

We next show that the spanner maintained by the algorithm is sparse. We will use the notation introduced in Section 8.3.1 for referring to various time moments. Specifically, for a vertex $v$ and a round $R$ of $v$, $\ell = \ell(v, R)$ (respectively, $k = k(v, R)$) is defined as the number of edges adjacent to $v$ that are processed by $v$ in the crash-loop (resp., scan-loop) of the round $R$ of $v$. For an index $i$, $1 \leq i \leq \ell + k$, the notation terms "$\{v, R, i\}$" and "$v.T\{R, i\}$" are defined exactly as in Section 8.3.1.

The following lemma shows that for every vertex $v$, the set $v.T$ of $T$-edges adjacent to $v$ is sparse.

**Lemma 9.12** *Consider a vertex $v$, a time moment $\{v, R, i\}$, and an edge $e' = (v, u')$ adjacent to $v$ such that $v.status(e')\{R, i\} = SCANNED$, $v.scan\_status(e')\{R, i\} = T$, $v.own(e')\{R, i\} = SELF$. Then the value $P' = v.label(e')\{R, i\}$ belongs to $v.A\{R, i\}$. (In other words, the label $P'$ is one of the active labels of the vertex $v$ at this time moment.) Moreover, $e'$ is the unique edge adjacent to $v$ with label $P'$ satisfying the conditions of the lemma.*

**Proof:** The proof is by induction on time moment $\{v, R, i\}$. At time $\{v, 1, 1\}$ there are no edges satisfying the assumptions of the lemma, and so the induction base holds vacuously.

Suppose that the lemma holds for a time moment $\{v, R, i\}$. Consider an edge $e' = (v, u')$ that satisfies the assumptions of the lemma for the time moment $\{v, R, i+1\}$. (We assume without loss of generality that $\{v, R, i+1\}$, and not $\{v, R+1, 1\}$, is the next time moment after $\{v, R, i\}$.) The analysis splits into two cases depending on whether the $i$th processed edge $e = (v, u)$ is processed in the crash-loop or in the scan-loop.

**Case 1:** The edge $e = (v, u)$ is processed in the crash-loop, that is, the index $i$ satisfies $1 \leq i \leq \ell$. The analysis splits into two subcases again, depending on whether $e = e'$ or not.

If $e = e'$ then as a part of the processing of the (super-crashing) $T$-edge $e$ on the $i$th iteration of the crash-loop, the variable $v.status(e')$ is set to either $NOTSCANNED$ or $CRASHED$ on line 4 of the Procedure *CrashItern* (Algorithm 7). Hence $v.status(e')\{R, i+1\} \neq SCANNED$, contradiction.

Hence we are left with the case that $e \neq e'$. By inspection of the pseudo-code of the Procedure *CrashItern* it is easy to verify that processing the edge $e$ may affect the set $v.A$ only if the edge $e$ is a $T$-edge, and consequently, the Procedures *Crash* and *Update_Label* are invoked. Let $P$ be the input parameter of the two procedures. Let $P' = v.label(e')\{R, i+1\}$.

If $P' < P$ then the processing has no effect on the record of the edge $e'$ local at $v$ (that is, on the variables $v.status(e')$, $v.scan\_status(e')$, $v.own(e')$, $v.label(e')$). Hence, in this case the edge $e'$ satisfies the assumption of the lemma at time $\{v, R, i\}$ as well, and by the induction hypothesis, $P' \in v.A\{R, i\}$. Since in this case the processing has no effect on the variable $v.A$ either, it follows that $v.A\{R, i+1\} = v.A\{R, i\} \ni P'$, as required. Also, the Procedure *Crash* could not possibly set the variables $v.status(e'')$, $v.scan\_status(e'')$, $v.own(e'')$, and $v.label(e'')$ to $SCANNED$, $T$, $SELF$, and $P'$, respectively, for some other edge $e'' \neq e'$, because the Procedure *Crash* never scans edges.

If $P' > P$ then the Condition I of the for-loop of the Procedure *Crash* is satisfied (see Algorithm 9), and so on line 4 the variable $v.status(e')$ is set to $NOTSCANNED$. Hence $v.status(e')\{R, i+1\} = NOTSCANNED$, contradicting the assumptions of the lemma at time $\{v, R, i+1\}$.

Finally, we are left with the case $P = P'$. However, in this case by the uniqueness part of the induction hypothesis, $e' = e$, contradiction.

**Case 2:** ($\ell < i \leq \ell + k$)
In this case the edge $e'$ is processed in the scan-loop. Consider first the case that $e' = e$ is the $i$th processed edge. In this case the first part of the lemma follows directly from lines 11-12 of the Procedure *Dyn_Rnd* (Algorithm 5). For the uniqueness part, suppose for contradiction that there is another edge $e''$ that satisfies $v.status(e'')\{R, i+1\} = SCANNED$, $v.scan\_status(e'')\{R, i+1\} = T$, $v.own(e'')\{R, i+1\} = SELF$, and $v.label(e'')\{R, i+1\} = P'$, and $e'' \neq e'$.

The $i$th iteration of the scan-loop could not possibly alter the record of the $T$-edge $e''$, and thus these variables had the same respective values at time $\{v, R, i\}$. Hence, by the induction hypothesis, $P' \in v.A\{R, i\}$. It follows that the label of $v$ at time $\{v, R, i\}$ is greater or equal to $P'$, i.e, $v.P\{R, i\} \geq P'$. However, on the $i$th iteration the scan-loop scans the $T$-edge $e'$, and as a result, the label of the vertex $v$ grows to $v.P\{R, i+1\} > v.P\{R, i\} \geq P'$. On the other hand, $v.P\{R, i+1\} = v.label(e')\{R, i+1\} = P'$, contradiction.

We are now left with the case that $e' \neq e$. In this case the first part of the lemma follows because processing the edge $e$ in the scan-loop cannot possibly alter one of the variables $v.status(e')$, $v.scan\_status(e')$, $v.own(e')$, or $v.label(e')$, for the $T$-edge $e'$, and neither could it cause $v$ to lose an active label $P'$. The uniqueness part follows by an argument similar to the one we use to prove the uniqueness in the case $e' = e$. ∎

Since for every vertex $v$ at any time the set $v.A$ contains at most one active label for each level, it contains overall at most $t$ active labels. Hence, by Lemma 9.12, $|v.T| \leq t - 1$ at any point of the execution. (Actually, Lemma 9.12 implies only the weaker bound of $|v.T| \leq t$, but it is easy to see that under the assumptions of the lemma, the label $P'$ has level at least 1.) We conclude that altogether the spanner contains at most $n \cdot (t-1)$ $T$-edges.

We next show that the set of $X$-edges is sparse as well. Consider a vertex $v$, and a time moment $\{v, R, i\}$, $i \leq \ell + k$, $\ell = \ell(v, R)$, $k = k(v, R)$. Let $v.A = v.A\{R, i\} = (P_0, P_1, \ldots, P_q)$, $P_0 < P_1 < \cdots < P_q$, be the sequence of active labels of the vertex $v$ at time $\{v, R, i\}$ in an increasing order.

Consider the vertex set $v.\hat{X}$ defined by the following process. Whenever the vertex $v$ $X$-scans an edge $e$ (in the scan-loop), this edge is inserted into the set $v.\hat{X}$. (At the same time the edge $e$ is inserted into the set $v.X$ too.) Whenever the vertex $v$ loses labels in the Procedure *Crash*, all edges $e$ that are removed from the set $v.X$ are removed from the set $v.\hat{X}$ as well. The difference between the two sets $v.X$ and $v.\hat{X}$ is that edges of $v.X$ may crash and be removed from $v.X$, and other (dropped) edges may be inserted into $v.X$ instead. These events have, however, no effect on the set $v.\hat{X}$. Consequently, there may be edges in $v.X$ that do not belong to $v.\hat{X}$, and edges that belong to $v.\hat{X}$ and do not belong to $v.X$. However, we next argue that the cardinality of the set $v.\hat{X}$ is always greater or equal to the cardinality of the set $v.X$. (See also Section 8.3.4 for an analogous argument.)

**Lemma 9.13** *At any time during an execution of the algorithm, $|v.\hat{X}| \geq |v.X|$.*

**Proof:** An edge $e'$ may be inserted into $v.X$ and not inserted into $v.\hat{X}$ only on line 11 of the Procedure *XReplace* (Algorithm 8). However, on this line some other edge $e$ is necessarily removed from $v.X$, and is not removed from $v.\hat{X}$, proving the lemma. ∎

Our strategy for proving upper bounds on $|v.X|$ is proving the desired bounds on $|v.\hat{X}|$.

Suppose that at a time moment $\{v, R, i\}$, $v.A = v.A\{R, i\} = (P_0, P_1, \ldots, P_q)$. For $j = 0, 1, \ldots, q$, let $v.\hat{X}_j\{R, i\}$ denote the subset of all edges $e$ of $v.\hat{X}\{R, i\}$ that satisfy $v.sec\_label(e) = P_j$. It is easy to see (by an induction on time moments $\{v, R, i\}$) that

$$v.\hat{X}\{R, i\} \;=\; \bigcup_{j=0}^{q} v.\hat{X}_j\{R, i\} \;.$$

We next show that each of the sets $v.\hat{X}_j\{R, i\}$ is sparse.

Suppose that the vertex $v$ acquired the label $P_j$ at time $\{v, R_j, i_j\}$, for some round $R_j$ of $v$ and index $i_j$ of a processed edge. Then the pair $\{R_j, i_j\}$ is called the *acquire-time* of the label $P_j$ (by the vertex $v$). For a vertex $v$, a time moment $\{v, R, i\}$, and an edge $e$, the *scan-time* of the edge $e$ (with respect to $\{v, R, i\}$) is the time moment when the vertex $v$ scanned the edge $e$ for the last time before the time $\{v, R, i\}$. Let $I_j = [\{R_j, i_j\}, \{R_{j+1}, i_{j+1}\})$ denote the time interval during which the vertex $v$ was labeled by $P_j$. The last interval $I_q$ is defined as $I_q = [\{R_q, i_q\}, \{R, i\})$. (The interval $I_j$ includes its left endpoint $\{R_j, i_j\}$, but does not include its right endpoint $\{R_{j+1}, i_{j+1}\}$. The right endpoint of the interval $I_q$ is the current time moment $\{v, R, i\}$.)

**Lemma 9.14** *For a vertex $v$, a round $R$ of $v$, an index $i$, $1 \leq i \leq \ell + k$, $\ell = \ell(v, R)$, $k = k(v, R)$, and an edge $e \in v.\hat{X}_j\{R, i\}$, the scan-time of $e$ belongs to the interval $I_j$.*

**Proof:** The proof is by induction on time moment $\{v, R, i\}$. The induction base $\{v, 1, 1\}$ holds vacuously. For the induction step suppose that $v.A\{R, i\} = (P_0, P_1, \ldots, P_q)$, and suppose that the statement of the lemma is true for $\{v, R, i\}$. We next prove it for $\{v, R, i+1\}$. (We assume without loss of generality that $\{v, R, i+1\}$ is the next time moment after $\{v, R, i\}$.)

The proof splits into two cases, according to the value of the index $i$.

**Case 1:** $(\ell + 1 \leq i \leq \ell + k)$
In this case between the time $\{v, R, i\}$ and $\{v, R, i+1\}$, the vertex $v$ scans an edge $e = (v, u)$. For the vertex $v$ to scan the edge $e = (v, u)$, it must hold that $P(u) \succ P(v)$, where $P(v)$ is the current label of $v$ (that is, $P(v) = P_q$), and $P(u)$ is the label of $u$ that $v$ has received from $u$ on its round $R$.

If the edge $e$ is either $T$- or $D$-scanned, then no new edge is inserted into $v.\hat{X}$, and we are done. Otherwise, if $e$ is $X$-scanned, then since $v.sec\_label(e)$ is set to $P(v) = P_q$ on line 15 of Algorithm 5, it follows that the edge $e$ joins the set $v.\hat{X}_q$. The scan-time of $e$, $\{v, R, i\}$, belongs to the interval $I_q = [\{R_q, i_q\}, \{R, i+1\})$, and we are done.

**Case 2:** $(1 \leq i \leq \ell)$
In this case between the time moments $\{v, R, i\}$ and $\{v, R, i+1\}$, the vertex $v$ processes a super-crash of an edge $e = (v, u)$. If $e$ is peer-scanned, or if it is either an $X$- or a $D$-edge, its super-crash has no effect on the set $v.\hat{X}$.

Hence we are left with the case that $e$ is a $T$-edge. In other words, we next show that an invocation $Crash(P^*)$ local at $v$, for some active label $P^* \in v.A$, does not ruin the induction invariant.

The Procedure *Crash* starts with removing from $v.A$ all labels greater or equal to $P^*$. Let $j$, $0 \leq j \leq q$, be the maximum index such that $P_j < P^*$ and $P_j \in v.A$. (Note that $P_j$ is equal to the label $\check{P}$ returned by the invocation of the Procedure *Update\_Label*.) The set $v.A$ becomes $v.A\{R, i+1\} = (P_0, P_1, \ldots, P_j)$. Moreover, the Procedure *Crash* removes from $v.X$ all $X$-edges owned by $v$ with a label greater or equal to $P_j$. (Recall the Condition III of the Procedure *Crash* (Algorithm 9).) By definition of $v.\hat{X}$, all these edges are also removed from the set $v.\hat{X}$. Finally, for an edge $e' \in v.\hat{X}_h$, for an index $h$, $0 \leq h < j$, the statement of the lemma holds by the induction hypothesis. ∎

We next show that the sets $v.\hat{X}_j$ are sparse.

**Lemma 9.15** *For a vertex $v$, a round $R$ of $v$, and an index $h$, $1 \leq h \leq \ell + k$, $\ell = \ell(v, R)$, $k = k(v, R)$, and an index $j$, $0 \leq j \leq q$, (such that $v.A\{R, h\} = (P_0, P_1, \ldots, P_q)$)*

$$|v.\hat{X}_j\{R, h\}| = O\left(\left(\frac{n}{t}\right)^{1/t} \cdot \log^{1-1/t} n\right),$$

*with high probability.*

**Proof:** By Lemma 9.14, $v.\hat{X}_j\{R, h\}$ is the set of edges inserted into $v.\hat{X}$ during the time interval $I_j$. In other words, edges of $v.\hat{X}_j$ were inserted there while the label of $v$ was equal to $P_j$. Let $\eta = (e_1 = (u_1, v), e_2 = (u_2, v), \ldots, e_k = (u_k, v))$ be the sequence of all edges inserted into $v.\hat{X}$ during the time interval $I_j$. Let $\sigma = (P^{(1)}, P^{(2)}, \ldots, P^{(k)})$, $P^{(i)} = P(u_i)$, $i \in [k]$, be the sequence of labels of the vertices $u_1, u_2, \ldots, u_k$. (The label $P^{(i)} = P(u_i)$ is the label of $u_i$ that $v$ received from $u_i$ right before scanning the edge $e_i$.)

The edge $e_i$ was inserted into $v.\hat{X}$ only if the base value $B^{(i)} = B(P^{(i)})$ appears in the sequence for the first time. Moreover, all the labels $P^{(i)}$ must be not selected. We can also assume that all these labels are of level $t - 2$ or smaller. To justify it recall that with high probability the overall number of labels of level $t - 1$ is $O\left(\frac{\log^{1-1/t} n}{t^{1/t}} \cdot n^{1/t}\right)$, and at most one edge is inserted into $v.\hat{X}$ for each such a label.

While the vertex $v$ was reading the edges of the sequence $\eta$, it encountered no edges $e' = (v, u')$ with $P(u') = P'$ being a selected label, and $P' \succ P_j$. In other words, $v$ encountered $k$ not selected labels with distinct base values *in a row*. However, for a positive constant $c$, the probability for this event with $k = c \cdot \left(\frac{n}{t}\right)^{1/t} \cdot \log^{1-1/t} n$ to hold is at most $\frac{1}{n^c}$, and so with high probability,

$$|v.\hat{X}_j\{R, h\}| = |\eta| = k = O\left(\left(\frac{n}{t}\right)^{1/t} \cdot \log^{1-1/t} n\right). \quad \blacksquare$$

The next corollary follows directly from Lemma 9.15 and Lemma 9.13.

**Corollary 9.16** *For a vertex $v$, and a round $R$ of $v$, and an index $h$, $1 \leq h \leq \ell + k$, $\ell = \ell(v, R)$, $k = k(v, R)$, with high probability,*

$$|v.X\{R, h\}| = O(n^{1/t} \cdot t^{1-1/t} \cdot \log^{1-1/t} n).$$

The next corollary shows that the overall set of $X$-edges used by the spanner maintained by our algorithm is sparse.

**Corollary 9.17** *For an asynchronous network, at a time moment $\alpha$, with high probability,*

$$\left|\bigcup_{v \in V} v.X\{\alpha\}\right| = O(n^{1+1/t} \cdot t^{1-1/t} \cdot \log^{1-1/t} n).$$

**Proof:** For a vertex $v$, let $R_v$ denote the round on which the vertex $v$ is at time $\alpha$, and $h_v$ denote the stage of the local execution of the vertex $v$ on round $R_v$ which $v$ executes at time $\alpha$. It follows that $v.X\{\alpha\} = v.X\{R_v, h_v\}$. By Corollary 9.16,

$$|v.X\{\alpha\}| = |v.X\{R_v, h_v\}| = O(n^{1/t} \cdot t^{1-1/t} \cdot \log^{1-1/t} n),$$

with high probability.

Hence, with high probability,

$$\left|\bigcup_{v\in V} v.X\{\alpha\}\right| = O(n^{1+1/t} \cdot t^{1-1/t} \cdot \log^{1-1/t} n) \ . \qquad \blacksquare$$

Since the set $\bigcup_{v\in V} v.T\{\alpha\}$ has cardinality at most $n \cdot (t-1)$, it follows that the set $Sp\{\alpha\} = \bigcup_{v\in V} v.T\{\alpha\} \cup \bigcup_{v\in V} v.X\{\alpha\}$ has, with high probability, $O(n^{1+1/t} \cdot t^{1-1/t} \cdot \log^{1-1/t} n)$ edges. We summarize the section with the following theorem.

**Theorem 9.18** *The presented algorithm is a fully dynamic distributed algorithm that maintains a $(2t-1)$-spanner for an unweighted graph in a synchronous or asynchronous network. The size of the maintained spanner is $O(n^{1+1/t} \cdot t^{1-1/t} \cdot \log^{1-1/t} n)$, with high probability over the coin tosses of the algorithm. The quiescence time of the algorithm is at most $3t$ time units.*

*Moreover, suppose that no edge crashes after time $\gamma$ (but edges are allowed to appear). Consider an edge $e$ present in the graph at some time $\beta$, $\beta \geq \gamma$. Then the spanner maintained by the algorithm provides stretch guarantee of $2t-1$ for this edge at time $\beta + 3t$ and later.*

*Furthermore, $t$ time units after the last hard crash all the results for the semi-decremental setting start to apply. Particularly, if the last hard crash occurs at time $\gamma$, and no edge crashes between time $\gamma$ and $\gamma + t$, then any edge present in the network at time $\beta \geq \gamma + t$ will be $(2t-1)$-spanned by the maintained spanner on or before time $\beta + 2t$.*

*In addition, if only soft crashes are allowed then the quiescence time of the algorithms is at most $2t$ (instead of $3t$) time units. Finally, if only incremental updates are allowed then every edge $e$ is $(2t-1)$-spanned by the spanner within at most $2t$ time units after it appears, even if incremental updates keep occurring in-between.*

We finish the section with a number of remarks.

**Remark:** The algorithm needs to know the number of vertices $n$ to compute the probability $p$ that it uses for selecting labels. However, it is easy to see that the same argument works if instead of $n$ we would use an upper bound $\hat{n} \geq n$ on the number of vertices. In this case the bound on the number of edges in the spanner would become $O(\hat{n}^{1+1/t} \cdot t^{1-1/t} \cdot \log^{1-1/t} \hat{n})$. If the estimate $\hat{n}$ is allowed to be smaller than $n$ then the bound becomes $O\left(\frac{n^2}{\hat{n}^{1-1/t}} \cdot t^{1-1/t} \cdot \log^{1-1/t} n\right)$.

We remark, however, that if one is interested in an $O(\log n)$-spanner then $p$ can be set to $1/2$, and, the algorithm does not need to know neither the number $n$ of vertices nor an estimate of $n$.

**Remark:** For a vertex $v$, and a label $P$ such that $v$ was labeled by $P$ during a time interval $I$ of an execution of our algorithm, let $v.\hat{X}(P, I)$ denote the set of edges $e = (v, u)$ that were inserted into $v.\hat{X}$ during this time interval.

Our previous argument shows that for a constant $c$, $c > 0$, with probability at least $1 - \frac{1}{n^c}$, it holds that $|v.\hat{X}(P, I)| \leq c \cdot \log^{1-1/t} n \cdot (n/t)^{1/t}$. Let $N$ denote the overall number of distinct triples $(v, P, I)$ such that $v$ was labeled by $P$ during the time interval $I$ on a certain execution $\varphi$ of the algorithm. (Moreover, for a triple $(v, P, I)$ to be counted, the interval $I$ must be maximal with respect to containment. Specifically, it must hold that there is no time interval $I'$ such that $I \subseteq I'$ and the triple $(v, P, I')$ satisfies the above condition as well.)

By union-bound it follows that the probability that the size of the spanner is at most $c \cdot \log^{1-1/t} n \cdot t^{1-1/t} \cdot n^{1+1/t}$ at every time moment during the execution $\varphi$ is at least $1 - \frac{N}{n^c}$. In the semi-decremental setting each vertex can be labeled by at most $t$ different labels during the execution $\varphi$, and so $N \leq n \cdot t$, and this probability is at least $1 - \frac{t}{n^{c-1}}$. In the fully dynamic setting (when hard crashes are allowed as well), the number $N$ may be larger than any function of $n$ and $t$.

Let $M$ denote the overall number of hard crashes occurring during the algorithm. Then the number $N$ of triples $(v, P, I)$ as above is at most $M \cdot n \cdot t$. (To justify this inequality we note that even if each hard crash makes each vertex to lose all its $t$ active labels, still there are at most $n \cdot t$ triples per crash.) Hence in this case the probability that the spanner has size at most $c \cdot \log^{1-1/t} n \cdot t^{1-1/t} \cdot n^{1+1/t}$ at every time moment during the execution is at least $1 - \frac{M}{n^{c-1/t}}$. Consequently, as long as the number $M$ of hard crashes during the execution is bounded by a polynomial in $n$, the spanner contains $O(\log^{1-1/t} n \cdot t^{1-1/t} \cdot n^{1+1/t})$ edges with high probability, where the constant hidden by the $O$-notation depends on the degree of this polynomial. Moreover, even if $M$ is quasi-polynomial in $n$ ($M = n^{\text{polylog }(n)}$), still the bound on the size of the spanner is $O(\text{polylog}(n) \cdot \log^{1-1/t} n \cdot t^{1-1/t} \cdot n^{1+1/t})$.

Finally, if a vertex $v$ detects that its set $v.X$ becomes too large as a result of either "poor luck" or too many hard crashes experienced in its neighborhood, $v$ may decide to toss fresh coins. Interestingly, the algorithm is sufficiently robust to ensure that tossing fresh coins does not require any synchronization between different vertices. Specifically, for the vertex $v$ to toss fresh coins it just crashes (actually, simulates a vertex crash), and emerges again. The network processes a crash of a vertex $x$ as crashes of all edges adjacent to $x$, and an appearance of a vertex $x$ as an appearance of all edges adjacent to it. (Here it is assumed that the vertex has a unique identifier. If this is not the case, the identifier is drawn at random from an appropriate probability distribution.)

**Remark:** In a weighted version of the algorithm each vertex $v$ actually runs in parallel $\log_{1+\epsilon} \hat{\omega}$ instances of the algorithm, one for each range of edges. (Recall that $\hat{\omega}$ is the aspect ratio of the network, that is, the ratio between the largest and the smallest edge weights.) If an edge $e$ changes weight then there are two possibilities. If it changes weight within the same weight range $[(1+\epsilon)^i, (1+\epsilon)^{i+1})$ for some integer $i$, then no action is needed. If its weight moves from the range $[(1+\epsilon)^i, (1+\epsilon)^{i+1})$ to the range $[(1+\epsilon)^j, (1+\epsilon)^{j+1})$ for some $j \neq i$, then the $i$th instance of the algorithm will consider this event as a crash of the edge $e$, and the $j$th instance of the algorithm will consider this event as an appearance of a new edge in the network.

## 10  Lower Bounds

In this section we show that for any $t$, any static algorithm for constructing sparse $(2t-1)$-spanners requires $\Omega(t)$ rounds even in the *synchronous* setting, even when arbitrarily large messages are allowed. We also show the same lower bound on the quiescence time of any algorithm that maintains sparse spanners in a distributed fully dynamic setting.

We need the following terminology. For a graph $G = (V, E)$, a vertex $v$, and a positive integer $i$, the *i-neighborhood* of $v$ in $G$ is the subgraph of $G$ induced by all the vertices at distance $i$ or less from $v$ in $G$. The length (either in terms of number of edges or vertices) of the longest cycle in a graph $G$ is called the *girth* of $G$.

Also, for a function $f : \mathbb{N} \to \mathbb{N}$, the notation $\tilde{O}(f(n))$ stands for $O(f(n) \cdot \text{polylog}(n))$.

The next lemma is central in our lower bounds.

**Lemma 10.1** *Let $g : \mathbb{N} \to \mathbb{N}$ be a monotone increasing function. For a positive integer $t$, suppose that for infinitely many positive integers $n$ there exist $n$-vertex graphs with girth greater or equal to $g(t)$ with $\Omega(n^{1+1/t})$ edges. Then any static distributed synchronous algorithm for constructing $(2t-1)$-spanners with $\tilde{O}(n^{1+1/t})$ edges for unweighted graphs requires at least $\left(\frac{g(t-1)}{2} - 1\right)$ rounds.*

**Proof:** Suppose that an algorithm $\Pi$ constructs a $(2t-1)$-spanner with $\tilde{O}(n^{1+1/t})$ edges for any input $n$-vertex graph $G$ within $\left(\frac{g(t-1)}{2} - 2\right)$ rounds. Choose a sufficiently large $n$ such that there exists an $n$-vertex graph $G'$ with $\Omega(n^{1+\frac{1}{t-1}})$ edges and with girth at least $g(t-1)$.

Consider an execution $\varphi'$ of the algorithm $\Pi$ on $G'$. For a vertex $v$ in $G'$, within $\left(\frac{g(t-1)}{2}-2\right)$ rounds the vertex $v$ is capable of collecting the topology of its $\left(\frac{g(t-1)}{2}-1\right)$-neighborhood in $G'$. (Note that in zero rounds, with no communication at all, the vertex $v$ knows which edges are adjacent to $v$.) Since the girth of $G'$ is at least $g(t-1)$, it follows that this neighborhood is a tree.

We next argue that based on this information the vertex $v$ has to include all edges adjacent to $v$ in the constructed spanner. Suppose for contradiction that this is not the case. Consider an execution $\varphi$ of the algorithm $\Pi$ on a tree $\tau$ in which the $\left(\frac{g(t-1)}{2}-1\right)$-neighborhood of the vertex $v$ is isomorphic to its neighborhood in $G'$. Hence the vertex $v$ behaves identically in the two executions $\varphi$ and $\varphi'$, and does not include in the spanner one or more of the edges adjacent to it in $\tau$. However, it follows that the spanner that the algorithm $\Pi$ constructs for the tree $\tau$ is not connected, contradiction.

Hence the spanner constructed by the algorithm $\Pi$ for the graph $G'$ contains $\Omega(n^{1+\frac{1}{t-1}})$ edges. This is a contradiction to the assumption that the algorithm constructs spanners of size $\tilde{O}(n^{1+1/t})$. ∎

It is easy to verify that Lemma 10.1 extends to randomized algorithms that (always) produce correct $(2t-1)$-spanners, but provide a guarantee on the size of constructed spanners only *in expectation*, or *with high probability*. (Our algorithm is of this kind.)

Under the Erdős girth conjecture, $g(t) = 2t+2$. Hence our lower bound is $\frac{g(t-1)}{2} - 1 = t-1$. The Erdős girth conjecture is proven for the values 1,2,3, and 5 of the parameter $(t-1)$. Hence the lower bound $(t-1)$ holds unconditionally for $t = 2, 3, 4$, and 6. Using the bounds on $g(t)$ for small values of $t$ due to [36, 19, 37] it is easy to verify that for values $t = 5, 7, 8, 9$, and 10 our lower bound $\left(\frac{g(t-1)}{2} - 1\right)$ is equal to $3, 5, 5, 6$, and 7, respectively.

Also, Lazebnik et al. [27] (based on the construction of Lubotzky et al. [28]) showed that for $t-1 = 3(r-1)$, $r \geq 5$, $g(t-1) = 4r$, and for $t-1 = 3r-1$, $r \geq 4$, $g(t-1) = 4r+2$. It follows that in the first case the lower bound is $\frac{2}{3}t + \frac{1}{3}$, and in the second case it is $\frac{2}{3}t$. Overall, the lower bound of $\lfloor \frac{2}{3}t \rfloor$ applies for all values of $t$, $t \geq 2$.

We summarize these lower bounds with the next theorem.

**Theorem 10.2** *For an integer constant $t \geq 2$, any static distributed synchronous algorithm for constructing $(2t-1)$-spanners with $\tilde{O}(n^{1+1/t})$ edges for unweighted $n$-vertex graphs requires at least $\lfloor \frac{2}{3}t \rfloor$ rounds. Moreover, slightly stronger bounds hold for some small constant values of $t$. Finally, under the Erdős girth conjecture, the lower bound becomes $t-1$.*

Note that in a fully dynamic setting all edges may crash at once, and a completely new set of edges may show up on instantly. Since a fully dynamic algorithm is supposed to handle such a situation efficiently (and our algorithm indeed does handle it within $3 \cdot t$ time units), it follows that the lower bound of Theorem 10.2 applies to the fully dynamic algorithms as well.

Like Lemma 10.1, Theorem 10.2 applies for randomized algorithms that always provide a correct spanner, but provide a guarantee on the size of the constructed spanner either *in expectation* or *with high probability*.

Note that the analysis above extends for super-constant values of $t$. Specifically, the lower bound applies as is as long as $1 \leq t = o\left(\sqrt{\frac{\log n}{\log \log n}}\right)$. A somewhat weaker lower bound $\Omega(t)$ applies to a wider range of parameter $t$, $1 \leq t = O\left(\frac{\log n}{\log \log n}\right)$. To verify it, suppose for contradiction that there exists an algorithm $\Pi$ that constructs $(2t-1)$-spanners of expected size $O(t \cdot n^{1+1/t})$ in $\alpha \cdot t$ rounds, for a sufficiently small constant $\alpha > 0$. Consider a graph $G$ of girth $2\alpha \cdot t + 4$ with $\Omega(n^{1+\frac{1}{\alpha t}})$ edges. By a previous argument, the algorithm $\Pi$ returns the entire edge set of $G$ as its $(2t-1)$-spanner, and thus constructs a spanner of size $\Omega(n^{1+\frac{1}{\alpha t}})$. However, $n^{\frac{1}{\alpha t}} > t \cdot n^{1/t}$ for a sufficiently small positive $\alpha$, and $t = O\left(\frac{\log n}{\log \log n}\right)$. Moreover, as long as $t = o\left(\frac{\log n}{\log \log n}\right)$, the lower bound is actually $(\frac{2}{3} - o(1))t$.

Finally, in the complementary range of $t$, $\omega\left(\frac{\log n}{\log \log n}\right) = t = O(\log n)$, the lower bound of $\Omega\left(\frac{\log n}{\log \log n}\right)$ follows by the same argument.

## 11 Open Questions

We present a sample of open questions that are related to our study.

1. Perhaps the most obvious open question is to close the constant gap between our upper and lower bounds of $3t$ and $\frac{2t}{3}$, respectively, on the best possible quiescence time of a dynamic algorithm for maintaining sparse spanners. Also, the lower bound of $\frac{2t}{3}$ applies only for $t = o\left(\frac{\log n}{\log \log n}\right)$, and it is desirable to extend it to the entire range of $t$, $1 \le t = O(\log n)$.

   One way to do it is to improve our bound on the expected size of constructed spanners from $O(t \cdot n^{1+1/t})$ to $O(n^{1+1/t})$.

2. Another avenue is to weaken the assumptions on computational model that are needed for our result. Specifically, in this paper the adversary that determines the topology updates is assumed to be non-adaptive and oblivious to the coin tosses of the algorithm. Devising an algorithm with properties similar to that of our algorithm in the presence of a more powerful adversary appears to be a challenging problem.

3. Our algorithm is randomized, and randomization is crucial for our analysis. Derandomizing our result is a challenging problem as well.

4. To this day there are only a handful of distributed algorithms for basic graph-theoretic problems that are provably applicable to a dynamic environment. We hope that our study will stimulate further development of this important area. Particularly, it would be very interesting to devise dynamic distributed algorithms for such basic problems as *MST*, *MIS*, and vertex cover.

## Acknowledgements

The author is grateful to Shlomi Dolev, Joan Feigenbaum, David Peleg, Seth Pettie, and Jian Zhang for helpful discussions. David Peleg introduced the author to the problem of devising a randomized distributed (static) algorithm for constructing sparse spanners some eight years, addressed the inquiries of the author regarding the paper [11], as well as some bibliographical inquiries. Joan Feigenbaum introduced the author to the streaming model of computation, and the author wishes to thank her for very inspiring discussions on the subject. Shlomi Dolev introduced the author to the problem of devising a self-stabilizing algorithm for maintaining sparse spanners. This problem, though not solved in this paper, provided the author a motivation to work on the problem in dynamic distributed models.## References

[1] Y. Afek, B. Awerbuch, and E. Gafni. Applying static network protocols to dynamic networks. In *Proc. 28th Symp. on Foundations of Computer Science*, pages 358–370, 1987.

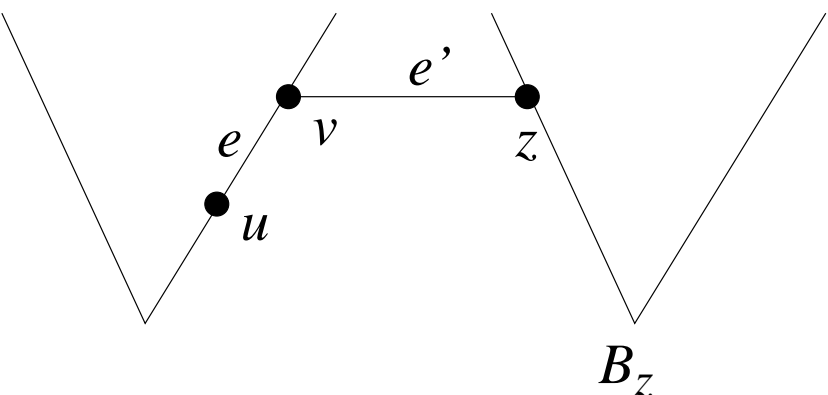